\documentclass[prx,twocolumn,english,
,superscriptaddress,nofootinbib,
floatfix,longbibliography]{revtex4-2}

\pdfoutput=1
\usepackage[utf8]{inputenc}
\usepackage[T1]{fontenc}
\usepackage{xcolor}
\usepackage{graphicx,enumerate,verbatim,bbold}
\usepackage{amsmath,amssymb,amsthm,float,mathrsfs}
\usepackage{dsfont}
\usepackage[normalem]{ulem}
\usepackage{lmodern}
\usepackage{hyperref}
\hypersetup{colorlinks,citecolor=blue,linkcolor=blue,urlcolor=blue}
\usepackage{braket}
\usepackage{multirow}
\usepackage{bbm}
\usepackage{physics}
\usepackage{bm}
\usepackage{tikz}
\usetikzlibrary{patterns}
\usetikzlibrary{fadings}
\usepackage[caption=false]{subfig}
\usepackage{dcolumn}
\usepackage{comment}
\usepackage{xspace}
\usepackage{makecell}

\newcommand{\bqa}{\begin{eqnarray}}
\newcommand{\eqa}{\end{eqnarray}}
\newcommand{\be}{\begin{equation}}
\newcommand{\ee}{\end{equation}}

\newcommand{\ctrlRG}{Ctrl$_\text{s}$RG\xspace}

\newcommand{\Utarget}{U_{\ast}}
\newcommand{\Useed}{U_s}
\newcommand{\dotUseed}{\dot{U}_s}
\newcommand{\Hseed}{H_s}
\newcommand{\UV}{U_V}
\newcommand{\HV}{H_V}
\newcommand{\Hacc}{H_{\rm acc}}
\newcommand{\Uacc}{U_{\rm acc}}
\newcommand{\Pacc}{\mathcal{P}}
\newcommand{\Pper}{{\mathcal P}_\perp}
\newcommand{\Kc}{K_{\rm c}}
\newcommand{\Rreg}{\mathcal{R}}
\newcommand{\Hper}{H_{\perp}}
\newcommand{\dotHper}{\dot{H}_{\perp}}
\newcommand{\Wnorm}{w_{\perp}}
\newcommand{\muf}{\mu_f}
\newcommand{\ipt}[2]{\langle #1,#2\rangle_t}
\newcommand{\Dec}{\mathcal{D}}
\newcommand{\Decmode}[1]{D_{#1}}
\newcommand{\Decval}[1]{d_{#1}} 
\newcommand{\Green}{G}
\newcommand{\Liou}{\mathcal{L}}

\newcommand{\hseed}{h_s}
\newcommand{\hacc}{h_{\rm acc}}
\newcommand{\hNN}{h_{\rm NN}}
\newcommand{\Unn}{U_{\rm NN}}
\newcommand{\epsU}{\epsilon_U}
\newcommand{\nc}{n_{\rm c}}
\newcommand{\Fnorm}[1]{\left\lVert #1\right\rVert_{\rm F}}

\newcommand{\HTFIM}{H_{\rm TFIM}}
\newcommand{\HCD}{H_{\rm CD}}
\newcommand{\HFF}{H_{\rm FF}}
\newcommand{\agp}{\mathcal{A}}
\newcommand{\epsad}{\epsilon_{\rm ad}}
\newcommand{\Facc}{F_{\rm acc}}

\newcommand{\Imperial}{Blackett Laboratory, Imperial College London, SW7 2AZ, United Kingdom}
\newcommand{\PKS}{Max Planck Institute for the Physics of Complex Systems, MPI-PKS, N\"othnitzerstr.~38, 01187 Dresden, Germany}
\newcommand{\Qolumbus}{Qolumbus SAS, 78 Boulevard de la République, 92100 Boulogne-Billancourt, France}
\newcommand{\SCNU}{Institute for Theoretical Physics, School of Physics, South China Normal University, Guangzhou 510006, China}
\newcommand{\SCNULAB}{Guangdong Provincial Key Laboratory of Quantum Engineering and Quantum Materials, Guangdong-Hong Kong Joint Laboratory of Quantum Matter, South China Normal University, Guangzhou 510006, China}

\begin{document}

\title{Quantum control as renormalization group flow}

\author{Zuo Wang}\email{zuowang@pks.mpg.de}\affiliation{\PKS}\affiliation{\SCNU}\affiliation{\SCNULAB}
\author{Paul M.~Schindler}\affiliation{\PKS}
\author{Modesto Orozco-Ruiz}\affiliation{\Imperial}
\author{Nguyen H. Le}\affiliation{\Imperial}\affiliation{\Qolumbus}
\author{Florian Mintert}\affiliation{\Imperial}
\author{Marin Bukov}\email{mgbukov@pks.mpg.de}\affiliation{\PKS}

\begin{abstract}

The Renormalization Group (RG) establishes relations between effective descriptions of physical systems at different scales; in Hamiltonian RG, these relations induce a flow in the space of Hamiltonians. 
In direct correspondence, we design an RG framework for quantum control by constructing a flow in the space of time-dependent Hamiltonians. 
A prescribed control subspace, selected to contain only experimentally accessible terms, forms the fixed-point manifold, while the flow preserves the target unitary and monotonically reduces the Hamiltonian weight outside this subspace. 
Exploiting the freedom in constructing the flow, we introduce flow-equation engineering to enhance the versatility of the framework for diverse control problems, including time-periodic and open-interval smooth protocols with prescribed endpoints.
We provide two disparate applications to exhibit the utility of our approach: 
(i) Nonperturbative Floquet engineering beyond the regime of validity of high-frequency expansions, which we demonstrate by constructing an implementable time-periodic protocol for the Haldane model; 
(ii) Accessible fast-forward driving from unit-fidelity counterdiabatic evolution, which we showcase for large transverse-field Ising chains away from the adiabatic regime by compressing the long-range multibody adiabatic gauge potential into the original time-dependent Ising control schedules.
Our control RG framework reveals a unifying perspective on Hamiltonian engineering and unitary quantum control, and can serve as an organizing principle behind seemingly disparate nonequilibrium quantum dynamics.

\end{abstract}

\maketitle

\section{Introduction}
\label{sec:intro}

The Renormalization Group (RG) establishes a relation between effective models of physical systems at different scales. 
In Hamiltonian RG, the coarse-graining of degrees of freedom alternates with a rescaling step; within an appropriate truncation, this recovers a structurally similar Hamiltonian, while tracking or discarding interactions with renormalized couplings. In the continuum limit, this irreversible procedure defines a nonlinear map on the space of Hamiltonians inducing the so-called RG flow~\cite{fisher1998renormalization}. 
Its fixed points are scale-invariant; linearizing the map around them identifies relevant, marginal, and irrelevant directions, and defines the concept of universality~\cite{wilson1975renormalization}.  
Here, we construct an RG flow over control Hamiltonians: a prescribed set of retained controls defines its fixed-point manifold, while the flow decimates inaccessible controls and preserves a target unitary.

We use a unitary flow~\cite{wegner1994flow,Glazek_Wilson_1994_PRD} that reorganizes the description of quantum systems by rotating away unwanted terms in the Hamiltonian.
Unlike Hamiltonian RG, this flow is determined by its fixed-point properties: the infinitesimal generator is characterized by the structure it decimates. Hence, the weight of the unwanted terms decays monotonically under the flow. 
A common application is Hamiltonian diagonalization, where unitary flows remove matrix elements in order of their energy separation, until the diagonal fixed point is reached~\cite{mielke1998flow,Kehrein_2006_book,long2023beyond,hahn2025predicting}. To leading order, the linearized flow reduces to the perturbative adiabatic elimination~\cite{brion2007adiabatic} and Schrieffer-Wolff transformations~\cite{Schrieffer_Wolff_1966_PR}. 

In dynamical systems, unitary flows have been used to reach an (almost) diagonal basis where time evolution reduces to phase accumulation; the inverse flow then maps the time-evolved observables backwards to the original basis~\cite{kehrein2005scaling,hackl2009unitary,thomson2020dynamics}. This approach treats physical time as a parameter independent of the flow.
In turn, stopping the flow early can exhibit timescale separation: partially converged flows define quasiconserved effective Hamiltonians whose validity is set by the residual off-diagonal weight, leading to prethermalization~\cite{berges2004prethermalization,moeckel2008interaction,moeckel2009real,abanin2017effective,Mori_2016_AP}. 
For periodically driven (Floquet) systems, the explicit time dependence becomes an interaction in the quasienergy operator~\cite{sambe1973steady}, which the flow gradually removes; requiring the fixed point to be simultaneously block-diagonal and time-independent, the original time-periodic Hamiltonian flows to its Floquet counterpart~\cite{Verdeny_etal_2013_PRL,Vogl_etal_2019_PRX,Novicenko_etal_2022_PRA,Thomson_etal_2021_SciPost,Claassen_2021_arXiv,Shi_Mintert_2024_PRR}. Hence, the flow parameter couples to physical time. 
In all of these approaches, the flow eliminates the time dependence: its fixed point is a time-independent effective Hamiltonian.

By contrast, unitary quantum control exhibits inherent time dependence, and hence a flow that converges to a time-independent fixed point cannot capture it.
Nonetheless, the unitary flow formalism allows the fixed point to be specified in advance, providing a framework to address existing open challenges in accessible quantum control.

\begin{figure}[t!]
\centering
\includegraphics{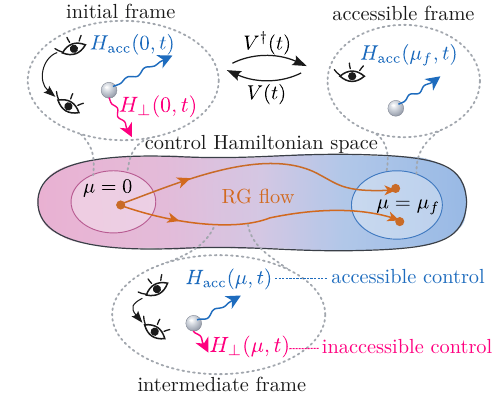}
\caption{
\textbf{Renormalization group flow in the space of control Hamiltonians.} 
The middle oval shows the space of time-dependent control Hamiltonians, shaded from the initial seed at $\mu=0$ (pink) to the subspace of purely accessible controls at $\mu=\muf\to\infty$ (blue). Each point is a time-dependent Hamiltonian; the orange curves are flow trajectories along the flow parameter $\mu$. The two trajectories shown share the same seed but end at different accessible control Hamiltonians, corresponding to different frames $V(\mu,t)$ and, equivalently, generators $K(\mu,t)$ through Eq.~\eqref{eq:KV}. 
Dashed ovals illustrate three stages of the flow. The initial frame at $\mu=0$ has a large experimentally inaccessible part $\Hper(0,t)$ (pink arrow); the intermediate frame has a smaller renormalized inaccessible part; at $\mu=\muf\to\infty$ only the renormalized accessible controls $\Hacc(\muf,t)$ (blue arrow) remain. 
Neighboring Hamiltonians along a trajectory are related by the frame transformation $V(\mu,t)$ through Eq.~\eqref{eq:Hmu}. Since $V(\mu,t)$ is fixed to be the identity at the two temporal endpoints, Eq.~\eqref{eq:V_pinning}, every Hamiltonian along the flow generates the same target unitary $\Utarget$, and the transformation between initial and final frames converges to $V(t)$, as indicated by the
two black arrows. 
}
\label{fig:cartoon}
\end{figure}

Generic problems of unitary control are defined in terms of a target unitary $U_\ast$ to be realized within a fixed duration $T$.
There is a large set of different time-dependent unitaries $U(t)$ that are consistent with the target $U_\ast$, i.e., $U(T)=U_\ast$, each generated by a different time-dependent Hamiltonian $H_U(t)=i\dot UU^\dagger$.
In principle, each Hamiltonian $H_U(t)$ is a solution to the original control problem; however, nearly all of them cannot be realized experimentally, since they contain long-range and multi-body interactions natively unavailable in any physical system.
The design of platform-agnostic methods for constructing \textit{accessible} control Hamiltonians $H_U(t)$ is therefore of paramount importance.

The state-of-the-art approach is constrained optimization within the accessible control space~\cite{Glaser_2015_EPJD,Koch_2022_EPJQT}. Under full controllability, the landscape is trap-free~\cite{rabitz2004quantum}, but accessibility constraints can produce rugged landscapes~\cite{day2019glassy,beato2025toward}, and optimization failure can be inconclusive. Here we take the opposite approach. Rather than searching the accessible set for an approximate representation of the target unitary $U_\ast$, we start from an exact but inaccessible representation; we then construct a flow that deforms it toward the accessible set while preserving the exact $U_\ast$ at every step.

The key conceptual contribution of this work is the design of a framework for accessible quantum control that bears a direct structural correspondence with the RG, see Fig.~\ref{fig:cartoon} and Table~\ref{tab:rg_correspondence}. 
Experimentally accessible Hamiltonians form a manifold of fixed points; the linearized flow around them has a spectrum of decay rates that removes inaccessible terms scale by scale, whereas marginal directions identify obstructions to completing the decimation.
Hence, our Control Similarity Renormalization Group (\ctrlRG) framework inherits the analytical structure of RG.
The flow equation is specified by a generator determined by the final accessible controls through a choice of boundary conditions on the (flow-parameter, physical-time) domain spanned by the flow parameter and physical time. 
We provide a numerical algorithm for finding its solutions.
Through \textit{flow-equation engineering} we generate target control unitaries that directly address the demands of contemporary experimental quantum control. 

We showcase two applications: nonperturbative Floquet engineering and accessible fast-forward driving.
First, we construct accessible time-periodic protocols whose stroboscopic evolution implements Haldane's topological Chern insulator using only time-periodic nearest-neighbor hoppings; crucially, the resulting protocol is nonperturbative in the inverse drive frequency and applies beyond the regime of validity of high-frequency expansions. 
We then implement exact counterdiabatic (CD) evolution in the transverse-field Ising chain away from the adiabatic regime using only accessible fast-forward controls. Remarkably, the flow compresses the long-range multibody terms from the exact adiabatic gauge potential into specific time-dependent schedules which can be directly implemented in the lab; importantly, the unit-fidelity property of CD driving is preserved by the flow. 

More generally, \ctrlRG provides a framework for constructing time-dependent unitary frame transformations that map an inaccessible target Hamiltonian to an accessible frame representation of it. In this respect, our work reveals a unifying perspective on Hamiltonian engineering and unitary quantum control as the two sides of the same coin, and positions \ctrlRG as an organizing principle for seemingly disparate nonequilibrium quantum dynamics.

\section{Renormalization group flow in control space}
\label{sec:theory}

Most approaches~\cite{Khaneja_2005_JMR, palao_2002_PRL, Doria_2011_PRL, Glaser_2015_EPJD, Koch_2022_EPJQT, Goerz_2019_SciPostPhys, Bukov_2019_PRX} to realize a desired unitary propagator $U_\ast$ start with the definition of the set of accessible Hamiltonians and aim at the direct identification of a Hamiltonian inside it, such that the resultant unitary dynamics yields a good approximation of the desired propagator $U_\ast$.

Instead, here we propose a fundamentally different approach.
Consider a family of time-dependent Hamiltonians $H_U(t)=i\dot U(t)U^\dagger(t)$ that all realize valid target unitaries: $U(T)=U_\ast$.
The majority of them, however, need not satisfy the accessibility constraints required for their practical implementation.
Therefore, we seek a Hamiltonian among them that satisfies all constraints.
Here, we demonstrate that flow equations provide a suitable framework to realize desired unitaries using accessible Hamiltonians.

A variety of ways of realizing desired unitaries have been developed, analytical~\cite{Khaneja_2001_PRA,carlini_2006_PRL,barnes_2012_PRL} and numerical~\cite{palao_2002_PRL,machnes_2018_PRL,niu_2019_npj}, often through Hamiltonians containing inaccessible terms.
It is thus safe to assume that a valid unitary $\Useed(t)$ -- {\it i.e.} a unitary such that $\Useed(T)=U_\ast$ --
is known.
This unitary will be referred to as the seed unitary,
and the corresponding infinitesimal generator as the seed Hamiltonian
\be
\Hseed=i\dotUseed\Useed^\dagger
\ .
\ee
For instance, $\Hseed$ can be a static target effective Hamiltonian~\cite{Bukov_2015_AdvPhys, Eckardt_2017_RMP, Rudner_Lindner_2020_NatRevPhys} or a time-dependent counterdiabatic Hamiltonian containing inaccessible gauge potential terms~\cite{Berry_2009_JPA, Sels_Polkovnikov_2017_PNAS, GueryOdelin_2019_RMP}, which we want to implement dynamically in the lab using accessible time-dependent drives. 

Starting with the seed unitary $\Useed(t)$, we now define an entire set of valid unitaries $\UV(t)$ via the ansatz $\UV(t)=V(t)\Useed(t)$ parametrized in terms of the auxiliary unitary $V(t)$.
Each unitary $\UV(t)$ is a valid choice, provided the boundary conditions $V(0)=V(T)=\mathbb 1$ are satisfied.
The Hamiltonian that generates the propagator $\UV(t)$ reads
\be
\HV=V \Hseed V^\dagger-iV\dot V^\dagger\ .
\label{eq:HV}
\ee

To ensure the generator $\HV(t)$ lies within the set of accessible controls, we want to construct a continuous deformation of the initial seed Hamiltonian $\Hseed$ using a family of successive frames $V$ by introducing a flow parameter $\mu$ to label them. As the deformation progresses, it gradually rotates away the inaccessible terms, leaving only those that can be implemented behind. 
If the condition $V(0)=V(T)=\mathbb 1$ is preserved, the representation of the target unitary remains valid throughout the deformation.

\subsection{Flow equation and canonical generator}
\label{subsec:flow_eq}

To make this idea handy, let us parametrize the unitary $V(t)\to V(\mu,t)$ with a scalar $\mu\ge0$, called the flow parameter.
Using Eq.~\eqref{eq:HV}, the transformed Hamiltonian $H_V\to H(\mu,t)$ obeys
\be
H(\mu,t)=V(\mu,t) \Hseed(t)V^\dagger(\mu,t)-iV(\mu,t)\dot V^\dagger(\mu,t)\ ,
\label{eq:Hmu}
\ee
where $\dot V=(\partial_t V)$.
Along the flow parameter $\mu$, the unitary $V(\mu,t)$ changes smoothly under the equation of motion~\cite{wegner1994flow}
\be
i\partial_\mu V(\mu,t)=K(\mu,t)V(\mu,t)\ 
\label{eq:KV}
\ee
with the initial condition $V(\mu{=}0,t){=}\mathbb 1$. Thus, instead of specifying $V(\mu,t)$, one may equally work with its Hermitian generator $K(\mu,t)$.

\textbf{Flow Equation.---}%
Taking the derivative of Eq.~\eqref{eq:Hmu} with respect to the flow parameter $\mu$, this parametrization leads to the equation of motion~\cite{Tomaras_Kehrein_2011_EPL}
\bqa
i\partial_\mu H(\mu,t) &=& [K(\mu,t),H(\mu,t)]+i\dot K(\mu,t)\ .
\label{eq:flow_equation}
\eqa
Given $K(\mu,t)$, this is an equation for the unknown Hamiltonian $H(\mu,t)$. 
The variable $t\in[0,T]$ describes the time dependence in the dynamical control problem of interest, e.g., Floquet engineering [Sec.~\ref{sec:floquet_engineering}] or fast-forward driving [Sec.~\ref{sec:ff_driving}], and should be distinguished from the flow parameter $\mu\in[0,\infty)$. 
In the limit $\mu\to\infty$, the flow Hamiltonian contains the sought-after time-dependent control schedules that implement the target unitary $U_\ast$.

\begin{figure}[t!]
\centering
\includegraphics{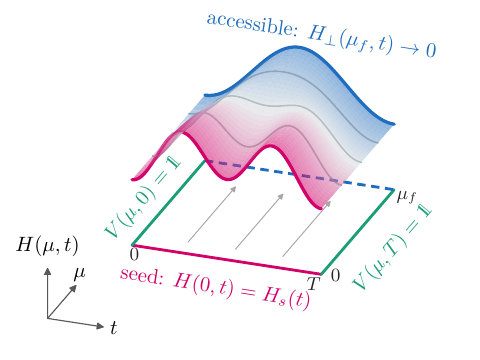}
\caption{
\textbf{Domain and boundary conditions of the control flow.} 
The generator couples the Hamiltonian at neighboring times through $\dot K(\mu,t)$, so the flow of Eq.~\eqref{eq:flow_equation} is a system of partial differential equations on the two-dimensional domain spanned by the physical time $t\in[0,T]$ and the flow parameter $\mu\ge0$, sketched as the horizontal plane. 
The surface above it is the Hamiltonian $H(\mu,t)$ with each line cut at a fixed $\mu$. The flow deforms the control schedule continuously as $\mu$ grows in the arrow's direction. 
The seed Hamiltonian fixes the $\mu=0$ edge, $H(\mu=0,t)=\Hseed(t)$ (red). On the two temporal edges (green) the frame transformation is fixed to the identity, $V(\mu,t=0)=V(\mu,t=T)=\mathbb 1$, for every value of $\mu$, which follows from the vanishing of the generator there, Eq.~\eqref{eq:K_pinning}, and guarantees that the generator decimates the undesired experimentally inaccessible term $\Hper$ monotonically and every flow Hamiltonian reproduces the same endpoint unitary $\Utarget$. The remaining edge (dashed, blue) is not imposed but reached: at large enough final $\mu=\muf$, $\Hper(\muf,t)\to0$, and the accessible controls alone reproduce $\Utarget$. See Sec.~\ref{subsec:bc} for details.
}
\label{fig:bc_cartoon}
\end{figure}

As we show below and in Sec.~\ref{sec:flow_eq_eng}, the generator $K(\mu,t)$ itself can be designed to induce desired structures and properties in the unknown late-flow Hamiltonian
\begin{equation}
\label{eq:H_infty}
    H_\infty(t) {=} \lim_{\mu\to\infty}H(\mu,t)\, .
\end{equation}
The essential difficulty here lies in finding a suitable choice of $K(\mu,t)$ that constrains properties of $H(\mu,t)$ down the flow as $\mu\to\infty$, using information contained in $H(\mu,t)$ at the \textit{current} flow parameter $\mu$.
This implies that the generator $K(\mu,t)$ is itself a function of the Hamiltonian $H(\mu,t)$ which, in turn, renders the flow equation, Eq.~\eqref{eq:flow_equation}, nonlinear in $H(\mu,t)$. 
Moreover, the flow equation does not treat each $t$-instant separately: the $ t$-derivative $\dot K(\mu,t)$ couples the terms in the Hamiltonian at neighboring times, so that the flow establishes a system of coupled partial differential equations (PDEs) in the two variables $\mu$ and $t$~\cite{Tomaras_Kehrein_2011_EPL}.
As a result, the entire control schedule flows under Eq.~\eqref{eq:flow_equation} as $\mu$ increases, see Fig.~\ref{fig:bc_cartoon}.
These features distinguish the time-dependent flow equation, Eq.~\eqref{eq:flow_equation}, from its commonly known time-independent variants, which define ordinary differential equations in $\mu$ alone~\cite{wegner1994flow,Glazek_Wilson_1994_PRD}. 

\textbf{Canonical Generator.---}%
Our goal is to find an operator $K(\mu,t)$ which generates a flow from the seed Hamiltonian $\Hseed(t)$ to an experimentally accessible Hamiltonian $H_\infty(t)$, such that both generate the same target unitary $U_\ast$. 

To construct such a generator, we first distinguish the accessible and inaccessible parts of $H(\mu,t)$. 
Denote by the superoperator $\Pper$ the projector onto the subspace of operators spanned by the experimentally inaccessible control terms, i.e., the terms that are meant to be attenuated under the flow, and by $\Pacc=1-\Pper$ its complement. These split the Hamiltonian $H(\mu,t)=\Hacc(\mu,t)+\Hper(\mu,t)$ along the flow into an accessible part $\Hacc=\Pacc H$ and an inaccessible part $\Hper=\Pper H$, which we use throughout.
The common choice of the canonical generator~\cite{Tomaras_Kehrein_2011_EPL}
\begin{eqnarray}
    \Kc(\mu,t) &=& i[\Hacc(\mu,t),\Hper(\mu,t)] + \dotHper(\mu,t)
\label{eq:canonical_generator}
\end{eqnarray}
decimates the undesired terms along the flow.
To see why, notice that the weight of the undesired terms obeys
\begin{equation}
\label{eq:canonical_contr}
    \tfrac12\partial_\mu\tr[\Hper^2(\mu,t)]=-\tr[\Kc^2(\mu,t)]+\partial_t\tr[\Hper \Kc(\mu,t)]
\end{equation}
locally for every $t\in[0,T]$ and $\mu\in[0,\infty)$.
Indeed, the first term on the right-hand side is manifestly negative and drives the decimation process; the second is a total time derivative and thus, upon integration over time, contributes only through its values at the two temporal boundaries $t{=}0,T$. 
The decimating property of the flow, as required by the physics of the control flow equation, thus translates directly to a specific boundary condition on the flow equation generator $K(\mu,t)$. We show in Sec.~\ref{subsec:reg} that the canonical generator $K_c$, in general, does not feature this property and thus requires modification. 

Up to this point, the flow equation and its canonical generator are established in the Wegner flow for the SRG~\cite{wegner1994flow,Glazek_Wilson_1994_PRD}, extended from static Hamiltonians to the time-dependent setting~\cite{Tomaras_Kehrein_2011_EPL}. 
In the rest of this section, we develop a framework for the accessible quantum control in which the flow preserves the target unitary and guarantees the monotonic decimation of the inaccessible terms. To this end, we impose boundary conditions at the two temporal edges and construct a regularized flow generator. The freedom in choosing the generator allows us to adapt this framework to different control problems via flow-equation engineering, which we discuss in Sec.~\ref{sec:flow_eq_eng}.

\subsection{Boundary conditions of the control problem}
\label{subsec:bc}
The flow Hamiltonian $H(\mu,t)$ is defined on the two-dimensional domain spanned by $t\in[0,T]$ and $\mu\in[0,\infty)$, illustrated in Fig.~\ref{fig:bc_cartoon}; hence, to solve the flow PDE, Eq.~\eqref{eq:flow_equation}, we have to specify boundary conditions, dictated by the control problem of interest: 

\begin{subequations}
\label{eq:BCs}
(1) At $\mu{=}0$ the flow starts from the seed Hamiltonian: 
\begin{equation}
\label{eq:seed_bdary}
    H(\mu=0,t)=\Hseed(t)\, .    
\end{equation}
This allows the flow, and hence the final control Hamiltonian $H_\infty$, to inherit specific properties of the evolution generated by $\Hseed(t)$, e.g., built-in guarantees for unit-fidelity state or gate preparation at the final time $T$. 

(2) To ensure that the resulting control Hamiltonian $H_\infty$ preserves the target unitary $U_\ast$, we impose a homogeneous Dirichlet condition
\be 
    K(\mu,0)=0=K(\mu,T)
\label{eq:K_pinning}
\ee
\end{subequations}
on the flow generator at the two temporal edges $t{=}0,T$. 
Indeed, since $V(\mu,t)$ is generated by $K(\mu,t)$ through Eq.~\eqref{eq:KV}, a generator vanishing at $t=0$ and $t=T$ leaves the frame transformation unchanged there along the entire flow, so that the frame is fixed at both ends,
\be 
    V(\mu,0)=\mathbb 1=V(\mu,T)
\label{eq:V_pinning}
\ee 
for all values of $\mu$. This guarantees that every flow Hamiltonian remains a representation of the same target unitary (see Appendix~\ref{app:pinned_frame}). 

Remarkably, Eq.~\eqref{eq:K_pinning} is precisely the decimation condition required for the flow to attenuate inaccessible control terms discussed around Eq.~\eqref{eq:canonical_contr}. To see why, notice that the time-integrated weight of the undesired terms decreases monotonically along the flow when the integrated total time derivative term vanishes, cf.~Eq.~\eqref{eq:canonical_contr}. This is exactly the case under the Dirichlet condition of Eq.~\eqref{eq:K_pinning}, which therefore also ensures the monotonic contraction property of the flow. 

Thus, the requirement of a monotonic contractive flow and the preservation of the target unitary $\Utarget$ impose the same condition, Eq.~\eqref{eq:K_pinning}, on the flow generator $K(\mu,t)$. 
Together with the seed boundary condition of Eq.~\eqref{eq:seed_bdary}, all boundary conditions for the control problem are specified: the seed Hamiltonian $\Hseed$ fixes the boundary at $\mu{=}0$, and the Dirichlet condition of Eq.~\eqref{eq:K_pinning} fixes the two temporal edges at $t{=}0,T$, see Fig.~\ref{fig:bc_cartoon}.

\subsection{Regularized generator}
\label{subsec:reg}

With the time derivative $\dotHper$ occurring in the canonical generator of Eq.~\eqref{eq:canonical_generator}, the flow equation, Eq.~\eqref{eq:flow_equation}, contains a second derivative in the physical time $t$, $\partial_\mu\Hper\sim\partial_t^2\Hper$, so that the undesired terms are not only attenuated, but also diffuse in the time direction during the flow (see concrete example in Appendix~\ref{app:2ls:flow}). The canonical generator of Eq.~\eqref{eq:canonical_generator}, however, does not vanish at the two ends of the protocol in general; the diffusive spreading typically leads to a violation of the boundary condition of Eq.~\eqref{eq:K_pinning} during the flow starting from a generic seed Hamiltonian. 
We now upgrade the canonical generator such that Eq.~\eqref{eq:K_pinning} is satisfied for all values of $\mu$. 

Rather than searching for a generator that decimates the undesired terms and vanishes at the two ends of the protocol at once, we retain the canonical generator and act on it with a superoperator $\Rreg$,
\begin{equation}
    K(\mu,\cdot)=\Rreg\Kc(\mu,\cdot), 
\label{eq:K_from_Kw}
\end{equation}
which we refer to as a regularizer for reasons that will become clear in Sec.~\ref{sec:flow_eq_eng}. The idea is to modify the temporal profile of the generator alone but leave the operator content of Eq.~\eqref{eq:canonical_generator} intact, so that the flow still decimates precisely the undesired terms selected by the projector $\Pper$. 

The simplest such regularizer is a pointwise multiplication in physical time, $K(\mu,t){=}f(t)\Kc(\mu,t)$, by a real-valued function $f(t){\ge}0$ that vanishes at the two temporal boundaries, $f(0){=}f(T){=}0$. The generator then inherits the zeros of this mask function $f(t)$. Moreover, since $f(t)\ge0$, each $K(\mu,t)$ is a nonnegative multiple of $\Kc(\mu,t)$, so that $\tr[\Kc K]=f(t)\tr[\Kc^2]\ge0$ and the decimation of Eq.~\eqref{eq:canonical_contr} is never reversed. In this way, Eq.~\eqref{eq:K_pinning} is satisfied by construction for any seed Hamiltonian $\Hseed$ and at every flow parameter $\mu$.

The decimation of the undesired terms survives under this modification. Repeating the argument that led to Eq.~\eqref{eq:canonical_contr} and collecting the local weights into
\begin{equation}
    \Wnorm(\mu)=\sqrt{\ipt{\Hper}{\Hper}/2},
\label{eq:inaccessible_weight}
\end{equation}
with $\ipt{A}{B}=\int_0^T\mathrm dt\,\tr[A(t)B(t)]$ the inner product on time-dependent operators, we obtain (see Appendix~\ref{app:ctrlsrg:contraction})
\begin{equation}
\partial_\mu[\Wnorm(\mu)^2]=-\ipt{\Kc}{\Rreg\Kc}\le0.
\label{eq:contraction}
\end{equation}
The boundary term encountered in Sec.~\ref{subsec:flow_eq} above has disappeared, since the generator now vanishes at the two ends of the protocol, and the remaining contribution is non-positive as long as $\Rreg$ is positive semidefinite; the latter is enforced by the condition $f(t)\ge0$, which gives $\partial_\mu[\Wnorm^2]=-\int_0^T\mathrm dt f(t) \tr[\Kc^2]\le0$. 

More general regularizers can also be used, provided they are positive semidefinite and self-adjoint with respect to $\ipt{\cdot}{\cdot}$, and they map onto operators that vanish at $t=0,T$.\footnote{Different choices of admissible regularizers may trace different paths through the space of time-dependent control Hamiltonians, potentially ending in different accessible protocols that all reproduce $\Utarget$.} 
In Sec.~\ref{sec:flow_eq_eng} we make use of this freedom to design generators for Floquet engineering and fast-forward driving. 
The time-integrated weight of the undesired terms $\Wnorm(\mu)$ is then guaranteed to decrease monotonically along the entire flow. 
Ultimately, at a large final flow parameter $\muf\gg 1$, where this weight has become negligible, the accessible terms alone reproduce the target unitary $\Utarget$, as illustrated in Fig.~\ref{fig:cartoon}.

Both boundary conditions of the control problem, Eqs.~\eqref{eq:BCs}, are thus fulfilled by the construction in Eqs.~\eqref{eq:canonical_generator} and~\eqref{eq:K_from_Kw}, and the flow equation, Eq.~\eqref{eq:flow_equation}, now defines a closed quadratic system of PDEs for the unknown flow Hamiltonian $H(\mu,t)$. 
In general, control problems may also require additional boundary conditions, such as prescribed derivatives of the Hamiltonian $H(\mu,t)$ at the temporal edges $t=0,T$, or strict time periodicity. 
We call this approach \textit{flow-equation engineering} and discuss it systematically in Sec.~\ref{sec:flow_eq_eng}.

\subsection{Control similarity renormalization group (\ctrlRG)}
\label{subsec:ctrlsrg}

A hidden structure underlying the flow becomes apparent once the flow equation is written in terms of the extended Hamiltonian $H(\mu,t)-i\partial_t$; its commutator with any time-dependent operator $A(t)$ reproduces both terms on the right-hand side of the flow equation, Eq.~\eqref{eq:flow_equation},
\begin{equation}
i[H(\mu,t)-i\partial_t,A(t)]=i[H(\mu,t),A(t)]+\dot A(t).
\label{eq:extended_commutator}
\end{equation}
The canonical generator of Eq.~\eqref{eq:canonical_generator} can then be compactly written as $\Kc=i[H-i\partial_t,\Hper]$. 
This form allows us to identify a formal analogy between the control flow equation, Eq.~\eqref{eq:flow_equation}, and Wegner's similarity renormalization group (SRG)~\cite{wegner1994flow,Glazek_Wilson_1994_PRD}. 
Whereas SRG was originally designed to construct flows that (block) diagonalize static Hamiltonians, as we now demonstrate, the underlying mathematical machinery can be extended to study time-dependent systems, such as the control problems of interest in this work.  
To complete the dictionary, note that the undesired inaccessible control terms $\Hper=\Pper H$ correspond to the off-diagonal part of the static Hamiltonian to be decimated by the Wegner flow, whereas the accessible controls $\Hacc(\mu,t)-i\partial_t$ replace the diagonal part of the static Hamiltonian. 

It is therefore instructive to analyze the small-$\Hper$ limit, where the flow can be linearized. To leading order, the flow is equivalent to a Schrieffer-Wolff transformation~\cite{Schrieffer_Wolff_1966_PR,Bravyi_etal_2011_AnnPhys} which removes $\Hper$ by a single rotation. The flow of the inaccessible terms is then given by
\begin{equation}
\partial_\mu \Hper(\mu,t)=-\Dec^\dagger\Dec\Hper(\mu,t),
\end{equation}
where the superoperator $\Dec\equiv\Rreg^{1/2}\Liou\Pper$ is defined in terms of the skew-symmetric Liouvillean $\Liou(\cdot)\equiv i[\Hacc(\mu,t)-i\partial_t,\cdot]$ (see Appendix~\ref{app:ctrlsrg}). Since the superoperator $\Dec^\dagger\Dec$ is self-adjoint and positive semidefinite by construction, we can diagonalize it and find its eigenmodes $\Decmode{\alpha}(t)$, which are time-dependent operators, corresponding to eigenvalues $\Decval{\alpha}^2\ge0$; the flow leaves the time-profile of $\Decmode{\alpha}(t)$ fixed and changes only their amplitudes. 

Writing $\Hper(\mu,t)=\sum_\alpha H_{\perp,\alpha}(\mu,t)$ with $H_{\perp,\alpha}(\mu,t)$ the component of $\Hper$ along $\Decmode{\alpha}$, each component is decimated exponentially in the flow parameter $\mu$ within the regime of validity of the linearization (see Appendix~\ref{app:ctrlsrg:linear})
\begin{equation}
H_{\perp,\alpha}(\mu,t)=e^{-(\mu-\mu_0)\Decval{\alpha}^2} H_{\perp,\alpha}(\mu_0,t).
\end{equation}
The flow thus eliminates undesired terms with larger eigenvalues first and those with smaller eigenvalues last; this removes inaccessible control terms scale by scale in much the same way the renormalization group flow removes ultraviolet degrees of freedom~\cite{Glazek_Wilson_1994_PRD}. 
This is the underlying reason for the name {\em  Control Similarity Renormalization Group} introduced in Sec.~\ref{sec:intro}; its structural correspondence with SRG and conventional Hamiltonian RG is summarized in Table~\ref{tab:rg_correspondence}. 

The decay spectrum $\{\Decval{\alpha}^2\}$ classifies the control modes into an irrelevant and a marginal part, in the same way as the classification in a conventional RG~\cite{wilson1975renormalization,fisher1998renormalization}. A mode with $\Decval{\alpha}^2>0$ has an amplitude that decays along the flow and is \emph{irrelevant}, so the decimation removes it. A mode with $\Decval{\alpha}=0$ is \emph{marginal}: the flow leaves it unchanged. Since $\Dec^\dagger\Dec$ is positive semidefinite, no eigenvalue is negative, and the inaccessible control terms contain no \emph{relevant} component. Marginal controls are therefore the only obstruction to a complete decimation, and we return to the plateaus they produce in Sec.~\ref{sec:discussion}.

\begin{table}[t]
\centering
\footnotesize
\setlength{\tabcolsep}{1.2pt}
\renewcommand{\arraystretch}{1.35}
\begin{tabular}{|l|l|l|l|}
\hline
 & \textbf{\ctrlRG} & \textbf{SRG} & \makecell{\textbf{Hamiltonian}\\\textbf{RG}} \\
\hline
\makecell[l]{\it flowing\\ \it object}
& \makecell[l]{Hamiltonian\\$H(\mu,t)$}
& \makecell[l]{Hamiltonian\\$H(\mu)$}
& \makecell[l]{effective\\Hamiltonian $H_\Lambda$} \\
\hline
\makecell[l]{\it running\\ \it scale}
& \makecell[l]{flow parameter $\mu$}
& \makecell[l]{flow parameter $\mu$}
& \makecell[l]{running cutoff $\Lambda$} \\
\hline
\makecell[l]{\it decimated\\ \it sector}
& \makecell[l]{unwanted terms\\$\Hper=\Pper H$}
& \makecell[l]{off-diagonal\\matrix elements}
& \makecell[l]{ultraviolet\\physics} \\
\hline
\makecell[l]{\it fixed-point\\ \it manifold}
& \makecell[l]{retained controls\\($\Hper=0$)}
& \makecell[l]{(block-)diagonal\\Hamiltonian}
& \makecell[l]{scale-invariant\\Hamiltonian} \\
\hline
\it flow map
& \makecell[l]{time-dep.~frame \\ transf.~$V(\mu,t)$}
& \makecell[l]{unitary rotation\\$V(\mu)$}
& \makecell[l]{coarse-graining,\\rescaling} \\
\hline
\makecell[l]{\it preserved\\ \it physics}
& \makecell[l]{target unitary\\$\Utarget$}
& \makecell[l]{spectrum}
& \makecell[l]{infrared\\physics} \\
\hline
\end{tabular}
\renewcommand{\arraystretch}{1}
\caption{\textbf{Correspondence between \ctrlRG, the similarity renormalization group (SRG), and conventional Hamiltonian RG.} Despite acting on different objects, all three frameworks decimate a selected sector along a flow toward a fixed point while preserving the physical information relevant to the problem. The comparison follows the standard constructions of SRG~\cite{wegner1994flow,Glazek_Wilson_1994_PRD} and Hamiltonian RG~\cite{wilson1975renormalization,fisher1998renormalization}.}
\label{tab:rg_correspondence}
\end{table}

\subsection{Numerical implementation}
\label{subsec:flow_eq-num}

Just like essentially any problem with a time-dependent Hamiltonian, the task of solving the \ctrlRG flow equation is, strictly speaking, an infinite-dimensional problem.
The flow takes place in the space of time-dependent control Hamiltonians
\begin{equation}
\mathcal{C}=\mathfrak{g}\otimes L^2[0,T]\ ,
\label{eq:control_space}
\end{equation}
which factorizes into an operator part $\mathfrak g$ that closes under commutation and the temporal part $L^2[0,T]$ given by the space of square-integrable real-valued functions of physical time.

Even for finite-dimensional quantum systems (with correspondingly finite-dimensional operator space $\mathfrak g$), the space $L^2[0,T]$ is infinite dimensional.
In practice, however, any system dynamics takes place on some finite minimal time-scale, so that a truncation to a finite dimension is possible at any given desired numerical accuracy.

For most interacting quantum systems the operator space $\mathfrak g$ is exponentially large in the number of subsystems.
Despite the finite dimension, it can thus be necessary to involve numerical approximations for explicit solutions.
The examples discussed further down (Secs.~\ref{sec:floquet_engineering} and~\ref{sec:ff_driving}) do not require such approximations of the operator space, because the system Hamiltonians of these examples give rise to a Lie algebra that scales polynomially in the number of subsystems, allowing numerical solutions without operator-space truncation for systems of up to $L=50$ sites discussed in Sec.~\ref{sec:ff_driving}.

With a numerically exact or approximate description, the solution to the \ctrlRG flow equation is comprised of the following five steps:
\begin{enumerate}
    \item fix the operator space $\mathfrak g$ and the inaccessible subspace defined by $\Pper$, and specify a discrete time grid of $N_t$ points; 

    \item initialize the flow from the seed, $H(\mu{=}0,t){=}\Hseed(t)$;

    \item build the canonical generator $\Kc$ using Eq.~\eqref{eq:canonical_generator};

    \item apply the regularizer, $K=\Rreg\Kc$, cf.~Eq.~\eqref{eq:K_from_Kw}; 

    \item propagate $H(\mu,\cdot)$ in the flow parameter $\mu$ according to the flow equation, Eq.~\eqref{eq:flow_equation}. 
\end{enumerate}
Repeating steps 3 to 5 until the weight $\Wnorm(\mu)$ of the undesired terms relative to that of the accessible terms has dropped below a pre-set threshold, we arrive at the accessible control schedule $\Hacc(\muf,t)\approx H_\infty$ at the final flow parameter $\muf$, which we refer to as converged. 
Since the weight of the inaccessible terms decreases monotonically, we can use a smaller threshold with a larger $\muf$ whenever a higher accuracy is required. In practice, the threshold need only be small compared with the experimental noise, such as miscalibrated control fields or decoherence during the protocol.

Every time-dependent Hamiltonian along the flow is a vector in $\mathcal{C}$, so that expanding it in a basis of $\mathfrak g$ turns the flow equation into a set of coupled nonlinear partial differential equations whose number is set by $\dim\mathfrak g$, rather than by the dimension of the Hilbert space. 
Notice the formal similarity to real-space RG whose flow is defined in the space of (static) Hamiltonians~\cite{fisher1998renormalization}.

For a two-level system, $\dim\mathfrak g=3$ and the flow equations are simple enough to be derived by hand\footnote{We are not aware of a closed-form analytical solution already for this simplest quantum system.}, which we do in Appendix~\ref{app:2ls} as an illustration of the entire construction. In the Haldane and transverse-field Ising model applications of Secs.~\ref{sec:floquet_engineering} and~\ref{sec:ff_driving}, $\dim\mathfrak g$ grows polynomially with system size; we present numerical solutions for the flow of up to $149$ coupled partial differential equations for a chain of $L=50$ sites. However, in generic interacting many-body systems the algebra grows exponentially with system size, so solving the flow equation becomes numerically expensive for large sizes; we discuss generalizations in the many-body regime in Sec.~\ref{sec:discussion}. 

The truncation of $L^2[0,T]$ to a finite-dimensional space can be realized in practice in terms of a discrete time grid. 
This converts the flow equation into a set of coupled nonlinear ordinary differential equations in $\mu$. The decay rates $\Decval{\alpha}^2$ introduced in Sec.~\ref{subsec:ctrlsrg} typically span many orders of magnitude, making the equations stiff; we therefore use an implicit solver. Here, a certain care is required, since the contraction argument of Eq.~\eqref{eq:contraction} relies on integration by parts in physical time. Implementing the derivative on the discrete time grid naively would preserve this integration-by-parts identity only up to a discretization error. However, the weight of the inaccessible terms then need not decrease at a large flow parameter, and we found this loss of monotonicity to become a practical obstacle for the computation of the flow equation of large systems. We therefore discretize the derivative $\partial_t$ and integral $\int_0^T\mathrm dt$ operations using a summation-by-parts scheme~\cite{Strand_1994_JCP,Svard_Nordstrom_2014_JCP} which reproduces the integration-by-parts identity exactly, and construct the regularizer on the same grid, as detailed in Appendix~\ref{app:reg:sbp}. The monotonic decrease of the undesired terms then holds at any number of grid points, rather than only in the continuum limit, which is what brings the large-system calculations to reality.

\section{Flow-equation engineering}
\label{sec:flow_eq_eng}

Employing \ctrlRG for realistic quantum control problems requires the regularizer $\Rreg$ for the flow equation generator $K(\mu,t)$ from Sec.~\ref{subsec:reg} to be positive semidefinite and self-adjoint, and to satisfy the boundary condition of Eq.~\eqref{eq:K_pinning}, as required for the contraction of Eq.~\eqref{eq:contraction} and linearized analysis of Sec.~\ref{subsec:ctrlsrg}.
This leaves a substantial residual freedom in choosing $\Rreg$. Different choices define different paths through control Hamiltonian space for the flow (see Fig.~\ref{fig:cartoon}), and hence generally lead to different accessible protocols, each of which reproduces the same target unitary $\Utarget$. We refer to the process of designing the regularizer $\Rreg$ as \textit{flow-equation engineering}.\footnote{In principle, flow-equation engineering can be regarded more generally as the process of designing the entire flow generator $K(\mu,t)$, since different generators may be better suited to studying other control problems.}

In practice, we typically have two requirements. 
The first is \emph{smoothness}: we want a smooth temporal profile of the generator $K$, and hence of the control schedules $H(\mu,t)$ it produces along the flow. The second is \emph{endpoint control}: we often have to impose additional boundary conditions at $t=0,T$ on the flow Hamiltonian $H(\mu,t)$ itself, beyond Eq.~\eqref{eq:BCs}. These endpoint conditions differ from one application to another: e.g., periodically driven systems require strictly time-periodic protocols, whereas fast-forward protocols on an open time interval require control fields that start and end at prescribed values. 
In the following, we introduce a class of regularizers that accommodates both \emph{smoothness} and \emph{endpoint control} in Sec.~\ref{subsec:eng_bvp}, and then design the \ctrlRG flow for the two classes of control problems addressed in this work, nonperturbative Floquet engineering in Sec.~\ref{subsec:eng_floquet} and accessible fast-forward driving in Sec.~\ref{subsec:eng_ff}.

\subsection{Regularizers from boundary value problems}
\label{subsec:eng_bvp}

A convenient class of regularizers is given by a convolution of the canonical generator with a kernel $\Green(t,s)$ in physical time,
\begin{equation}
    K(\mu,t)=(\Rreg K_c)(\mu,t) = \int_0^T\mathrm ds\,  \Green(t,s) \Kc(\mu,s),
\label{eq:kernel}
\end{equation}
where $\Green(t,s)$ is symmetric and positive semidefinite, so that $\Rreg$ is self-adjoint and positive semidefinite as required for the contraction of Eq.~\eqref{eq:contraction}, and satisfies $\Green(0,s)=\Green(T,s)=0$, so that the generator obeys the Dirichlet condition of Eq.~\eqref{eq:K_pinning}. 
The mask function $f(t)$ introduced in Sec.~\ref{subsec:reg} corresponds to a diagonal kernel: $\Green(t,s)=f(t)\delta(t-s)$; thus, Eq.~\eqref{eq:kernel} is a strict generalization of $K(\mu,t){=}f(t)\Kc(\mu,t)$. 
Although not every valid regularizer takes this form, those that do are particularly useful, as we will see, and we restrict our attention to them in what follows. 
With $\Kc$ given by Eq.~\eqref{eq:canonical_generator}, the flow equation
\begin{align}
\partial_\mu H(\mu,t) = \int_0^T\mathrm \mathrm d s\Big( 
& i\,\Green(t,s)\left[H(\mu,t),\Kc(\mu,s)\right] \nonumber\\
& +\left[\partial_t\Green(t,s)\right]\Kc(\mu,s)\Big),
\label{eq:closed_flow}
\end{align}
then becomes an explicit and closed equation for the flow Hamiltonian, in which time derivatives act on the kernel alone.

A kernel of this kind arises naturally as the Green's function of a differential operator in physical time. 
Inspired by this analogy, we define the regularizer $\Rreg_4$ implicitly, through the fourth-order differential equation it solves,
\begin{equation}
\left(1-\gamma_4^2\partial_t^2\right)^{2}K=\Kc,
\label{eq:filter4}
\end{equation}
so that $\Rreg_4$ is the Green's function of the operator on the left, (see Appendices~\ref{app:reg:positivity} and \ref{app:reg:green}). 
To see why this can be useful, notice that, in the frequency domain, the regularizer acts as a low-pass filter 
\begin{equation}
\hat\Rreg_4=\frac{1}{(1+\gamma_4^2\omega^2)^{2}}\, . 
\end{equation}
Therefore, temporal structures faster than $\omega\sim\gamma_4^{-1}$ are suppressed, and the control schedules produced along the flow stay smooth on time scales below $\gamma_4$, which is the \emph{smoothness} requirement. 

\emph{Endpoint control} is met by the ability to impose four boundary conditions on the fourth-order differential equation, Eq.~\eqref{eq:filter4}. The first two are fixed by the Dirichlet conditions of Eq.~\eqref{eq:K_pinning}. In turn, the remaining two are chosen according to the control problem of interest; they let us prescribe the behavior of the flow Hamiltonian at the two ends of the protocol. Indeed, since the generator vanishes there, the flow equation reduces at $t=0,T$ to
\begin{equation}
\partial_\mu H(\mu,t)\big|_{t=0,T}=\dot K(\mu,t)\big|_{t=0,T},
\label{eq:bc_translation}
\end{equation}
so that any condition imposed on the flow of the Hamiltonian, $\partial_\mu H(\mu,t)$, at the temporal edges translates directly into a condition on the time-derivative of the generator there. 

In the following, we work out the conditions for the two classes of control problems addressed in this work, summarized in Table~\ref{tab:bc}.

Since the regularizer acts on the temporal profile of the generator while the projector $\Pacc=1-\Pper$ acts on its operator content, they commute, and we may assign different filters to different operator sectors. For instance, combining a fourth-order filter on the accessible controls with a second-order one, defined in the same way through 
\begin{equation}
    \left(1-\gamma_2^2\partial_t^2\right)K=\Kc,
\end{equation}
on the terms to be decimated gives
\begin{equation}
\Rreg=\Rreg_4\Pacc+\Rreg_2\Pper,
\label{eq:hybrid_filter}
\end{equation}
which constrains the accessible controls at $t=0,T$ while leaving the decimated terms free there. 

For the same reason, the individual operator components of $K$ may carry different positive constant weights, which is useful whenever the flow has to distinguish between different control directions (see Sec.~\ref{subsec:Haldane_model} and Appendix~\ref{app:reg:sector}).

\begin{table}[t]
\centering
\footnotesize
\setlength{\tabcolsep}{1.2pt}
\renewcommand{\arraystretch}{1.35}
\begin{tabular}{|l|l|l|}
\hline
 & \textbf{time-periodic} & \textbf{open interval} \\
\hline
{\it required of $H$} & $H|_{t=0}=H|_{t=T}$ & $\partial_\mu\left(\Pacc H\right)|_{t=0,T}=0$ \\
\hline
{\it imposed on $K$} & $\dot K|_{t=0}=\dot K|_{t=T}$ & $\Pacc\dot K|_{t=0,T}=0$ \\
\hline
{\it regularizer} & $\Rreg_4$ & $\Rreg_4\Pacc+\Rreg_2\Pper$ \\
\hline
{\it BC} & Eq.~\eqref{eq:Floquet_BC} & Eq.~\eqref{eq:FF_BC} \\
\hline
{\it application} & \makecell[l]{Floquet engineering \\
(Sec.~\ref{sec:floquet_engineering})} & \makecell[l]{fast-forward driving \\(Sec.~\ref{sec:ff_driving})} \\
\hline
\end{tabular}
\renewcommand{\arraystretch}{1}
\caption{\textbf{Flow-equation engineering for the two classes of control problems.} In both cases the generator obeys the Dirichlet condition $K|_{t=0,T}=0$, which guarantees the preservation of the target unitary $\Utarget$ and the monotonic decrease of the weight of the inaccessible terms. This fixes two boundary conditions of the boundary value problem defined by the regularizer, while the remaining ones are chosen through Eq.~\eqref{eq:bc_translation}, depending on the requirements of the control problem on the flow Hamiltonian $H$ at the two ends of the protocol. A time-periodic protocol requires the same boundary conditions for every operator component, and hence a single fourth-order regularizer. On an open time interval, by contrast, only the accessible controls must keep their values at the two ends. This costs one further condition on the slope of the generator, and hence a fourth-order regularizer. The inaccessible terms obey the Dirichlet condition alone, for which a second-order regularizer suffices. Together they give the hybrid choice of Eq.~\eqref{eq:hybrid_filter}.
See Secs.~\ref{subsec:eng_floquet} and~\ref{subsec:eng_ff} for details.
}
\label{tab:bc}
\end{table}

\subsection{Time-periodic protocols}
\label{subsec:eng_floquet}

We first consider control problems whose protocol is required to be time-periodic, $H(\mu,0)=H(\mu,T)$.\footnote{We do not impose periodicity of the higher-order time derivatives here; each of them would constrain one further derivative of the generator and hence call for a higher-order regularizer. Periodicity of $H$ is all the Floquet problem requires, and the smoothness of the drive is set by $\gamma_4$ regardless.} If we initialize the flow from a seed that is itself periodic, $\Hseed(0)=\Hseed(T)$, the task of the flow is to preserve this periodicity during the flow. Using Eq.~\eqref{eq:bc_translation}, periodicity of $H$ translates into a condition on the slope of the generator, $\partial_tK(\mu,0)=\partial_tK(\mu,T)$, so we choose a fourth-order regularizer that allows us to impose this as a boundary condition. 
Evaluating $K=\Rreg_4\Kc$ then requires solving the differential equation
\begin{equation}
\left(1-\gamma_4^2\partial_t^2\right)^2K(\mu,t) = \Kc(\mu,t)
\label{eq:Floquet_BVP}
\end{equation}
on the interval $t\in[0,T]$, subject to the boundary conditions
\begin{subequations}
\label{eq:Floquet_BC}
\begin{align}
K(\mu,t)\big|_{t=0} &= 0\, , \label{eq:Floquet_BC_a}\\
K(\mu,t)\big|_{t=T} &= 0\, , \label{eq:Floquet_BC_b}\\
\dot K(\mu,t)\big|_{t=0} &= \dot K(\mu,t)\big|_{t=T}\, , \label{eq:Floquet_BC_c}\\
\ddot K(\mu,t)\big|_{t=0} &= \ddot K(\mu,t)\big|_{t=T}\, . \label{eq:Floquet_BC_d}
\end{align}
\end{subequations}
The fourth-order equation, Eq.~\eqref{eq:Floquet_BVP}, requires four such conditions. The first two, Eqs.~\eqref{eq:Floquet_BC_a} and~\eqref{eq:Floquet_BC_b}, are the two temporal edges of Eq.~\eqref{eq:K_pinning}, while the third, Eq.~\eqref{eq:Floquet_BC_c}, maintains the periodicity of the flow Hamiltonian along the \ctrlRG flow. The fourth, Eq.~\eqref{eq:Floquet_BC_d}, completes the fourth-order boundary value problem, and matching the second derivative is the natural choice, as it leaves the generator periodic together with its derivatives.

With these four conditions, the regularizer is fully specified and the flow always preserves the periodicity. Protocols of this kind arise in Floquet engineering, where a periodic drive is designed to reproduce the stroboscopic evolution of a Hamiltonian that the system does not possess. We consider such a problem in Sec.~\ref{sec:floquet_engineering}.

\subsection{Protocols on an open time interval}
\label{subsec:eng_ff}

We now turn to control problems on an open time interval, where the protocol starts and ends at prescribed values of the control fields. In an experiment, these values are set before the protocol begins, as in the fast-forward control of Sec.~\ref{sec:ff_driving}, where a field of definite strength is prepared before the sweep. Any deviation of the flow Hamiltonian from the prescribed boundary values at $t=0$ or $t=T$ would have to be realized by a sudden quench of the control fields, which we want to avoid. 

We therefore require the accessible controls to retain the values of the seed at the two temporal ends throughout the flow, $\partial_\mu\Pacc H|_{t=0,T}=0$. The inaccessible terms, by contrast, need not be constrained there, as they are decimated along the flow anyway. 

Invoking Eq.~\eqref{eq:bc_translation}, the two sectors thus require a different number of boundary conditions, which we accommodate by the hybrid regularizer of Eq.~\eqref{eq:hybrid_filter}. Evaluating $K=\left(\Rreg_4\Pacc+\Rreg_2\Pper\right)\Kc$ requires solving the two differential equations, which decouple because $\Rreg$ commutes with the projectors,
\begin{align}
\left(1-\gamma_4^2\partial_t^2\right)^2\Pacc K(\mu,t) &= \Pacc\Kc(\mu,t), \label{eq:FF_BVP_acc}\\
\left(1-\gamma_2^2\partial_t^2\right)\Pper K(\mu,t) &= \Pper\Kc(\mu,t), \label{eq:FF_BVP_per}
\end{align}
on the interval $t\in[0,T]$, subject to the boundary conditions
\begin{subequations}
\label{eq:FF_BC}
\begin{align}
K(\mu,t)\big|_{t=0} &= 0, \label{eq:FF_BC_a}\\
K(\mu,t)\big|_{t=T} &= 0, \label{eq:FF_BC_b}\\
\Pacc\dot K(\mu,t)\big|_{t=0} &= 0, \label{eq:FF_BC_c}\\
\Pacc\dot K(\mu,t)\big|_{t=T} &= 0. \label{eq:FF_BC_d}
\end{align}
\end{subequations}
As in the periodic case, the first two conditions, Eqs.~\eqref{eq:FF_BC_a} and~\eqref{eq:FF_BC_b}, are the two temporal edges of Eq.~\eqref{eq:K_pinning}, which applies to both sectors and completes the second-order problem of the inaccessible terms. 
The remaining two, Eqs.~\eqref{eq:FF_BC_c} and~\eqref{eq:FF_BC_d}, fix the time derivative of the accessible component of the generator at the two ends, and thereby complete the fourth-order problem of the accessible sector. 

The hybrid regularizer of Eq.~\eqref{eq:hybrid_filter} thus fixes the accessible controls at the two ends while leaving the decimated sector free there.
Such protocols can be used for fast-forward driving, where a schedule of accessible controls is designed to reproduce the final evolution of a counterdiabatic Hamiltonian between the two ends of the protocol. We discuss such a problem in Sec.~\ref{sec:ff_driving}.

\section{Nonperturbative Floquet engineering}
\label{sec:floquet_engineering}

Floquet engineering refers to the study of how periodic drives can be used to re-design the dynamical properties of physical systems~\cite{Weitenberg_Simonet_2021_NatPhys}.
The most common approach is based on high-frequency expansions~\cite{Goldman_Dalibard_2014_PRX,Bukov_2015_AdvPhys, Eckardt_2017_RMP, Eckardt_Anisimovas_2015_NJP,Rudner_Lindner_2020_NatRevPhys}, with drive frequency $\omega$ much larger than the natural energy scales $J$ of the nondriven system, $\omega\gg J$. 

Although Floquet engineering is, by now, part of the standard toolbox for quantum simulation, it is not straightforward to use: the design of the periodic drive requires both creativity and ingenuity and, for many interesting applications, remains involved and convoluted guesswork; this often precludes its on-demand use. 
In addition, high-frequency Floquet engineering is, by construction, limited to perturbative regimes. In many applications, the interesting engineered terms appear at order $\omega^{-1}$ and require the frequency $\omega$ to be neither too large nor too small: a too-large drive frequency decreases the prominence of the engineered effects in the dynamics, whereas a frequency too low risks that the engineered physics is polluted by unwanted higher-order corrections; a prime example for such a behavior is the realization of Haldane's model, where the Floquet-engineered topological gap appears at order $\omega^{-1}$ in the effective Hamiltonian~\cite{Oka_Aoki_2009_PRB, Jotzu_etal_2014_Nature, Eckardt_Anisimovas_2015_NJP}. 
Finally, inverse-frequency expansions, such as the Floquet Magnus or the van Vleck series, have been shown to be asymptotic, i.e., they diverge beyond a certain model-dependent expansion order~\cite{Mori_2016_AP}. 
These issues point at the necessity of devising alternative nonperturbative approaches to Floquet engineering. 

Below we demonstrate that the \ctrlRG flow paves a natural \textit{nonperturbative} road to \textit{on-demand} Floquet engineering. 
Unlike previous work on Floquet systems using flow equations~\cite{Verdeny_etal_2013_PRL, Vogl_etal_2019_PRX, Thomson_etal_2021_SciPost, Claassen_2021_arXiv, Novicenko_etal_2022_PRA, Shi_Mintert_2024_PRR} that started from a lab-frame Hamiltonian to find the effective Floquet Hamiltonian using a procedure similar to Wegner's SRG in Sambe space, here we are interested in the inverse control problem: we start from a given effective Hamiltonian to be engineered, and seek a time-periodic lab-frame Hamiltonian containing only accessible terms that implements it via Floquet theory. 
We emphasize that our approach is exact within numerical tolerance and does not rely on any approximations applied to the flow equations present in Refs.~\cite{Vogl_etal_2019_PRX, Thomson_etal_2021_SciPost}. Therefore, it allows us to (i) circumvent the perturbative deficiency of inverse-frequency expansions, extending Floquet engineering into nonperturbative regimes, and (ii) remove the obstacle with the intricate procedure of designing the periodic drive, thus preparing the Floquet engineering toolbox for on-demand use.

\subsection{Micromotion as \ctrlRG frame transformation}
\label{subsec:micromotion_CtrlRG}

Consider an inaccessible effective Hamiltonian $H_\text{eff}$ that we want to Floquet-engineer using some accessible periodic drive $H(t)$. These two Hamiltonians are constrained by Floquet's theorem to obey
\begin{equation}
\label{eq:inf_Floquet_thm}
    H(t) = P(t)H_\text{eff}P^\dagger(t) - i P(t) \dot P^\dagger(t)\, ,
\end{equation}
where $P(t)=P(t+T)$ is the time-periodic unitary micromotion operator. 
We seek a micromotion $P(t)$ which produces an accessible lab-frame implementation $H(t)$.  

To apply the \ctrlRG framework, we first identify the target unitary
\begin{equation}
    U_\ast=\exp(-i T H_\text{eff})
\end{equation}
with the one-period evolution of the static effective Hamiltonian; thus, we can use the latter as a time-independent seed Hamiltonian, i.e., $\Hseed=H_\text{eff}$.  
Next, observing the structural similarity between Eq.~\eqref{eq:inf_Floquet_thm} and Eq.~\eqref{eq:Hmu}, we interpret the change-of-frame transformation $V(\mu,t)$ as micromotion; periodicity in time $t$ is automatically enforced by the boundary conditions of Eq.~\eqref{eq:V_pinning}.
We emphasize that, although Eq.~\eqref{eq:Hmu} produces formally a family of lab-frame Hamiltonians $H(\mu,t)$ which all implement the desired Floquet evolution, they are in general inaccessible, since they retain an inaccessible part $\Hper(\mu,t)$. However, as we showed in Sec.~\ref{sec:theory}, the \ctrlRG flow gradually decimates the inaccessible components and produces an accessible lab-frame Hamiltonian in the limit $\mu_f\to\infty$. It follows that the \ctrlRG flow fixed point $H_\infty(t)$ of Eq.~\eqref{eq:H_infty}, by construction, reproduces the target unitary at stroboscopic times. In practice, we stop the flow at a large final flow parameter $\muf$ and use $H(\muf,t)\approx H_\infty(t)$.

Imposing the properties of the infinitesimal version of Floquet's theorem, Eq.~\eqref{eq:inf_Floquet_thm}, on Eq.~\eqref{eq:Hmu} requires that the flow Hamiltonian $H(\mu,t)$ is periodic with period $T=2\pi/\omega$. We therefore apply the formalism of Sec.~\ref{subsec:eng_floquet}, and require the boundary conditions $H(\mu,0)=H(\mu,T)$ for every fixed flow parameter $\mu$, see Table~\ref{tab:bc}. At the temporal boundaries $t{=}0,T$, the generator vanishes, fixing the Floquet gauge~\cite{Bukov_2015_AdvPhys}, so Eq.~\eqref{eq:flow_equation} reduces to $\partial_\mu H=\partial_tK$; enforcing the time-periodicity of the drive therefore translates to a condition on the temporal slope of $K$, $\partial_tK(\mu,0)=\partial_tK(\mu,T)$. As we explained in Sec.~\ref{subsec:eng_floquet}, this can be implemented using a fourth-order temporal filter $\Rreg_4$, which we realize numerically on a periodic time grid (see Appendix~\ref{app:reg:periodic}). 

This formulation of Floquet engineering as a \ctrlRG flow problem is generic. To illustrate it in practice, we now show a concrete example of nonperturbative Floquet engineering using the flow approach.

\begin{figure}[t!]
\centering
\includegraphics[width=\linewidth]{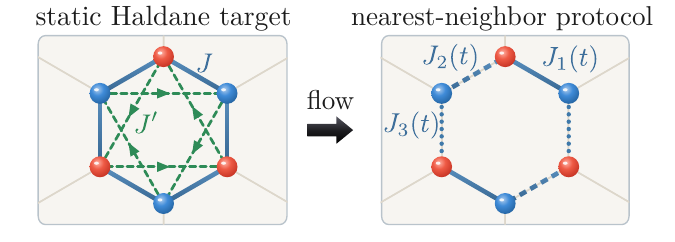}
\caption{
\textbf{Haldane's model as a control target, and its flow to an implementable protocol.} 
The left panel illustrates the static Haldane Hamiltonian of Eq.~\eqref{eq:haldane_seed}, which has a real-valued nearest-neighbor hopping $J$ (solid), a complex-valued next-nearest-neighbor hopping $J'$ (dashed), and a potential offset $M$ between the red and blue sublattices. 
The right panel illustrates the fixed point of the flow, i.e., the accessible implementation protocol of Eq.~\eqref{eq:haldane_acc} which has one complex-valued time-dependent hopping $J_\alpha(t)$ per bond direction; both the next-nearest-neighbor hopping and the sublattice offset have been decimated by the flow. The resulting nearest-neighbor protocol reproduces the same one-period unitary as the static Haldane Hamiltonian. See Sec.~\ref{subsec:Haldane_model} for details.
}
\label{fig:haldane_setup}
\end{figure}

\subsection{Nonperturbative Haldane model engineering}
\label{subsec:Haldane_model}

\textbf{\ctrlRG flow setup.---}
To engineer the Haldane model on the honeycomb lattice~\cite{Haldane_1988_PRL}, cf.~Fig.~\ref{fig:haldane_setup}, we encode it as the seed Hamiltonian
\begin{equation}
    \Hseed = - J\sum_{\langle ii'\rangle}c^\dagger_ic_{i'} - \sum_{\langle\langle ii'\rangle\rangle}J'_{ii'}c^\dagger_ic_{i'} + \mathrm{h.c.} + M\sum_i\xi_ic^\dagger_ic_i, 
\label{eq:haldane_seed}
\end{equation}
where $c^\dagger_i$ is the single particle creation operator on site $i$, the sums $\langle ii'\rangle$ and $\langle\langle ii'\rangle\rangle$ run over nearest- and next-nearest-neighbor bonds, and $\xi_i=+1$ ($-1$) on sublattice $A$ ($B$) makes the staggered sublattice potential term of strength $M$. The nearest-neighbor hopping $J$ is real-valued; the next-nearest-neighbor hopping $J'_{ii'}$ carries a phase, $\arg J'$, along the direction marked in Fig.~\ref{fig:haldane_setup} (left panel). Hence, each sublattice triangle [green dashed lines] encloses a flux $\pm 3 \arg J'$ of opposite sign for the two sublattices; this breaks time-reversal symmetry and allows for topologically nontrivial states of nonzero Chern number $C$. 

Using the standard rhombic supercell for the honeycomb lattice with periodic boundary conditions, we transform the seed Hamiltonian to momentum space, 
\begin{equation}
    \Hseed=\sum_{\mathbf k}c_{\mathbf k}^\dagger\hseed(\mathbf k)c_{\mathbf k}\ ,
\end{equation}
with
\begin{equation}
    \hseed(\mathbf k)=-J\left[\mathrm{Re}f(\mathbf k) \sigma_x-\mathrm{Im}f(\mathbf k) \sigma_y\right]+f_z(\mathbf k) \sigma_z\, . 
\label{eq:haldane_bloch_seed}
\end{equation}
Here $c_{\mathbf k}=(c_{\mathbf k,A},c_{\mathbf k,B})^{\mathsf T}$ is the sublattice spinor, the Pauli matrices $\sigma_x$, $\sigma_y$, and $\sigma_z$ act on the sublattice index, $f(\mathbf k)=\sum_\alpha e^{i\mathbf k\cdot\boldsymbol\delta_\alpha}$, $\boldsymbol\delta_\alpha$ define the three nearest-neighbor vectors, and $f_z(\mathbf k)=M+2 \mathrm{Im}(J')\sum_\alpha\sin(\mathbf k\cdot\mathbf b_\alpha)$ combines the sublattice offset and the next-nearest-neighbor hopping along the three next-nearest-neighbor vectors $\mathbf b_\alpha$. 

In numerical simulations, we use the natural discrete $N_k\times N_k$ momentum grid provided by the lattice quantization condition.

Having specified the seed Hamiltonian, we now choose as accessible controls the three nearest-neighbor hopping terms
\begin{equation}
    \Hacc(\mu,t) = -\sum_{\alpha=1}^{3}J_\alpha(\mu,t)\sum_{\langle ii'\rangle_\alpha}c^\dagger_i c_{i'}+\mathrm{h.c.},
\label{eq:haldane_acc}
\end{equation}
where $\langle ii'\rangle_{\alpha}$ runs over the nearest-neighbor bonds along direction $\alpha$, see Fig.~\ref{fig:haldane_setup} (right panel). Here we take $J_\alpha(\mu,t){\in}\mathbb{C}$, so that every other operator component in the target seed Hamiltonian of Eq.~\eqref{eq:haldane_seed} is part of the inaccessible subspace; in particular, we assume that the staggered potential breaking sublattice inversion is also inaccessible. 
The flow has to reproduce the effect of these unwanted terms using the nearest-neighbor hopping terms of Eq.~\eqref{eq:haldane_acc} alone. 

In the momentum basis, the accessible controls are given by $\Hacc(\mu,t) = \sum_{\mathbf k} c_{\mathbf k}^\dagger\hacc(\mu,t,\mathbf k) c_{\mathbf k}$, with
\begin{equation}
    \hacc(\mu,t,\mathbf k)=- \sum_{\alpha=1}^{3} \left( J_\alpha(\mu,t) e^{i\mathbf k\cdot\boldsymbol\delta_\alpha} \sigma_++\mathrm{h.c.} \right), 
\label{eq:haldane_bloch_acc}
\end{equation}
and $\sigma_\pm=(\sigma_x\pm i\sigma_y)/2$. They carry no $\sigma_z$ component at any instant during the flow, so the entire $f_z$ term of the target Hamiltonian in Eq.~\eqref{eq:haldane_bloch_seed} has to be generated by the late-flow micromotion.

\begin{figure*}[t!]
\centering
\includegraphics{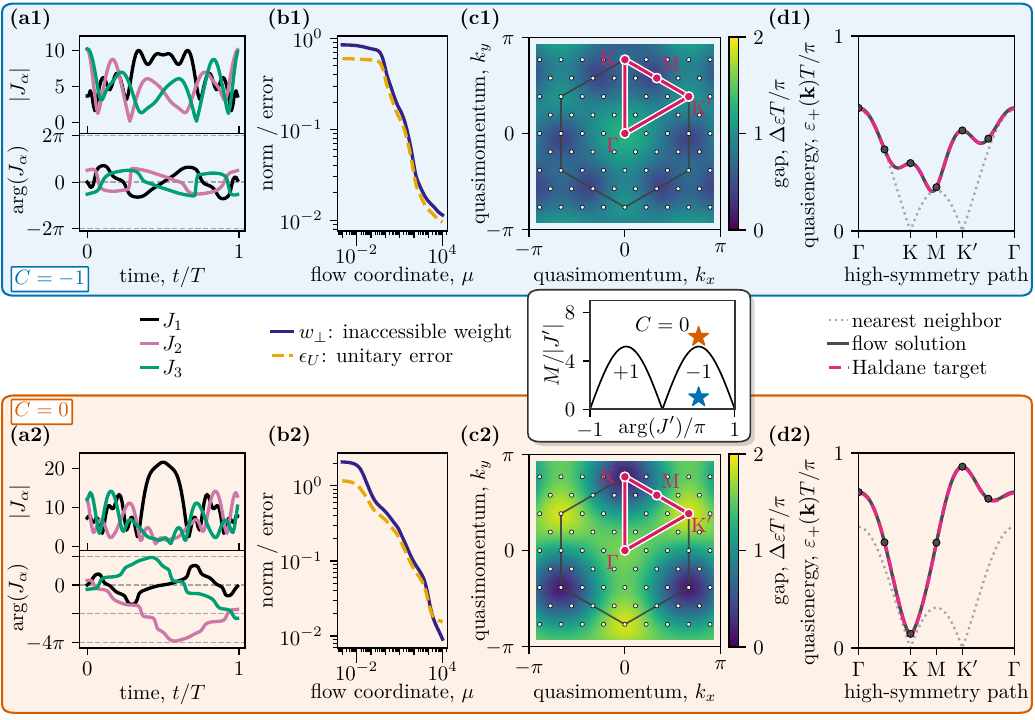}
\caption{
\textbf{\ctrlRG flow behavior and validation of the flow-engineered Haldane model.} The panels are grouped by topology: the blue group above is the topological case, $M/|J'|=1$ and $C=-1$, and the orange group below is the trivial one, $M/|J'|=6$ and $C=0$. The Haldane phase diagram at the center gives the lower-band Chern number of the static target Hamiltonian in each region, with stars marking the parameters for the two targets. 
\textbf{(a1,a2)} Magnitude and phase of the three converged hopping schedules over one drive period [legend between top and bottom panels]. Three schedules break the $C_3$ symmetry. 
\textbf{(b1,b2)} Inaccessible-control weight $\Wnorm(\mu)$ in Eq.~\eqref{eq:inaccessible_weight} (blue) and unitary error $\epsU(\mu)$ of the accessible drive in Eq.~\eqref{eq:haldane_epsU} (orange) decrease along the flow parameter $\mu$. 
\textbf{(c1,c2)} Quasienergy gap $\Delta\varepsilon T/\pi$ of the converged flow-engineered drive over the Brillouin zone. The gap is open everywhere, including at the Dirac points. White dots are the momentum grid points at which the flow constrains the unitary; pink lines denote the $\Gamma$--$\mathrm{K}$--$\mathrm{M}$--$\mathrm{K}'$--$\Gamma$ path (cf.~the horizontal axis of panels (d1,d2)). 
\textbf{(d1,d2)} Quasienergy along the path in (c1,c2) for the converged flow-engineered drive (black solid), the static Haldane target (pink dashed), and the nondriven nearest-neighbor term (gray dotted): the first two coincide, while the third stays gapless at the Dirac points. 
Simulation parameters use a drive period $JT=0.65$, purely imaginary $J'/J=0.4\,i$, and a filter scale $\gamma_4 J=0.1$, on a $6\times6$ momentum grid and a periodic time grid of $100$ points. 
The regularizer is the periodic fourth-order filter in Eq.~\eqref{eq:haldane_regularizer}. 
See Sec.~\ref{subsec:Haldane_model} for details. 
    }
\label{fig:haldane_flow}
\end{figure*}

\textbf{Breaking $C_3$-lattice symmetry in the flow.---}%
Notice that the accessible controls of Eq.~\eqref{eq:haldane_bloch_acc} with equal hoppings, $J_\alpha(\mu,t){=}J(\mu,t)$, would necessarily leave the two Dirac points gapless as a result of a $C_3$ lattice symmetry, see Appendix~\ref{app:haldane:algebra}. When this symmetry is present for all times within the period, the gap remains closed also in the effective Hamiltonian.
By contrast, the Haldane model of Eq.~\eqref{eq:haldane_seed} that we want to engineer exhibits finite gaps. 
Therefore, if the accessible Hamiltonian from \ctrlRG is to reproduce the target unitary at the Dirac points for $\mu_f\gg 1$ faithfully, it has to break the threefold lattice $C_3$-symmetry. This justifies the anisotropic ansatz of Eq.~\eqref{eq:haldane_bloch_acc}. 

However, a $C_3$-symmetry-preserving flow starting from the $C_3$-symmetric Haldane model as an initial seed can never break this symmetry, since the flow equation then maps $C_3$-symmetric Hamiltonians onto $C_3$-symmetric ones\footnote{In practice, we have observed symmetry breaking in the numerical solution due to amplified machine precision fluctuations, but this mechanism is unstable.}. Hence, the $C_3$-symmetry has to be broken explicitly in the flow equation. To achieve this, we use flow-equation engineering by weighting the regularizer $\Rreg$ along one bond direction with a positive $\mu$- and $t$-independent weight. 
This immediately breaks the $C_3$ symmetry in the generator of the flow while preserving the positivity of the regularizer and the boundary condition of Eq.~\eqref{eq:K_pinning}. Consequently, the inaccessible-control weight still contracts by Eq.~\eqref{eq:contraction}, and every Hamiltonian along the flow still generates the target unitary $\Utarget$ at $t=T$ by Eqs.~\eqref{eq:Hmu} and~\eqref{eq:V_pinning}.
Yet, it unlocks a path in control space to connect the initial seed to the $C_3$-symmetry-breaking fixed-point Hamiltonian of interest. 

In practice, we use the operator basis of the underlying algebra to expand the flow equation in, cf.~Sec.~\ref{subsec:flow_eq-num}.
Every component of this algebra shares the fourth-order regularizer $\Rreg_4$ of Eq.~\eqref{eq:filter4} with the periodic boundary conditions of Eqs.~\eqref{eq:Floquet_BC}. To break the threefold $C_3$-symmetry using weights in the generator, we define the regularizer
\begin{equation}
\Rreg=\Rreg_4\left[4 \Pi_1+\left(\mathbb 1-\Pi_1\right)\right]=\Rreg_4\left[3 \Pi_1+\mathbb 1\right],
\label{eq:haldane_regularizer}
\end{equation}
where $\Pi_\alpha$ projects onto the accessible controls of bond direction $\alpha$. Using this specific choice of weights, the first bond direction is driven four times stronger than the other two, while the contraction property of Eq.~\eqref{eq:contraction} remains intact. For details, see Appendices~\ref{app:reg:sector} and~\ref{app:haldane:algebra}.

\textbf{Results.---}%
We solve numerically the flow for two Haldane targets with trivial (bottom panel) and nontrivial (top panel) band topologies shown in the schematic phase diagram in the middle panel of Fig.~\ref{fig:haldane_flow}. 
At large flow parameter $\mu$, the flow converges to a protocol built from nearest-neighbor hopping alone, see Figs.~\ref{fig:haldane_flow}(a1,a2). 
As guaranteed by the contractive property, the inaccessible weight $\Wnorm(\mu)$ decreases monotonically and falls by approximately two orders of magnitude, as shown in Figs.~\ref{fig:haldane_flow}(b1,b2). 
Truncating the fixed-point protocols to the harmonics $|m|\le10$ ($11$) for the trivial (topological) case preserves the target unitary within the precision of the numerical solution, see Appendix~\ref{app:haldane:harmonics}. 

At each fixed flow parameter $\mu$, the nearest-neighbor (NN) accessible part of the drive generates the Floquet unitary
\begin{equation}
\Unn(\mu,\mathbf k)=\mathcal{T} e^{-i\int_0^T\mathrm dt\, \hacc(\mu,t,\mathbf k)}. \label{eq:haldane_Unn}
\end{equation}
Notice that the inaccessible terms are excluded from $\Unn(\mu,\mathbf k)$.
On the other hand, the full flow Hamiltonian reproduces $\Utarget(\mathbf k)=e^{-iT\hseed(\mathbf k)}$ exactly at every $\mu$. 
Hence, the distance between $\Unn$ and the target unitary $\Utarget(\mathbf k)$ can be measured by the one-period Frobenius norm error\footnote{The unitaries at different $\mathbf k$ act on the same many-body state; their relative phases are physical and included in $\epsU$. The overall phase shared by every $\mathbf k$ is unobservable, which makes $\epsU$ an upper bound on the physical error.}
\begin{equation}
\epsU(\mu)=\left[\frac{1}{2N_k^2}\sum_{\mathbf k}\Fnorm{\Unn(\mu,\mathbf k)-\Utarget(\mathbf k)}^2\right]^{1/2}. \label{eq:haldane_epsU}
\end{equation}
The error $\epsU(\mu)$ drops as the flow progresses, together with $\Wnorm(\mu)$, see Fig.~\ref{fig:haldane_flow}(b1,b2).
At the end of the flow we thus find an accessible drive that generates the target unitary with a small error, set by the residual inaccessible weight and by the tolerance of the flow equation solver. 

The fixed-point unitary $\Unn(\mu_f,\mathbf k)$ gives the spectrum of the implemented effective target Hamiltonian. Its two eigenvalues are $e^{-iT\varepsilon_\pm(\mathbf k)}$, which define the flow-engineered quasienergies $\varepsilon_\pm(\mathbf k)$ modulo $2\pi/T$, shown over the Brillouin zone in Figs.~\ref{fig:haldane_flow}(c1,c2) and along a high-symmetry path in Figs.~\ref{fig:haldane_flow}(d1,d2).
In turn, the flow-induced Chern number $C$ of the lower effective band $\varepsilon_{-}(\mathbf k)$ is computed from the corresponding Floquet eigenvectors. 

As expected, we observe that the quasienergy gap $\Delta\varepsilon=\varepsilon_+-\varepsilon_-$ in Figs.~\ref{fig:haldane_flow}(c1,c2) is open at the Dirac points. Along the high-symmetry path of Fig.~\ref{fig:haldane_flow}(d1,d2), the quasienergies of the converged drive lie on top of those of the static Haldane target; by contrast, the non-driven nearest-neighbor model at the same hopping stays gapless at the Dirac points, so the difference between the two dispersions is induced by the micromotion. 

For the flow-engineered topological target, the Chern number induced by the drive is $C{=}-1$, as in the corresponding Haldane model. Hence, the flow transfers the properties of the Haldane band topology onto the time dependence of the nearest-neighbor hopping alone. 
By contrast, the same calculation gives $C{=}0$ for the converged drive of the topologically trivial target. Thus, the flow reproduces both target Chern numbers using nearest-neighbor hopping alone.

\textbf{Comparison with perturbative Floquet engineering.---}
The celebrated Floquet protocols to engineer the Haldane Hamiltonian use high-frequency circular shaking of the honeycomb lattice~\cite{Rudner_Lindner_2020_NatRevPhys,Weitenberg_Simonet_2021_NatPhys}. 
As we alluded to above, however, these constructions are perturbative. 
The zeroth-order time-average term contains no next-nearest-neighbor hopping; the topological NNN-hopping term occurs as a first-order correction and scales as $|J'|\sim J^2/\omega$~\cite{Eckardt_Anisimovas_2015_NJP, Mikami_etal_2016_PRB}. 
Truncated to first order, the effective Hamiltonian provides a clean description as long as the drive stays weak and fast, so that higher-order terms in the inverse-frequency fall off quickly. 
Strong or slow drives, however, inevitably leave this perturbative regime of validity.

To display the nonperturbative aspect of the \ctrlRG fixed-point solution, we now compare the converged flow-engineered drives with a perturbative $\omega^{-1}$-series. Applying the Floquet-Magnus\footnote{We do not consider the van Vleck expansion, since we work with the stroboscopic Floquet unitary.} expansion to the fixed-point drive $\hacc(\muf,t,\mathbf k)$, which generates $\Unn(\muf,\mathbf k)$ of Eq.~\eqref{eq:haldane_Unn}, produces an effective static Hamiltonian order by order: the period-averaged Hamiltonian is nearest-neighbor, so the next-nearest-neighbor and staggered potential terms of the target can only occur in higher-order corrections, produced by the micromotion. 
For the topologically trivial target, we find that the norm of the terms in the Floquet-Magnus series grows at higher orders, while the truncated unitary error remains far above that of the flow solution (see Fig.~\ref{fig:magnus} in Appendix~\ref{app:haldane:pert}). 
For the topological target, the Floquet-Magnus series also stops decreasing and turns upward at higher orders; its truncated unitary remains several times less accurate than the flow fixed-point solution (see Fig.~\ref{fig:magnus}(b) in Appendix~\ref{app:haldane:pert}). 
In summary, adding higher-order terms does not reach the accuracy of the flow solution in either case.
The \ctrlRG flow, by contrast, constructs these drives nonperturbatively within the numerical tolerance. The full flow Hamiltonian is guaranteed to represent the target unitary exactly at every flow parameter $\mu$, so the accuracy of the accessible drive is limited only by the residual weight of the inaccessible terms and by the solver tolerance.

\section{Accessible fast-forward driving}
\label{sec:ff_driving}

Recently, there has been a surge of interest in counterdiabatic (CD) driving for quantum control~\cite{GueryOdelin_2019_RMP, Sels_Polkovnikov_2017_PNAS, Kolodrubetz_etal_2017_PhysRep}. In a nutshell, CD driving is designed to accelerate exact adiabatic dynamics by adding the adiabatic gauge potential (AGP) as an additional term to the original control Hamiltonian. The AGP compensates exactly for the diabatic excitations that would occur at fast ramp speeds, and hence the CD-evolved states follow the instantaneous eigenstates at any driving speed~\cite{Berry_2009_JPA, GueryOdelin_2019_RMP}. 
Due to the large success of the approach, several advanced algorithms have been proposed to compute the AGP approximately~\cite{Sels_Polkovnikov_2017_PNAS,Claeys_etal_2019_PRL,Hatomura_2021_PRA,Ieva_2023_PRXQ,Takahashi_2024_PRX,Morawetz_2024_PRB,Hegade_2021_PRApplied,Schindler_Bukov_2024_PRL}.
However, both exact and approximate gauge potentials often contain difficult-to-implement imaginary-valued current-like terms; moreover, generically the exact AGP contains long-ranged and multi-body terms. Therefore, the adiabatic gauge potential (and with it counterdiabatic driving) is usually inaccessible or challenging to apply in experiments.

To mitigate the problem, it was noted that the exact AGP operator can be entirely generated by the dynamical Lie algebra of the terms present in the original control Hamiltonian~\cite{petiziol2018fast,Claeys_etal_2019_PRL}. This implies the existence of a so-called fast-forward (FF) protocol which only reproduces the target unitary at the two temporal boundaries at $t=0,T$~\cite{Masuda_Nakamura_2010_PRSA, Torrontegui_2012_PRA, Patra_Jarzynski_2017_NJP,Bukov_2019_PRX};\footnote{For instance, in a two-level system with accessible $x$ and $z$-control fields, the AGP points along the $y$-direction; the statement about the existence of a fast-forward protocol then reduces to the Euler angle theorem which allows us to implement any rotation, including those along the $y$-axis, using only the accessible $x$- and $z$-controls.} the state itself is free to make nonadiabatic ``excursions'' away from the instantaneous eigenbasis at intermediate times. 
The study of FF protocols has attracted significant attention because they inherit the highly desirable final unit-fidelity property of the underlying CD dynamics, while using only accessible control terms. 
However, the explicit construction of these potentially experimentally very useful FF protocols has remained elusive besides simple few-level systems and brute-force cost-function optimization.

Beyond brute-force constrained optimization~\cite{Caneva_2009_PRL,Caneva_2011_PRA,Doria_2011_PRL,Bukov_2019_PRX}, in which the schedules are parametrized and the endpoint state fidelity is maximized by some variant of gradient ascent, the state-of-the-art approach makes use of Floquet engineering~\cite{Claeys_etal_2019_PRL}. However, as we saw in Sec.~\ref{sec:floquet_engineering}, the latter is usually limited to the perturbative high-frequency regime; moreover, because the quality of the resulting protocols is controlled by increasing the Floquet frequency, it is not optimal for implementations in complex systems where large frequencies can induce unwanted resonances (e.g., direct higher-band transitions in quantum simulators based on optical lattices). 

As an alternative, we now show that FF protocols can be naturally constructed using the \ctrlRG framework. Importantly, the approach introduces no auxiliary frequency scales and is fully nonperturbative. Moreover, it inherits some benefits from the RG framework, such as monotonic decimation of inaccessible terms.

\begin{figure}[t!]
\centering
\includegraphics{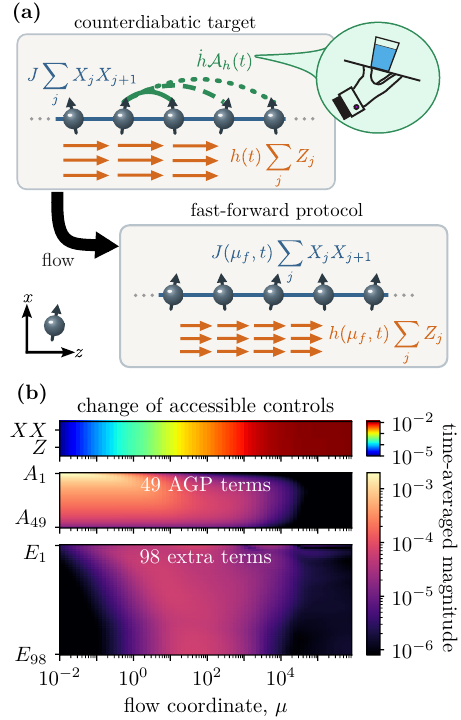}
\caption{
\textbf{\ctrlRG flow to a fast-forward (FF) protocol in the transverse-field Ising chain.} 
\textbf{(a)} Schematic illustration of the flow from the seed counterdiabatic (CD) Hamiltonian to the fast-forward protocol. The seed contains the static Ising interaction $J$, the transverse field $h(t)$, and the nonlocal adiabatic gauge potential (AGP) term $\dot h(t)\agp_h(t)$, which enforces adiabatic evolution but is inaccessible. The flow decimates the AGP term and produces a fast-forward protocol with only the renormalized Ising interaction $J(\muf,t)$ and transverse field $h(\muf,t)$. 
\textbf{(b)} Time-averaged magnitude of every operator coefficient as a function of the flow parameter $\mu$. For the two accessible controls, nearest-neighbor $XX$ and $Z$, the upper band shows instead the time-averaged change from the seed schedule; the middle band shows the 49 inaccessible AGP schedules, and the lower band the 98 schedules of additional inaccessible terms. Both inaccessible bands are ordered by operator range ($A_\ell$ is a string of range $\ell$, $E_{2\ell-1}$ and $E_{2\ell}$ have range $\ell+1$ and $\ell$, respectively). 
The inaccessible schedules are decimated as the flow parameter increases, while the accessible coefficients change to form the fast-forward protocol. 
The simulation uses $L{=}50$ spins, a field ramp from $h/J{=}0.7$ to $h/J{=}1.4$, $JT{=}43.79$, and a physical-time grid of $201$ points. The regularizer is the hybrid filter of Eq.~\eqref{eq:hybrid_filter} with $\gamma_4 J{=}1$ and $\gamma_2 J{=}2$.
See Secs.~\ref{subsec:FF_CtrlRG} and~\ref{subsec:FF_TFIM} for details.
}
\label{fig:tfim_cartoon}
\end{figure}

\subsection{Accessible realization of counterdiabatic driving within the \ctrlRG framework}
\label{subsec:FF_CtrlRG}

Consider an adiabatic control problem defined by the control Hamiltonian
\begin{equation}
H_{\rm ctrl}(\lambda) = H_0+\lambda H_1, 
\label{eq:H_ctrl}
\end{equation}
where $H_0$ and $H_1$ are the accessible control terms, and $\lambda(t)$ is a fixed time-dependent control schedule. In the adiabatic limit, the evolved state follows the instantaneous eigenstates $|n[\lambda]\rangle$ of $H_{\rm ctrl}(\lambda)$ with eigenenergies $E_n(\lambda)$; however, the system gets excited away from the adiabatic regime. 

Counterdiabatic driving counteracts these diabatic excitations by adding the adiabatic gauge potential (AGP) $\agp_\lambda$ to the control Hamiltonian, 
\begin{equation}
\HCD(t) = H_{\rm ctrl}(\lambda(t))+\dot\lambda(t)\agp_{\lambda}(t)\, ;
\label{eq:H_CD}
\end{equation}
the AGP is defined by its matrix elements in the instantaneous eigenbasis~\cite{Kolodrubetz_etal_2017_PhysRep},
\begin{equation}
\agp_\lambda = i\sum_{m\neq n}\frac{|m\rangle\langle m|\partial_\lambda H_{\rm ctrl}|n\rangle\langle n|}{E_n-E_m}.
\label{eq:agp_general}
\end{equation}

Equivalently, it is sometimes useful to expand the AGP in nested commutators of the control Hamiltonian~\cite{Claeys_etal_2019_PRL},
\begin{equation}
\agp_\lambda = i\sum_{k\ge1}\alpha_k \text{ad}_{H_{\rm ctrl}}^{2k-1}\left(\partial_\lambda H_{\rm ctrl}\right),
\label{eq:agp_commutator}
\end{equation}
with $\text{ad}_{H}(O)\equiv[H,O]$. Truncating this series and determining the coefficients $\alpha_k$ variationally produces approximate AGPs. Each term of Eq.~\eqref{eq:agp_commutator} is a repeated commutator of $H_0$ and $H_1$, so the AGP and $\HCD(t)$ live in the Lie algebra $\mathfrak g$ generated by the accessible controls $H_0, H_1$. General theorems of optimal control for systems on compact Lie groups~\cite{Jurdjevic_Sussmann_1972_JDE,Bukov_2019_PRX} guarantee that, for a long enough duration $T$, the target unitary 
\begin{equation}
\label{eq:Utgt_FF}
    \Utarget=\mathcal T e^{-i\int_0^T\mathrm dt \HCD(t)}    
\end{equation}
can be reproduced by a fast-forward Hamiltonian 
\begin{equation}
\label{eq:H_FF}
    H_\text{FF}(t) = a_\text{FF}(t) H_0 + b_\text{FF}(t) H_1
\end{equation}
built from the accessible controls $a_\text{FF}(t), b_\text{FF}(t)$ alone. However, these existence theorems are non-constructive, and the duration they ask for need not be compatible with the finite coherence times in experiments; the practical difficulty is therefore how to find such protocols constructively at a prescribed duration and schedule smoothness.

We designed the \ctrlRG flow with exactly this setup in mind, since the counterdiabatic Hamiltonian already generates the target unitary $U_\ast$, see Eq.~\eqref{eq:Utgt_FF}, and only its representation has to be modified to ensure an accessible implementation of CD driving. We therefore take the CD Hamiltonian of Eq.~\eqref{eq:H_CD} as the seed, $\Hseed(t)=\HCD(t)$, and let $H_0$ and $H_1$ be the accessible operators selected by the projector $\Pacc$. The flow equation, Eq.~\eqref{eq:flow_equation}, then deforms the seed while preserving $\Utarget$; its fixed point is a fast-forward Hamiltonian, $\HFF(t)=\Hacc(\muf,t)$. The frame transformation of Eqs.~\eqref{eq:Hmu} and \eqref{eq:V_pinning} then generates the nonadiabatic excursions between the two temporal boundaries $t=0,T$. 

Unlike the Floquet target of Sec.~\ref{sec:floquet_engineering}, the fast-forward protocol lives on an open time interval; hence, we require the accessible controls to remain fixed at the two ends, see Sec.~\ref{subsec:eng_ff} and Table~\ref{tab:bc}. 
The accessible part of the generator must therefore have vanishing slope at the two temporal boundaries, $\Pacc\partial_tK(\mu,0)=\Pacc\partial_tK(\mu,T)=0$, which supplies the additional conditions required to apply the fourth-order regularizer of Eq.~\eqref{eq:filter4}. Since the inaccessible terms need not be fixed at the ends, we choose the second-order filter $\Rreg_2$ for them, giving in total the hybrid regularizer of Eq.~\eqref{eq:hybrid_filter}.

In the following, we show that the \ctrlRG flow produces a fast-forward protocol for transverse-field Ising chains of up to $L=50$ sites, built only out of the accessible Ising interaction and transverse field.

\begin{figure*}[t!]
\centering
\includegraphics{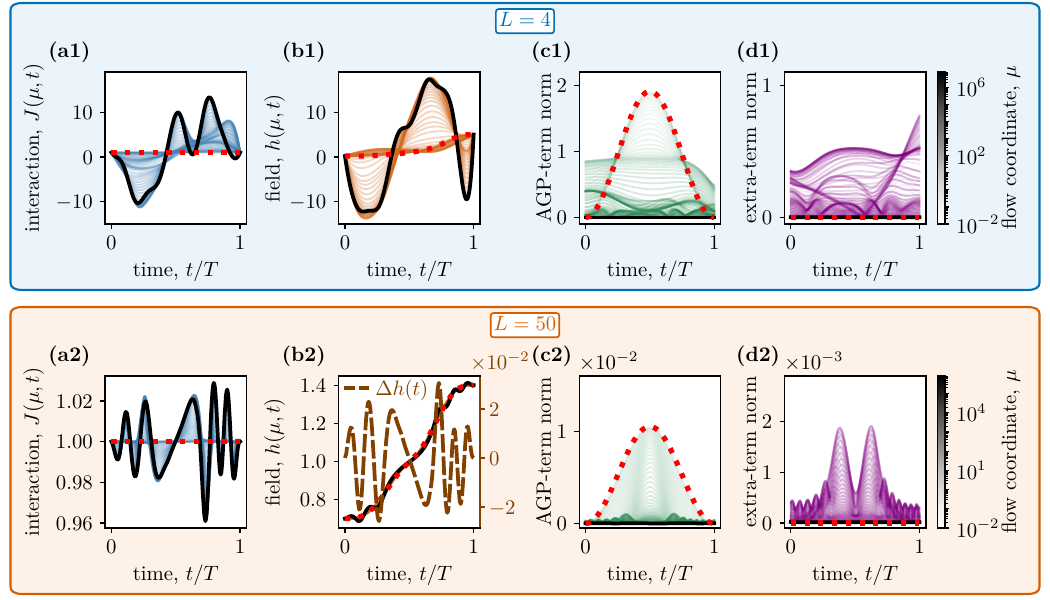}
\caption{\textbf{Flow histories of the two nonadiabatic Ising protocols.} 
First row: $L=4$ protocol with field ramp $h/J=0.2\to5.0$ and duration $JT=0.32$. Second row: $L=50$ protocol with $h/J=0.7\to1.4$ and $JT=43.79$. 
\textbf{(a1,a2)} Interaction schedule $J(\mu,t)$ along the flow. The constant seed interaction $J$ (red, dotted) becomes time-dependent during the flow, while its values at $t=0,T$ are fixed. 
\textbf{(b1,b2)} Transverse-field schedule $h(\mu,t)$ along the flow; panel (b2) also shows $\Delta h(t)=h(\muf,t)-h(0,t)$ (brown, dashed) on the right axis. The flow reshapes the seed ramp, with corrections to the field being of the same order of magnitude as those to the interaction schedule in (a1,a2). 
\textbf{(c1,c2)} Instantaneous magnitude $w_{\mathfrak a}(\mu,t)$ of the gauge-potential sector against physical time, defined in Eq.~\eqref{eq:tfim_sector_norms}. The flow decimates the gauge-potential term by more than three orders of magnitude.
\textbf{(d1,d2)} The same for the additional-term sector, $w_{\mathfrak e}(\mu,t)$. These terms are absent in the initial seed, generated at intermediate flow parameters, and decimated again as the flow converges. 
Red dotted curves are the seed protocol at $\mu=0$, colored curves show intermediate flow values, and the black curve is the converged FF protocol; color bars give the flow parameter $\mu$. 
In both rows, the inaccessible norms are decimated by the end of the flow while the accessible schedules change. 
All quantities are in units of the seed Ising interaction $J$. All calculations use $201$ points in time grid, and the hybrid regularizer of Eq.~\eqref{eq:hybrid_filter} with $(\gamma_4 J,\gamma_2 J){=}(0.4,0.8)$ in the first row and $(\gamma_4 J,\gamma_2 J){=}(1.0,2.0)$ in the second.
See Sec.~\ref{subsec:FF_TFIM} for details.
}
\label{fig:tfim_history}
\end{figure*}

\subsection{Fast-forward driving of the transverse-field Ising chain}
\label{subsec:FF_TFIM}

\textbf{Inaccessible counterdiabatic driving.---}
To illustrate how to compute FF Hamiltonians in practice using the \ctrlRG flow, consider the transverse-field Ising chain 
\begin{equation}
\HTFIM(t) = -J\sum_{i=1}^{L}X_iX_{i+1} - h(t)\sum_{i=1}^{L}Z_i,
\end{equation}
where $X_i$, $Y_i$, and $Z_i$ are Pauli operators at site $i$, $L$ is the number of sites, $J$ is the Ising interaction, and $h(t)$ is the time-dependent transverse field; we use periodic boundary conditions. 
The transverse-field Ising chain is the standard benchmark for counterdiabatic driving and shortcuts to adiabaticity~\cite{delCampo_etal_2012_PRL,Sels_Polkovnikov_2017_PNAS,GueryOdelin_2019_RMP}. It is an ideal playground for \ctrlRG: whereas its exact gauge potential is known in closed form~\cite{delCampo_etal_2012_PRL,Kolodrubetz_etal_2017_PhysRep}, it is long-ranged and multibody~\cite{Sels_Polkovnikov_2017_PNAS}, which makes exact CD driving inaccessible.

We choose as the control parameter the transverse field, $\lambda=h$; then, the counterdiabatic Hamiltonian of Eq.~\eqref{eq:H_CD} is $\HCD(t)=\HTFIM(t)+\dot h(t)\agp_h(t)$, and the AGP of Eq.~\eqref{eq:agp_general} can be exactly expressed as 
\begin{align}
\agp_h(t) &= \sum_{\ell=1}^{L-1}\frac{J}{2}\left[\alpha_\ell^{+}(t)+\alpha_\ell^{-}(t)\right]A_\ell, \label{eq:tfim_agp_sectors}\\
\alpha_\ell^{\pm}(t) &= \sum_{k\in{\rm BZ}^{\pm}}\frac{1}{4L}\frac{i\sin k e^{ik\ell}}{\left[h(t)-J\cos k\right]^2+J^2\sin^2k}, \label{eq:tfim_alpha}\\
A_\ell &= \sum_{j=1}^{L}\left(\prod_{j<m<j+\ell}Z_m\right)\left[Y_jX_{j+\ell}+X_jY_{j+\ell}\right], \label{eq:tfim_agp_string}
\end{align}
where ${\rm BZ}^{\pm}$ in Eq.~\eqref{eq:tfim_alpha} are the momentum grids of the two fermion-parity sectors of the Jordan-Wigner transformation spanning the Brillouin zone~\cite{delCampo_etal_2012_PRL}. 
Notice the multibody string operators $A_\ell$ in Eq.~\eqref{eq:tfim_agp_string} couple two distant spins through a product of $Z$ operators on every site in between. The exact gauge potential contains all of them up to $\ell=L-1$, and hence implementing $\HCD$ exactly would require the control of $L-1$ multi-body couplings in real time. No present-day quantum simulation platform can do that.

The protocols we consider lie outside the adiabatic regime: bare ramps of the same durations have endpoint infidelities $0.7$ and $0.2$ for $L=4$ and $L=50$, respectively [cf.~Fig.~\ref{fig:tfim_history}(a,b) red-dotted lines]. We therefore take the counterdiabatic Hamiltonian as the seed Hamiltonian, and seek a fast-forward protocol with only the accessible Ising interaction and transverse field as time-dependent control schedules.

\textbf{\ctrlRG flow setup.---}
The \ctrlRG flow offers a construction of accessible FF protocols by starting from the seed Hamiltonian, cf.~Sec.~\ref{subsec:FF_CtrlRG}. 
At every value of the flow parameter, the accessible part of the Hamiltonian contains only protocols of the Ising interaction and the transverse field,
\begin{equation}
\Hacc(\mu,t)=-J(\mu,t)\sum_{i=1}^{L}X_iX_{i+1}-h(\mu,t)\sum_{i=1}^{L}Z_i,
\end{equation}
where the schedules $J(\mu,t)$ and $h(\mu,t)$ reduce to $J$ and $h(t)$ at $\mu{=}0$ by the boundary condition of Eq.~\eqref{eq:seed_bdary}. In turn, the fast-forward Hamiltonian is the endpoint of the flow, $\HFF(t)\equiv\Hacc(\muf,t)$, see Fig.~\ref{fig:tfim_cartoon}(a). 

The two accessible controls and the gauge potential generate an algebra
\begin{equation}
\mathfrak g=\mathfrak c\oplus\mathfrak a\oplus\mathfrak e, \label{eq:tfim_algebra}
\end{equation}
whose accessible-control ($\mathfrak c$), gauge-potential ($\mathfrak a$), and additional-term ($\mathfrak e$) subspaces have dimensions $2$, $L-1$, and $2L-2$, respectively (see Appendix~\ref{app:ising:basis} for the explicit operator basis used). 
The \ctrlRG flow equation, Eq.~\eqref{eq:flow_equation}, for this problem is therefore a set of $3L-1$ coupled nonlinear PDEs, i.e., $149$ equations at $L=50$. Figure~\ref{fig:tfim_cartoon}(b) traces all of them along the flow, grouped into three bands. The gauge-potential schedules in the middle band fall by more than three orders of magnitude [see colorbar], and the remaining inaccessible schedules in the lower band, generated at intermediate flow parameters, are decimated as well, while the change of the two accessible controls in the upper band saturates as they converge to the sought-after fast-forward protocols.
Hence, as it progresses, the flow transfers weight from the inaccessible terms onto the accessible schedules.
Crucially, unlike optimization algorithms, the converged \ctrlRG flow inherits the unit fidelity guarantees of the seed CD Hamiltonian within numerical tolerance.

\textbf{Results.---}
We solve the flow at two chain lengths: 
(i) a small one of $L=4$ where every diagnostic is still available in the full spin basis; and 
(ii) a large one of $L=50$ where the flow is still tractable since the computation is limited by the size of the algebra rather than that of the spin Hilbert space. Both protocols ramp the transverse field across the critical point $h/J{=}1$, with the narrower ramp at $L=50$ keeping the counterdiabatic amplitude bounded. 
For each chain, we fix the endpoint infidelity of the bare protocol and solve for the duration $T$ that achieves it (see Appendix~\ref{app:ising:schedule}). In this way, the duration is set by how far the bare protocol lies outside the adiabatic regime, which places protocols of different length on a common footing since the same $T$ can give very different bare infidelities at different $L$. 

\begin{figure*}[t!]
\centering
\includegraphics{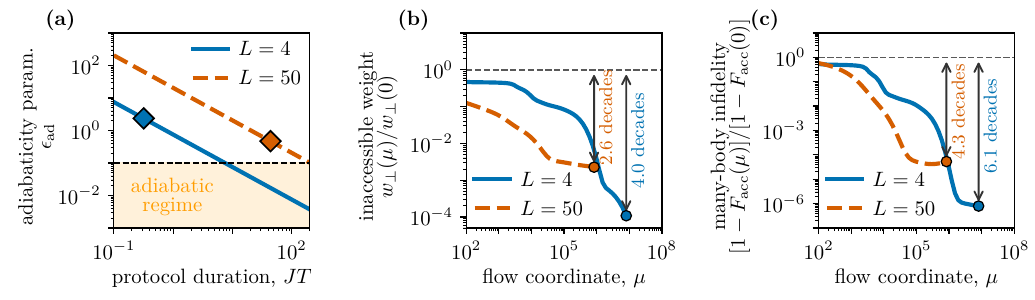}
\caption{\textbf{Adiabaticity and flow diagnostics of the two Ising protocols.} 
\textbf{(a)} Adiabaticity parameter $\epsad$ of the seed protocol against its duration $T$. Diamonds mark the two protocols used in Fig.~\ref{fig:tfim_history}, with $\epsad=2.35$ for $L=4$ and $0.47$ for $L=50$, respectively. 
The dashed line at $\epsad=0.1$ marks the upper edge of the shaded region, shown as a guide to the adiabatic limit $\epsad\ll1$. Both seed protocols stay outside the adiabatic regime. 
\textbf{(b)} Relative inaccessible-control norm $\Wnorm(\mu)/\Wnorm(0)$ dropping by several orders of magnitude along the flow. The decrease is monotonic, as guaranteed by the contraction property of Eq.~\eqref{eq:contraction}. 
\textbf{(c)} Endpoint infidelity relative to the bare seed protocol, $[1-\Facc(\mu)]/[1-\Facc(0)]$, evaluated with the accessible schedules alone, which falls by more than four orders of magnitude. The converged accessible schedules alone thus reproduce the target ground state. 
The duration of each protocol is calibrated so that the bare seed protocol has endpoint infidelity $1{-}\Facc(0)=0.7$ at $L=4$ and $0.2$ at $L=50$, respectively. The remaining parameters are listed in Fig.~\ref{fig:tfim_history}.
See Sec.~\ref{subsec:FF_TFIM} for details.}
\label{fig:tfim_diagnostics}
\end{figure*}

Figure~\ref{fig:tfim_history} shows the complete flow histories of the two protocols. The interaction and the field schedules of Figs.~\ref{fig:tfim_history}(a1,a2) and \ref{fig:tfim_history}(b1,b2) deform continuously from the seed protocol to the converged one. Figures~\ref{fig:tfim_history}(c1,c2) and (d1,d2) trace the two inaccessible sectors through their instantaneous magnitudes
\begin{equation}
w_{\mathfrak a}(\mu,t)=\frac{\lVert\mathcal P_{\mathfrak a}H(\mu,t)\rVert}{\mathcal N_L}, \quad w_{\mathfrak e}(\mu,t)=\frac{\lVert\mathcal P_{\mathfrak e}H(\mu,t)\rVert}{\mathcal N_L}, \label{eq:tfim_sector_norms}
\end{equation}
where $\mathcal P_{\mathfrak a}$ and $\mathcal P_{\mathfrak e}$ project onto the gauge-potential and the additional-term subspace of Eq.~\eqref{eq:tfim_algebra}; the normalization factor $\mathcal N_L = \sqrt{L 2^L}$ puts both on the same order of magnitude as $J$ and $h$. As the flow progresses, $w_{\mathfrak a}(\cdot, t)$ and $w_{\mathfrak e}(\cdot, t)$ fall towards zero as functions of the physical time $t$. 
The two rows differ in scale and show the same overall behavior: at $L=4$ the flow reshapes both schedules strongly, whereas at $L=50$ the corrections to the interaction and to the field are both within the order of $10^{-2}$, comparable to the gauge-potential magnitude they absorb, see Fig.~\ref{fig:tfim_history}(c2). The converged schedules are smooth and bounded, and they meet the seed values (i.e., the values of the original adiabatic schedules) at the two ends by construction.

Note that the corrections of order $10^{-2}$ in the flow of $L=50$ spins are nevertheless necessary to keep the state transitionless, since both seed protocols are driven outside the adiabatic regime. 
To quantify this, notice that under the Jordan-Wigner transformation the chain decomposes into independent two-level problems, one per momentum, with instantaneous gap $\Delta_k(t)$. The adiabaticity of a ramp can be measured by 
\begin{equation}
\epsad=\max_{t,k}\frac{|\dot h(t)|\left|\langle e_k(t)|\partial_hH_k(t)|g_k(t)\rangle\right|}{\Delta_k(t)^2},
\label{eq:tfim_adiabaticity}
\end{equation}
where $|g_k(t)\rangle$ and $|e_k(t)\rangle$ are the instantaneous ground and excited states in the momentum sector $k$ (see Appendix~\ref{app:ising:agp}). Intuitively, $\epsad$ measures the worst-case breakdown of adiabatic evolution across the entire Brillouin zone and protocol duration. For instance, an adiabatic ramp has $\epsad\ll1$; the shaded region $\epsad<0.1$ in Fig.~\ref{fig:tfim_diagnostics}(a) is a guide to this regime. Figure~\ref{fig:tfim_diagnostics}(a) shows $\epsad$ against the protocol duration for the two chain lengths; the markers label the durations we use, where $\epsad$ is non-negligible compared with unity in both cases.

The construction of FF protocols using the \ctrlRG flow equation succeeds once the inaccessible weight of Eq.~\eqref{eq:inaccessible_weight} drops below the preset stopping threshold value of inaccessible weight; its accuracy can then be read off from the endpoint fidelity of the accessible schedules alone,
\begin{equation}
\Facc(\mu)=\left|\left\langle\psi_{\rm f}\left|\mathcal{T}e^{-i\int_0^T\mathrm dt \Hacc(\mu,t)}\right|\psi_{\rm i}\right\rangle\right|^2,
\label{eq:tfim_fidelity}
\end{equation}
where $|\psi_{\rm i/f}\rangle$ are the ground states of $\HTFIM$ at the initial/final transverse field. Notice that the flow constrains the full propagator; hence the above ground-state fidelity is only a convenient probe. 
The relative inaccessible weight $\Wnorm(\mu)/\Wnorm(0)$ decreases monotonically, by about $4$ orders of magnitude at $L=4$ and $2.6$ orders of magnitude at $L=50$ [Fig.~\ref{fig:tfim_diagnostics}(b)]. Thus, the decimation of the inaccessible terms is guaranteed at every flow parameter $\mu$, even on the discrete time grid and for the $149$ coupled equations at $L=50$. 
The endpoint infidelity $1-\Facc(\mu)$ drops by $6.1$ and $4.3$ orders of magnitude, to $6\times10^{-7}$ at $L=4$ and $1\times10^{-5}$ at $L=50$ [Fig.~\ref{fig:tfim_diagnostics}(c)] --- a remarkable improvement for many-body state control where a small mismatch between the final and target wavefunctions is amplified exponentially in the number of spins $L$.  

\textbf{Comparison with perturbative Floquet-engineered fast-forward driving.---}%
The state-of-the-art implementation of CD driving by accessible controls uses Floquet engineering~\cite{Claeys_etal_2019_PRL}. There, the AGP is expanded in the nested-commutator series Eq.~\eqref{eq:agp_commutator}, truncated to a few terms when dealing with large systems, and then reproduced stroboscopically by a fast periodic modulation of the accessible controls. This construction thus relies on two approximations: 
(i) the variational form of the truncated AGP should approximate the exact one to a good accuracy; and 
(ii) the drive frequency of the Floquet modulation should be large enough for the stroboscopic Hamiltonian to reproduce the truncated CD Hamiltonian. 

We apply this Floquet engineering construction to the two Ising protocols of Fig.~\ref{fig:tfim_history} and compare it with the \ctrlRG flow in Appendix~\ref{app:ising:floquet_cd}. To have a meaningful comparison, we fix the drive frequency by the maximum frequency, $\omega_{\rm c}$, contained in the converged \ctrlRG schedules in Fig.~\ref{fig:tfim_history}, since the Floquet engineering construction improves with increasing drive frequency but excessively large drive frequencies are also inaccessible. For the small chain with $L=4$, the variational AGP of second order already coincides with the exact one, so the bottleneck is the drive frequency: at the frequency $\omega_{\rm c}$, the Floquet-engineered drive remains almost five orders of magnitude less accurate than the flow; it improves only as the drive frequency and amplitude are increased further beyond the values required for the \ctrlRG-flow FF schedules, and it still stays above $10^{-4}$ at $\omega=10^3\omega_{\rm c}$ (see Fig.~\ref{fig:fe_frequency}(a) in Appendix~\ref{app:ising:floquet_cd}). For the large chain with $L=50$, by contrast, the bottleneck is the truncated AGP itself, and the Floquet-engineered drive saturates near the infidelity of the bare ramp (see Fig.~\ref{fig:fe_frequency}(b) in Appendix~\ref{app:ising:floquet_cd}). 

To separate the high-frequency approximation and the AGP truncation, we also start the \ctrlRG flow from the same truncated CD Hamiltonian, so that both constructions have the same target unitary. We observe that even then, the Floquet-engineered drive remains less accurate than the flow at the same drive frequency, by more than one order of magnitude at $L=4$ and by four orders of magnitude at $L=50$ (see Fig.~\ref{fig:fe_truncated_seed} in Appendix~\ref{app:ising:floquet_cd}). 

Therefore, the Floquet-engineered drive is limited by both of its approximations. The \ctrlRG flow, by contrast, introduces no auxiliary frequency scale, and its accuracy is limited only by the residual weight of the inaccessible terms, set by the stopping threshold and by the PDE solver tolerance.

\section{Discussion and outlook}
\label{sec:discussion}

Given a target unitary and its seed-Hamiltonian representation containing terms that cannot be implemented directly, we construct a time-dependent frame in which experimentally accessible controls generate the same unitary. Fixing the change-of-frame transformation to the identity at the beginning and end of the protocol is sufficient to guarantee this equivalence. We design this construction using the Control Similarity Renormalization Group (\ctrlRG) flow, which acts on time-dependent Hamiltonians over the full protocol duration. Within this framework, Hamiltonians composed of the prescribed control terms form a fixed-point manifold; linearizing the flow around it gives a spectrum of decay rates that remove unwanted terms scale by scale, in direct correspondence with RG, cf.~Table~\ref{tab:rg_correspondence}. In the applications we consider, experimentally accessible control determines which Hamiltonian terms to retain; in turn, terms outside the experimentally accessible control set form the sector decimated by the flow. 

To improve the versatility of our \ctrlRG framework, we introduce flow-equation engineering: the design of suitable flow generators with the help of boundary value problems. Different generators select different paths through control Hamiltonian space while preserving the monotonic decrease of the inaccessible-control norm that ensures the decimating properties of the flow. We show instances of how this freedom can be exploited to enforce temporal-edge boundary conditions (periodic and open drives), smooth time dependence of the final protocols, and symmetry requirements. 
Beyond these instances, the same freedom can be deformed to impose further requirements, e.g., controlling the amplitude growth of the accessible control, flowing the decimated terms to a prescribed time-dependent schedule, or letting the protocol duration itself flow~\cite{Zenodo_2026}. 
It is an open problem to determine the limitations of flow equation engineering in implementing physical constraints of the final drive schedules.
    
Once the flow equation has been engineered for a concrete application, its numerical implementation is a direct nonlinear evolution in the flow parameter $\mu$.
Expanding the Hamiltonian in a basis of the operator algebra $\mathfrak g$ gives $\dim\mathfrak g$ coupled nonlinear partial differential equations, rather than a number set by the Hilbert-space dimension; discretizing physical time then converts them into a system of ordinary differential equations in $\mu$. 
The monotonic decimation of inaccessible control terms relies on integration by parts in physical time, which we show can be preserved by a summation-by-parts discretization. The contraction of the flow is therefore exact on the time grid to any finite resolution; additional numerical challenges can arise due to stiffness, time-grid convergence, and operator algebra growth. 
We note that the monotonic contraction of the inaccessible-control norm does not, by itself, guarantee convergence to the prescribed control subspace: a plateau in this norm may arise from limited temporal resolution, since both a sparse time grid and a strongly smoothing regularizer suppress high-frequency temporal directions. Because the full Hamiltonian generates the target unitary at every flow parameter $\mu$ up to this discretization error, if a plateau persists upon varying the grid and filter strength, it instead signals a (near-)marginal RG direction of the flow itself; then, the flow decay spectrum provides an explicit diagnosis when the flow becomes marginal. 
Unlike optimal control, \ctrlRG does not involve variational optimization over control parameters and therefore avoids local minima in the parameter space. 

The \ctrlRG framework offers a new tool for Floquet engineering. Unlike schemes based on high-frequency expansions, the flow equation is nonperturbative and can be applied on-demand, i.e., with arbitrary effective Hamiltonians.  
For instance, for a target unitary generated by a static Haldane Hamiltonian, the flow constructs a finite-period protocol using only three time-dependent, bond-resolved nearest-neighbor hopping schedules, without directly implementing the next-nearest-neighbor hopping or sublattice offset of the target Hamiltonian. The resulting drive reproduces the target stroboscopic evolution, quasienergy-gap structure, and band topology for both topological and trivial targets within numerical accuracy. Because the flow targets the one-period Floquet unitary directly, it applies at preset finite driving periods without relying on a perturbative high-frequency expansion; indeed, the corresponding Magnus expansion either diverges or stops improving before reaching the same accuracy as the flow solution. 
The Haldane application also reveals how symmetries constrain the flow: an instantaneously $C_3$-symmetric isotropic nearest-neighbor protocol lacks the control directions required to open the gap at the Dirac points. We show that breaking this instantaneous $C_3$ symmetry through flow-equation engineering enables the flow to reproduce both targets. 
Beyond the results presented here, further design of flow equations can impose additional experimental requirements such as real-valued hopping schedules and flatter quasienergy bands. 

As a complementary application in quantum control, the \ctrlRG framework allows us to resolve an open outstanding problem with the inaccessibility of counterdiabatic (CD) driving, providing a toolbox to explicitly construct fast-forward protocols. 
For instance, in the transverse-field Ising model, it converts the nonlocal multibody gauge potential (AGP) terms into experimentally accessible schedules of the original Ising interaction and transverse field. We demonstrate this construction for two nonadiabatic protocols with spin chain length $L=4$ and $L=50$: in both cases the accessible endpoint evolution approaches the target as the inaccessible terms are decimated. Notably, for $L=50$ spins, the flow consists of $149$ coupled nonlinear partial differential equations, one for each operator component, which we solve on a discrete physical-time grid that preserves contractivity. 
Beyond FF driving, the flow-equation approach has built-in guarantees for unit-fidelity state preparation and gate implementation inherited from the seed adiabatic gauge potential, providing an alternative to the conventional brute-force searches for fast-forward drives~\cite{Bukov_2019_PRX}. Moreover, eigensubspace-specific gauge potentials, such as those used in hemidiabatic driving~\cite{gjonbalaj2025shortcuts}, and approximate AGPs obtained from variational principles~\cite{Sels_Polkovnikov_2017_PNAS,Kolodrubetz_etal_2017_PhysRep} can be used as seeds, with resulting accessible protocols inheriting the corresponding accuracy.

Beyond the two applications we presented, there are a number of quantum control problems that can be solved using the \ctrlRG framework. Whenever a target unitary has a known seed Hamiltonian representation and a desired final Hamiltonian can be specified, \ctrlRG provides a framework to dynamically connect the two. 
Extending the approach to generic interacting many-body systems, where the operator algebra scales exponentially with system size, will require flow-equation engineering together with suitable truncations of operator space that properly handle the corresponding information loss. Notice the close analogy with the truncation step in standard Hamiltonian RG and continuous flow-equation methods of wide applicability~\cite{wilson1975renormalization,Glazek_Wilson_1994_PRD,Kehrein_2006_book}; we therefore expect that truncation schemes originating in this well-established area of statistical mechanics can be further developed and adapted for \ctrlRG, making its flow irreversible and bringing the framework away from the original unitary Wegner-flow paradigm.
At the opposite extreme of few-level systems, \ctrlRG already provides a platform-agnostic route to speed up adiabatic gate implementations for emerging quantum computing and simulation architectures using fast-forward driving. Moreover, a digital version based on discrete temporal bases, such as the Walsh basis, could generate protocols converged to circuit-friendly control Hamiltonians~\cite{votto2024universal,walkling2025walsh}. Developing these directions will establish control-space RG flows as a common toolbox for analog and digital quantum dynamics.

Motivated by direct applications, in this work we have designed \ctrlRG to meet concrete experimental control requirements, with flow-equation engineering providing the versatility required for different platforms and control tasks. Yet accessibility presents only one possible criterion for selecting the fixed-point manifold: any prescribed operator subspace can define the retained sector, while the flow decimates its complement. For example, the flow may be used to retain any chosen family of few-body operators while decimating all other terms, irrespective of whether they are experimentally available. 
This perspective reveals a hidden link in the structures behind Hamiltonian engineering and unitary quantum control, which emerge as the two facets of the same core dynamical principle.  
An instance of such an interconnection was recently identified for periodically driven systems within Geometric Floquet theory~\cite{Schindler_Bukov_2025_PRX}. 
Formalizing the intricate underlying structure to arbitrary time dependence, \ctrlRG provides a generic framework for identifying Hamiltonian representations of a prescribed evolution, and can therefore serve as an organizing principle for nonequilibrium quantum dynamics.

\section*{Acknowledgments}

We thank Anatoli Polkovnikov for valuable discussions.
The authors used Claude Opus 4.6 and Claude Opus 5.5 during the preparation of this manuscript to assist with code optimization and to improve the readability of the manuscript. The authors assume full responsibility for the scientific content, accuracy, and integrity of this paper.

Funded by the European Union (ERC, QuSimCtrl, 101113633). Views and opinions expressed are, however, those of the authors only and do not necessarily reflect those of the European Union or the European Research Council Executive Agency. Neither the European Union nor the granting authority can be held responsible for them. 
This work was supported by the U.K. Engineering and
Physical Sciences Research Council via the EPSRC Hub in
Quantum Computing and Simulation (EP/W524311/1).
Z.W. acknowledges financial support from the China Scholarship Council (CSC, No. 202506750006) and the NKRDPC (Grant No. 2022YFA1405304).

\section*{Data availability}
The data and code associated with this manuscript are deposited on Zenodo~\cite{Zenodo_2026}, DOI: \href{https://doi.org/10.5281/zenodo.23168463}{10.5281/zenodo.23168463}.
Automatic access is restricted; access credentials will be provided to human researchers upon reasonable request via Zenodo for academic and verification purposes.

\clearpage

\tableofcontents

\appendix

\section{\ctrlRG flow for a two-level system}
\label{app:2ls}

Here, we apply the \ctrlRG flow to the simplest system of a single qubit with two accessible controls. Let $S^a=\sigma^a/2$ for $a=x,y,z$. They span the operator space
\begin{equation}
\mathfrak g=\text{span}\{S^x,S^y,S^z\},
\quad
\dim\mathfrak g=3,
\end{equation}
which closes under commutation, $i[S^a,S^b]=-\epsilon^{abc}S^c$, so a control Hamiltonian is
\begin{equation}
H(\mu,t)=x(\mu,t)S^x+y(\mu,t)S^y+z(\mu,t)S^z .
\label{eq:2ls_H}
\end{equation}

We consider a fast-forward problem of the type from Sec.~\ref{sec:ff_driving}, where a field is swept in the $xz$ plane and the gauge potential lives along the $y$ direction. The bare control Hamiltonian is 
\begin{equation}
H_{\rm ctrl}(t)=x_{\rm ctrl}(t)S^x+z_{\rm ctrl}(t)S^z,
\end{equation}
and its adiabatic gauge potential is 
\begin{equation}
\agp(t)
=\frac{\boldsymbol a_{\rm ctrl}\times\dot{\boldsymbol a}_{\rm ctrl}}{\left|\boldsymbol a_{\rm ctrl}\right|^{2}}\cdot\boldsymbol S
=\frac{z_{\rm ctrl}\dot x_{\rm ctrl}-x_{\rm ctrl}\dot z_{\rm ctrl}}{x_{\rm ctrl}^{2}+z_{\rm ctrl}^{2}} S^y ,
\end{equation}
with $\boldsymbol a_{\rm ctrl}=(x_{\rm ctrl},0,z_{\rm ctrl})$. The counterdiabatic seed $\Hseed=H_{\rm ctrl}+\agp$ is the initial condition of the flow, and fixes the three schedules at $\mu=0$,
\begin{align}
x(0,t)&=x_{\rm ctrl}(t),\qquad z(0,t)=z_{\rm ctrl}(t),\nonumber\\
y(0,t)&=\frac{z_{\rm ctrl}\dot x_{\rm ctrl}-x_{\rm ctrl}\dot z_{\rm ctrl}}{x_{\rm ctrl}^{2}+z_{\rm ctrl}^{2}} .
\label{eq:2ls_seed}
\end{align}
We suppose that $S^y$ is inaccessible,
\begin{equation}
\Pacc H=xS^x+zS^z,
\quad
\Hper=(1-\Pacc)H=yS^y ,
\end{equation}
so that the inaccessible weight of Eq.~\eqref{eq:inaccessible_weight} reduces to
\begin{equation}
[\Wnorm(\mu)]^2=\frac{1}{4}\int_0^T \mathrm dt y(\mu,t)^2.
\label{eq:2ls_weight}
\end{equation}
The task is to decimate $y$ along the flow while keeping the endpoint unitary fixed. 

\subsection{Component flow equations}

Expanding the generator as $K(\mu,t)=\sum_a\kappa^a(\mu,t)S^a$, the component flow equation is
\begin{align}
\partial_\mu x &= -\left(y\kappa^z-z\kappa^y\right)+\partial_t\kappa^x,\nonumber\\
\partial_\mu y &= -\left(z\kappa^x-x\kappa^z\right)+\partial_t\kappa^y,\label{eq:2ls_flow}\\
\partial_\mu z &= -\left(x\kappa^y-y\kappa^x\right)+\partial_t\kappa^z,\nonumber
\end{align}
which closes once the generator is built from the flow Hamiltonian itself. Writing the canonical generator of Eq.~\eqref{eq:canonical_generator} as $\Kc=\sum_a\kappa_{\rm c}^a S^a$, the regularizer of Eq.~\eqref{eq:K_from_Kw} acts component by component,
\begin{equation}
\kappa^a=\Rreg_a\kappa_{\rm c}^a ,
\label{eq:2ls_closure}
\end{equation}
and with $\Hper=yS^y$ the three canonical components are simply
\begin{equation}
\kappa_{\rm c}^x=z\,y,
\qquad
\kappa_{\rm c}^y=\partial_t y,
\qquad
\kappa_{\rm c}^z=-x\,y .
\label{eq:2ls_canonical}
\end{equation}
Equations~\eqref{eq:2ls_flow}--\eqref{eq:2ls_canonical} are three coupled nonlinear partial differential equations in $(\mu,t)$.

\begin{figure*}[t]
\centering
\includegraphics{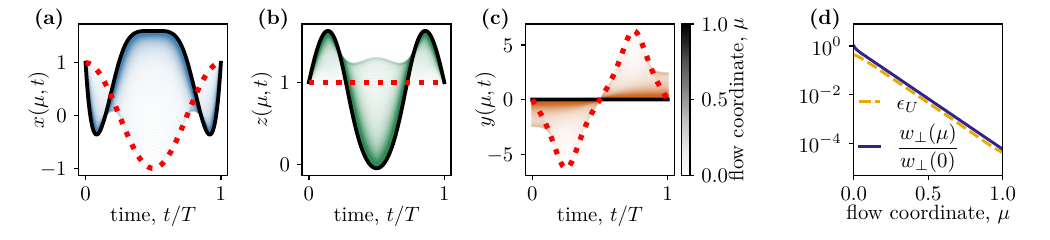}
\caption{
\textbf{\ctrlRG flow for the two-level system.} \textbf{(a--c)} The three schedules of Eq.~\eqref{eq:2ls_H} along the flow, shaded by the flow parameter $\mu$ as in the colorbar, with the seed of Eq.~\eqref{eq:2ls_seed} dotted in red and the converged protocol in black. \textbf{(d)} Inaccessible weight $\Wnorm(\mu)/\Wnorm(0)$ of Eq.~\eqref{eq:2ls_weight} and unitary error $\epsU$ of Eq.~\eqref{eq:2ls_epsU} against $\mu$. The calculation uses $T=1$, $\gamma_2=\gamma_4=0.1$, a time grid of $401$ points, and the hybrid regularizer of Eq.~\eqref{eq:hybrid_filter}. See Appendix~\ref{app:2ls:flow} for details.
}
\label{fig:2ls}
\end{figure*}

\subsection{Diffusion of the canonical flow and its regularization}
\label{app:2ls:flow}

Setting $\Rreg_a=1$ in Eq.~\eqref{eq:2ls_closure}, so that $K=\Kc$, the inaccessible component satisfies,
\begin{equation}
\partial_\mu y
=\partial_t^2 y-\left(x^2+z^2\right)y .
\label{eq:2ls_diffusion}
\end{equation}
This is a heat diffusion equation, plus a decay term set by the strength of the accessible field. If treating coefficient $x^2+z^2$ as constant, each temporal mode decays as 
\begin{equation}
y_\omega(\mu)\propto e^{-\Decval{\omega}^2\mu},\quad \Decval{\omega}^2=\omega^2+x^2+z^2 ,
\end{equation}
which exhibits the scale-by-scale decimation of Sec.~\ref{subsec:ctrlsrg}. However, the rate is unbounded in $\omega$, so the flow is stiff and the numerical integration of Eq.~\eqref{eq:2ls_diffusion} is difficult. In particular, refining the time grid to make the discretization more accurate makes the flow stiffer.

The positive regularizer removes this difficulty. The unbounded rate comes from $\kappa_{\rm c}^y=\partial_ty$ of Eq.~\eqref{eq:2ls_canonical}, which responds to arbitrarily fast temporal modes in $y$. We therefore replace it by the filtered generator $\kappa^y=\Rreg_2\partial_ty$, with the second-order filter $\Rreg_2=(1-\gamma_2^2\partial_t^2)^{-1}$. The filter suppresses structure on time scales below $\gamma_2$ while leaving slow modes essentially untouched, so the mode $e^{i\omega t}$ now decays as $\omega^2/(1+\gamma_2^2\omega^2)+x^2+z^2$. The largest possible rate now saturates at $\gamma_2^{-2}$, and the stiffness will not increase for whatever the resolution of the time grid.

The regularizer is also necessary for the temporal boundary conditions. The components of $\Kc$ in Eq.~\eqref{eq:2ls_canonical} do not vanish at $t=0,T$ for a generic seed, so the Dirichlet condition of Eq.~\eqref{eq:K_pinning} is violated, and the endpoint unitary drifts along the flow. Here the two sectors need different endpoint conditions. We want the accessible $x$ and $z$ to start and end at the values in the bare protocol, which by Eq.~\eqref{eq:2ls_flow} requires $\partial_t\kappa^{x,z}=0$ at $t=0,T$ and thus the fourth-order filter $\Rreg_4$; the inaccessible $y$ is left free at the two ends, so its generator component $\kappa^y$ need only obey the Dirichlet condition of Eq.~\eqref{eq:K_pinning}, for which the second-order filter suffices. We therefore take the hybrid regularizer of Eq.~\eqref{eq:hybrid_filter}, with $\Rreg_4$ on $x$ and $z$ and the Dirichlet filter $\Rreg_2$ on $y$.

Figure~\ref{fig:2ls} shows the resulting flow for the sweep
\begin{equation}
x_{\rm ctrl}(t)=\cos\frac{2\pi t}{T},
\qquad
z_{\rm ctrl}(t)=1 ,
\label{eq:2ls_protocol}
\end{equation}
at $T=1$ and $\gamma_2=\gamma_4=0.1$, on a uniform time grid of $N_t=401$ points, integrated in $\mu$ up to $\muf=1$ with the implicit \emph{TRBDF2} method at relative tolerance $10^{-10}$ and absolute tolerance $10^{-12}$. The gauge potential of Eq.~\eqref{eq:2ls_seed} decays exponentially towards zero while the two accessible schedules deform to absorb it. Alongside the weight of Eq.~\eqref{eq:2ls_weight} we monitor the unitary error of the accessible controls alone,
\begin{align}
\epsU(\mu)&=\frac{1}{\sqrt2}\Fnorm{\Uacc(\mu)-\Utarget},\nonumber\\
\Uacc(\mu)&=\mathcal{T}e^{-i\int_0^T\mathrm dt\,\Hacc(\mu,t)},
\label{eq:2ls_epsU}
\end{align}
which is the two-level case of Eq.~\eqref{eq:haldane_epsU}, with $\Fnorm{\cdot}$ the Frobenius norm. Both fall exponentially in $\mu$, with the same rate and by about four orders of magnitude in total.

\section{Frame fixed at the temporal endpoints}
\label{app:pinned_frame}

Here, we show that $V$ leaves the target unitary untouched once it equals $\mathbb 1$ at the two ends of the protocol, and the flow preserves this property because it is enforced at each flow parameter $\mu$. 

Let $\Useed(t,t')$ be the propagator generated by the seed Hamiltonian $\Hseed(t)$ between an initial time $t'$ and a later time $t$, so that 
\begin{equation}
i\partial_t\Useed(t,t')=\Hseed(t)\Useed(t,t'),
\quad
\Useed(t',t')=\mathbb 1 .
\label{eq:app_seed_propagator}
\end{equation}
The frame transformation $V(t)$ turns the seed into the Hamiltonian
\begin{equation}
\HV(t)=V(t)\Hseed(t)V^\dagger(t)-iV(t)\dot V^\dagger(t),
\label{eq:app_HV}
\end{equation}
and the corresponding propagator is transformed as 
\begin{equation}
\UV(t,t')=V(t)\Useed(t,t')V^\dagger(t').
\label{eq:app_general_transformed_propagator}
\end{equation}
Indeed, we can directly obtain
\begin{align}
i\partial_t\UV(t,t')
&=\left[i\dot VV^\dagger+V\Hseed V^\dagger\right]\UV(t,t')\nonumber\\
&=\HV(t)\UV(t,t'),
\label{eq:app_transformed_schrodinger}
\end{align}
which means that $\UV$ is indeed the propagator generated by $\HV$. 

The seed, by construction, realizes the target unitary over full protocol, $\Useed(T,0)=\Utarget$, so Eq.~\eqref{eq:app_general_transformed_propagator} gives
\begin{equation}
\UV(T,0)=V(T)\Utarget V^\dagger(0).
\label{eq:app_endpoint_unitary}
\end{equation}
Any frame that is the identity at the two ends of the protocol, $V(0)=V(T)=\mathbb 1$, leaves the target unitary unchanged, $\UV(T,0)=\Utarget$. 

The flow of Sec.~\ref{subsec:flow_eq} uses a specific class of frames, $V(t)\to V(\mu,t)$, one for each flow parameter $\mu$. Since Eq.~\eqref{eq:V_pinning} imposes $V(\mu,0)=V(\mu,T)=\mathbb 1$ for every $\mu$, the argument above applies at each $\mu$ separately.

\section{Contraction and linearized decimation of the \ctrlRG flow}
\label{app:ctrlsrg}

\subsection{Control space and superoperators}

The flow happens in the space of time-dependent control Hamiltonians $H(\mu,t)$, which factorizes into an operator part and a temporal part. Let $\mathfrak{g}$ denote the operator space closed under commutation, the control space is then 
\begin{equation}
\mathcal{C}=\mathfrak{g}\otimes L^2[0,T], 
\label{eq:app_control_space}
\end{equation}
where the temporal factor $L^2[0,T]$ is the space of real, square-integrable functions of physical time, and a time-dependent Hamiltonian at each $\mu$ is a vector in $\mathcal{C}$. The inner product is the one inherited from the two factors, 
\begin{equation}
\ipt{A}{B}=\int_0^T \mathrm dt \tr[A(t)B(t)].
\label{eq:app_inner_product}
\end{equation}
This space $\mathcal{C}$ is the arena of the flow, where \ctrlRG moves a control Hamiltonian along a trajectory $\mu\mapsto H(\mu,\cdot)$ from one time-dependent Hamiltonian to another, as illustrated in Fig.~\ref{fig:cartoon}. 

The frame generator lives in a subspace of $\mathcal{C}$. Write $\mathcal{K}\subset\mathcal{C}$ for the Dirichlet subspace, the vectors that vanish at the two time endpoints, $\mathcal{K}=\left\{ A\in\mathcal{C}:A(0)=A(T)=0 \right\}$.

Three superoperators act on $\mathcal{C}$: the projector $\Pacc$, the regularizer $\Rreg$, and the Liouvillean $\Liou$ of the accessible controls.

The projector $\Pacc$ only acts on the operator sector and commutes with the regularizer $\Rreg$. 
The regularizer $\Rreg$ is, by construction, self-adjoint and positive semidefinite with respect to $\ipt{\cdot}{\cdot}$, and its range lies in $\mathcal{K}$, so the generator $K=\Rreg\Kc$ satisfies the Dirichlet condition of Eq.~\eqref{eq:K_pinning} automatically. From the spectral decomposition $\Rreg=\sum_n r_n\Pi_n$, with $r_n\ge0$ and every eigenvector of $r_n>0$ in $\mathcal{K}$, the positive square root $\Rreg^{1/2}=\sum_n r_n^{1/2} \Pi_n$ is again self-adjoint and positive, and it maps into $\mathcal{K}$. 

The Liouvillean $\Liou(\cdot)=i[\Hacc(\mu,t)-i\partial_t,\cdot]$ satisfies, for any $A\in\mathcal{C}$ and $B\in\mathcal{K}$,
\begin{align}
\ipt{A}{\Liou B}
=&\int_0^T\mathrm dt \tr\left[A\left(i[\Hacc,B]+\partial_tB\right)\right]\nonumber\\
=&-\int_0^T\mathrm dt \tr\left[\left(i[\Hacc,A]\right)B\right]
+\left[\tr(AB)\right]_0^T\nonumber\\
&-\int_0^T\mathrm dt \tr\left[\left(\partial_tA\right)B\right]\nonumber\\
=&-\ipt{\Liou A}{B},
\label{eq:app_skew}
\end{align}
where the boundary term $\left[\tr(AB)\right]_0^T$ vanishes because $B$ obeys the Dirichlet condition. The Liouvillean is therefore skew-symmetric, $\Liou^\dagger=-\Liou$, whenever one of its two arguments vanishes at the two temporal ends.

\subsection{Monotonic contraction of the inaccessible weight}
\label{app:ctrlsrg:contraction}

The flow equation Eq.~\eqref{eq:flow_equation} gives 
\begin{equation}
\partial_\mu H=\Liou K+i[\Hper,K].
\label{eq:app_flow_split}
\end{equation}
The inaccessible weight of Eq.~\eqref{eq:inaccessible_weight} then contracts along the flow,
\begin{align}
\partial_\mu\left[\Wnorm^2\right]
&=\ipt{\Hper}{\partial_\mu \Hper}=\ipt{\Hper}{\partial_\mu H}\nonumber\\
&=\ipt{\Hper}{\Liou K}+i\ipt{\Hper}{[\Hper,K]} \nonumber\\
&=-\ipt{\Liou\Hper}{K}=-\ipt{\Kc}{\Rreg\Kc}\le0 .
\label{eq:app_contraction}
\end{align}
The rate vanishes when $\Rreg^{1/2}\Kc=0$. Every fully accessible control, $\Hper=0$, is such a point. So is every $\Hper\ne0$ with $\Rreg^{1/2}\Liou\Hper=0$. This happens when $\Hper$ satisfies $\Liou\Hper=i[\Hacc,\Hper]+\dotHper=0$, and also when $\Liou\Hper$ falls in the kernel of the regularizer. The contraction therefore guarantees that the weight of the inaccessible term never grows, but not that it reaches zero.

\subsection{Linearization and exponential decay}
\label{app:ctrlsrg:linear}

Any control that is already fully accessible, $\Hper=0$, is a fixed point of the flow. There are many such fixed points, one for each accessible Hamiltonian, and they form a fixed-point manifold. When the flow converges to one of them, we can linearize around it and, as in conventional RG, classify each inaccessible direction by how fast it is removed.

Since the canonical generator of Eq.~\eqref{eq:canonical_generator} is built from $\Hper$, the generator $K$ is itself first order in $\Hper$, so the nonlinear term $i[\Hper,K]$ of Eq.~\eqref{eq:app_flow_split} is second order and drops in the linear regime. Projecting Eq.~\eqref{eq:app_flow_split} onto the inaccessible sector gives $\Pper\Liou K$, and to the leading order we fix $\Liou$ at its fixed-point value. Using $K=\Rreg\Kc$ and $\Kc=\Liou\Hper=\Liou\Pper\Hper$, we obtain 
\begin{equation}
\partial_\mu \Hper=\Pper \Liou \Rreg \Liou \Pper \Hper + O(\|\Hper^2\|_\mathrm{F}).
\label{eq:app_linear}
\end{equation}
We can split $\Rreg=\Rreg^{1/2}\Rreg^{1/2}$ and define
\begin{equation}
\Dec\equiv\Rreg^{1/2} \Liou \Pper. \label{eq:app_decimator}
\end{equation}
It satisfies
\begin{align}
\ipt{\Dec A}{B}
&=\ipt{\Liou\Pper A}{\Rreg^{1/2}B}
=-\ipt{\Pper A}{\Liou\Rreg^{1/2}B} \nonumber\\
&=\ipt{A}{-\Pper\Liou\Rreg^{1/2}B},
\end{align}
and thus 
\begin{equation}
\Dec^\dagger=-\Pper \Liou \Rreg^{1/2}.
\end{equation}
So the linearized flow of Eq.~\eqref{eq:app_linear} can be written as
\begin{equation}
\partial_\mu \Hper=-\Dec^\dagger\Dec \Hper .
\end{equation}

The operator $\mathcal{M}\equiv\Dec^\dagger\Dec$ is self-adjoint and positive semidefinite, $\ipt{A}{\mathcal{M}A}=\ipt{\Dec A}{\Dec A}\ge0$. On the discrete time grid of Appendix~\ref{app:reg:sbp}, where $\mathcal{M}$ is a finite symmetric matrix, its eigenvalues are therefore nonnegative and its eigenvectors orthogonal,
\begin{align}
\mathcal{M}\Decmode{\alpha}&=\Decval{\alpha}^2 \Decmode{\alpha},
\quad
\Decval{\alpha}\ge0, \\
\ipt{\Decmode{\alpha}}{\Decmode{\beta}}&=\delta_{\alpha\beta}.
\end{align}
The eigenvalue problem lives in the control space $\mathcal{C}$ of Eq.~\eqref{eq:app_control_space}, so each normalized eigenmode $\Decmode{\alpha}=\Decmode{\alpha}(t)$ is a complete time-dependent operator, carrying both an operator content and a temporal profile. Expanding the inaccessible part in this basis,
\begin{equation}
\Hper(\mu,t)=\sum_\alpha H_{\perp,\alpha}(\mu,t),
\quad
H_{\perp,\alpha}(\mu,t)=c_\alpha(\mu) \Decmode{\alpha}(t),
\end{equation}
with $c_\alpha(\mu)=\ipt{\Decmode{\alpha}}{\Hper(\mu,\cdot)}$, the modes decouple, $\partial_\mu c_\alpha=-\Decval{\alpha}^2 c_\alpha$, so each component is damped exponentially in the flow parameter,
\begin{equation}
H_{\perp,\alpha}(\mu,t)=e^{-\Decval{\alpha}^2(\mu-\mu_0)} H_{\perp,\alpha}(\mu_0,t),
\end{equation}
where $\mu_0$ is a large flow parameter where the linearization is valid. So the flow removes the inaccessible terms scale by scale, the modes with the largest eigenvalues first and those with the smallest last, in the same manner as a renormalization group removes ultraviolet degrees of freedom. The marginal modes, with $\Decval{\alpha}=0$, are not removed, so $\Wnorm(\mu)$ develops a plateau at the value they provide. The marginal control commutes with the accessible Hamiltonian in the extended sense of Eq.~\eqref{eq:extended_commutator}, so the generator produces no rotation along $\Decmode{\alpha}$, and diagonalizing $\Dec^\dagger\Dec$ gives the height of the plateau and the control term responsible for it. A variational optimization that falls in a local minimum has no such structure, as where it stops depends on the initialization.

\section{Temporal regularization and time discretization}
\label{app:reg}

\subsection{Positivity of the temporal filters}
\label{app:reg:positivity}

The positivity of $\Rreg$ ensures the contraction of the flow, while the self-adjointness is required for the linearized analysis. Our filters have both properties.

The filters $\Rreg_2$ and $\Rreg_4$ are inverses of differential operators, 
\begin{equation}
\Rreg_2^{-1}=1-\gamma_2^2\partial_t^2,
\quad
\Rreg_4^{-1}=(1-\gamma_4^2\partial_t^2)^2.
\end{equation}
We consider the boundary conditions 
\begin{align}
\text{(i)}\quad & K|_{0,T}=0, && (\Rreg_2),
\label{eq:bc_dirichlet}\\
\text{(ii)}\quad & K|_{0,T}=0, \partial_tK|_{0,T}=0, && (\Rreg_4),
\label{eq:bc_positivity_slope}\\
\text{(iii)}\quad & K|_{0,T}=0, \partial_t^{j=1,2}K|_{0}=\partial_t^{j=1,2}K|_{T}, && (\Rreg_4).
\label{eq:bc_periodic}
\end{align}
All three impose the Dirichlet condition of Eq.~\eqref{eq:K_pinning} and are therefore subspaces of $\mathcal{K}$. Integrating by parts once for $\Rreg_2^{-1}$ and twice for $\Rreg_4^{-1}$ gives the symmetric forms
\begin{align}
\ipt{A}{\Rreg_2^{-1}B} &=\ipt{A}{B}+\gamma_2^2\ipt{\partial_tA}{\partial_tB},
\label{eq:app_form2}\\
\ipt{A}{\Rreg_4^{-1}B} &=\ipt{A}{B}+2\gamma_4^2\ipt{\partial_tA}{\partial_tB}+\gamma_4^4\ipt{\partial_t^2A}{\partial_t^2B},
\label{eq:app_form4}
\end{align}
where the boundary terms vanish. The right-hand sides are symmetric under $A\leftrightarrow B$ and positive for $A=B\ne0$, so $\Rreg_2^{-1}$ and $\Rreg_4^{-1}$ are self-adjoint and positive definite.

\subsection{Boundary conditions and Green's functions}
\label{app:reg:green}

The filters give boundary value problems under one of the three conditions, which can be solved by the Green's functions
\begin{equation}
K(t)=\int_0^T \mathrm d s\Green_n(t,s) \Kc(s).
\label{eq:app_convolution}
\end{equation}
For the Dirichlet condition (i), the Green's function is 
\begin{equation}
\Green_2(t,s) = \frac{\sinh\left(t_</\gamma_2\right) \sinh\left[(T-t_>)/\gamma_2\right]} {\gamma_2 \sinh\left(T/\gamma_2\right)},
\end{equation}
with $t_<=\min(t,s)$ and $t_>=\max(t,s)$. For $\Rreg_4^{-1}$ the homogeneous solutions are $(a+bt)e^{\pm t/\gamma_4}$, and for the vanishing-slope condition (ii) the Green's function reads
\begin{equation}
\Green_4(t,s)=\sum_{i,j=1}^{2}\left[\mathcal{N}^{-1}\right]_{ij}\phi_i(t_<) \phi_j(T-t_>),
\end{equation}
where
\begin{align}
\phi_1(t)&=2\sinh(t/\gamma_4)-\frac{2t}{\gamma_4}e^{-t/\gamma_4},\\
\phi_2(t)&=2t\sinh(t/\gamma_4),
\end{align}
and $\mathcal{N}$ is a constant $2\times2$ matrix, depending on $T$ and $\gamma_4$. The periodic condition (iii) gives a kernel that is again symmetric and positive, and can be found in closed form as well. For simple applications of the flow equation, these Green's functions can be sampled directly on a time grid to construct the generator. However, in the numerical simulation of the cases discussed in the main manuscript, naively implementing these Green's functions and computing the convolution on a discrete time grid does not usually give results with satisfactory accuracy. We therefore used additional numerical tricks for the discretization to enhance the precision, as discussed in detail in Appendices~\ref{app:reg:sbp} and~\ref{app:reg:discrete}.

\subsection{Sector-resolved regularizers}
\label{app:reg:sector}

A regularizer that applies different temporal filters to different operator sectors is also self-adjoint and positive. Let $\{\Pi_\alpha\}$ be orthogonal projectors on $\mathfrak{g}$ with $\sum_\alpha\Pi_\alpha=1$, let each $\Rreg^{(\alpha)}$ be one of the filters above, and let $s_\alpha>0$ be a constant weight. Each filter $\Rreg^{(\alpha)}$ acts on the temporal sector of $\mathcal{C}$ and each projector on the operator sector, so every term commutes with every projector, and
\begin{align}
\Rreg&=\sum_\alpha s_\alpha\Rreg^{(\alpha)}\Pi_\alpha,\\
\ipt{A}{\Rreg A}&=\sum_\alpha s_\alpha\ipt{\Pi_\alpha A}{\Rreg^{(\alpha)}\Pi_\alpha A}\ge0.
\end{align}
The contraction of Eq.~\eqref{eq:contraction} is still satisfied.

Two choices are used in this work. The hybrid filter of Eq.~\eqref{eq:hybrid_filter} takes $\Pi_1=\Pacc$ and $\Pi_2=1-\Pacc$, with $\Rreg^{(1)}=\Rreg_4$ on the accessible sector and $\Rreg^{(2)}=\Rreg_2$ on the inaccessible one, and $s_1=s_2=1$. It is the choice used for the Ising calculation. The directional filter has one projector per control direction,
\begin{equation}
\Rreg=\sum_{a=1}^{d}s_a \Rreg^{(a)} \Pi_a ,
\label{eq:app_directional_filter}
\end{equation}
where the filters $\Rreg^{(a)}$ may have different filtering strength $\gamma_a$ and different boundary conditions of Eqs.~\eqref{eq:bc_dirichlet}--\eqref{eq:bc_periodic}. The Haldane calculation uses Eq.~\eqref{eq:app_directional_filter}, with the index $a$ running over the three bond directions and a fourth projector onto everything else, so that the $\Pi_a$ resolve the identity. All sectors share one filter $\Rreg^{(a)}=\Rreg_4$, the periodic fourth-order one of condition (iii) with a single uniform $\gamma$. We introduce inhomogeneity by giving the first bond direction the larger weight $s_1=4$, while every other sector keeps $s_{a\neq1}=1$. This is Eq.~\eqref{eq:haldane_regularizer} of the main text, where the projector $\Pi_1$ covers the whole components of that bond direction, which in the real operator basis means both the cosine and the sine control, so the real and imaginary parts of $J_1$ are driven equally hard. Since all $s_a$ are positive, the contraction is untouched, but the flow equation is no longer invariant under the threefold rotation that permutes the three bond directions. This explicitly breaks the threefold symmetry, which is what we want to impose in Sec.~\ref{sec:floquet_engineering}.

\subsection{Summation-by-parts discretization}
\label{app:reg:sbp}

Numerically, it is convenient to discretize time on a uniform grid. But a naive discretization of the derivative $\partial_t$ and the integral $\int \mathrm dt$ would break the contractive flow, so we use a summation-by-parts (SBP) discretization that preserves the contraction exactly. Here, we describe the open-interval grid used for the Ising calculation, which includes both endpoints; the periodic grid used for the Haldane calculation is in Appendix~\ref{app:reg:periodic}.

Let $\boldsymbol A=(A_1,\ldots,A_{N_t})^{\mathsf T}$, where $A_n=A(t_n)$ is the operator at the $n$th grid point. We introduce the real $N_t\times N_t$ matrices $M_t$ and $D$ acting only on the time-grid index, 
\begin{align}
    (M_t\boldsymbol A)_n & = w_nA_n,\\
    (D\boldsymbol A)_n & = \sum_{m=1}^{N_t}D_{nm}A_m,
\end{align}
where $M_t=\mathrm{diag}(w_1,\ldots,w_{N_t})$ and $D\boldsymbol A$ approximates $\partial_tA$. The discrete inner product therefore takes the form
\begin{equation}
\ipt{A}{B}
=\sum_{n,m=1}^{N_t}(M_t)_{nm}\tr[A_nB_m]
=\sum_{n=1}^{N_t}w_n\tr[A_nB_n].
\end{equation}
The discrete counterpart of integration by parts reads
\begin{align}
    &\ipt{A}{DB}+\ipt{DA}{B} \nonumber\\
    = &\sum_{n,m=1}^{N_t}\left(M_tD+D^{\mathsf T}M_t\right)_{nm}\tr[A_nB_m],
\end{align}
where $D^{\mathsf T}$ denotes the transpose in the time-grid space, $(D^{\mathsf T})_{nm}=D_{mn}$. We choose $M_t$ and $D$ so that this expression reduces to two endpoint values,
\begin{equation}
M_tD+D^{\mathsf T}M_t =e_{N_t}e_{N_t}^{\mathsf T}-e_1e_1^{\mathsf T},
\label{eq:sbp}
\end{equation}
where $e_j$ is the $N_t$-component unit vector at grid point $j$. The right-hand side of Eq.~\eqref{eq:sbp} is a matrix with value $-1$ at entry $(1,1)$ and $+1$ at entry $(N_t,N_t)$. Hence we obtain 
\begin{equation}
\ipt{A}{DB}+\ipt{DA}{B}=\tr[A_{N_t}B_{N_t}]-\tr[A_1B_1].
\end{equation}
A pair $(M_t,D)$ satisfying Eq.~\eqref{eq:sbp} is a summation-by-parts operator~\cite{Svard_Nordstrom_2014_JCP}. 

We use the so-called SBP$(4,2)$ pair of Ref.~\cite{Strand_1994_JCP}, which is fourth-order accurate in the interior, second-order accurate at the two ends, and has a diagonal $M_t$. In the interior, $D$ is the fourth-order centered difference,
\begin{equation}
(DA)_n=\frac{A_{n-2}-8A_{n-1}+8A_{n+1}-A_{n+2}}{12 \Delta t},
\end{equation}
where $\Delta t$ is the time-grid spacing, and the weights are uniform, $w_n=\Delta t$. Close to the two ends, the first four rows of $D$ read
\begin{align}
(DA)_1&=\frac{1}{\Delta t}\left(-\frac{24}{17}A_1+\frac{59}{34}A_2-\frac{4}{17}A_3-\frac{3}{34}A_4\right),
\\
(DA)_2&=\frac{1}{\Delta t}\left(-\frac{1}{2}A_1+\frac{1}{2}A_3\right),
\\
(DA)_3&=\frac{1}{\Delta t}\left(\frac{4}{43}A_1-\frac{59}{86}A_2+\frac{59}{86}A_4-\frac{4}{43}A_5\right),
\\
(DA)_4&=\frac{1}{\Delta t}\left(\frac{3}{98}A_1-\frac{59}{98}A_3+\frac{32}{49}A_5-\frac{4}{49}A_6\right),
\end{align}
while the first four weights are adjusted to
\begin{equation}
(w_1,w_2,w_3,w_4)=\Delta t\left(\frac{17}{48},\frac{59}{48},\frac{43}{48},\frac{49}{48}\right).
\end{equation}
The last four rows and weights mirror these, $D_{N_t+1-n,N_t+1-m}=-D_{nm}$ and $w_{N_t+1-n}=w_n$, and all remaining entries of $D$ vanish. These coefficients are exactly what makes Eq.~\eqref{eq:sbp} hold.

The contraction then holds on the grid in the same form. Replacing $\partial_t\to D$ and $\int \mathrm dt\to\sum_nw_n$ in the action, $\Liou A=i[\Hacc,A]+DA$, the commutator cancels under the trace, and Eq.~\eqref{eq:sbp} leaves
\begin{equation}
\ipt{A}{\Liou B}+\ipt{\Liou A}{B}
=\tr\left[A_{N_t}B_{N_t}\right]-\tr\left[A_1B_1\right],
\end{equation}
so $\Liou$ is exactly skew-symmetric as soon as one argument vanishes at the two endpoints, just as in Eq.~\eqref{eq:app_skew}. The remaining step of Eq.~\eqref{eq:app_contraction} is just algebraic identities that the grid leaves untouched. But another crucial aspect that the time grid does not guarantee is the vanishing of $K$ at the two endpoints and the positivity of $\Rreg$. These are properties of the regularizer, and we construct its discrete version in the next part so that they are also fixed exactly at any $N_t$. The contraction is then exact on the time grid. 

\subsection{Discretized regularizer}
\label{app:reg:discrete}

The contraction of Eq.~\eqref{eq:app_contraction} rests on three facts, and Eq.~\eqref{eq:sbp} provides only the first of them. Turning $\ipt{\Hper}{\Liou K}$ into $-\ipt{\Liou\Hper}{K}$ is exact, but it still leaves the boundary term there, which goes away only if $K$ vanishes exactly. And in the last step, the term $-\ipt{\Kc}{\Rreg\Kc}\le0$ is true only if the discrete version of $\Rreg$ is positive semidefinite. These facts are part of the properties of the regularizer, so they have to be built into this discretization. 

Applying the regularizer needs solving the boundary value problem (BVP) $\Rreg_n^{-1}K=\Kc$ in Appendix~\ref{app:reg:green}, and discretizing this differential equation naively, $K=(\mathbb 1-\gamma_2^2D^2)^{-1}\Kc$ for the second-order filter, gives a $K$ that is nonzero at the two ends. The resulting matrix is neither symmetric nor positive, so the sign of $\partial_\mu\Wnorm^2$ is no longer guaranteed. The inaccessible weight may then grow, which gives a practical obstacle at large system size. 

We therefore discretize the \emph{variational form} of the BVP instead of the equation itself, which is the standard construction of finite-element methods~\cite{Brenner_Scott_2008_book}. In physicists' language, this is simple: instead of discretizing the equation of motion, we discretize the action functional and let the discrete equation of motion be derived from it, so that the symmetry and positivity can pass directly to the discrete equation of motion. 

\subsubsection{The functional behind the boundary value problem}

The starting point is that $\Rreg_n^{-1}$ is self-adjoint and positive on the \emph{admissible} controls, by which we mean those obeying the desired boundary conditions of the filter (Appendix~\ref{app:reg:positivity}). The continuum solution $K=\Rreg_n\Kc$ can be treated as the stationary point of functional~\cite{Brenner_Scott_2008_book}
\begin{equation}
\mathcal{J}_n[K]=\tfrac12\ipt{K}{\Rreg_n^{-1}K}-\ipt{K}{\Kc},
\label{eq:app_reg_functional}
\end{equation}
over \emph{admissible} $K$. Because this functional is quadratic, its second variation is a constant bilinear form, and shifting $K$ by two \emph{admissible} controls $\Phi$ and $\Psi$ satisfying the same boundary conditions gives
\begin{equation}
\delta^2\mathcal{J}_n[\Phi,\Psi]=\ipt{\Phi}{\Rreg_n^{-1}\Psi}.
\label{eq:app_reg_hessian}
\end{equation}
The self-adjointness of $\Rreg_n^{-1}$ makes Eq.~\eqref{eq:app_reg_hessian} symmetric under $\Phi\leftrightarrow\Psi$, and its positivity makes $\delta^2\mathcal{J}_n[\Phi,\Phi]>0$ for each nonzero \emph{admissible} $\Phi$. The two properties of $\Rreg_n^{-1}$ are therefore encoded in the curvature of $\mathcal{J}_n$ and any discretization that represents that curvature faithfully inherits them~\cite{Brenner_Scott_2008_book}.

To see that its stationary point solves our problem, replacing $K\to K+\epsilon\Phi$ gives 
\begin{equation}
\mathcal{J}_n[K+\epsilon\Phi]=\mathcal{J}_n[K]+\epsilon\left(\ipt{\Phi}{\Rreg_n^{-1}K}-\ipt{\Phi}{\Kc}\right)+O(\epsilon^2),
\label{eq:app_reg_variation}
\end{equation}
so $\mathcal{J}_n$ is stationary at $K$ when
\begin{equation}
\ipt{\Phi}{\Rreg_n^{-1}K}=\ipt{\Phi}{\Kc}.
\label{eq:app_weak_form}
\end{equation}
Eq.~\eqref{eq:app_weak_form} is the \emph{variational form} of the BVP~\cite{Brenner_Scott_2008_book}. Collecting its two sides, it reads
\begin{equation}
\ipt{\Phi}{\Rreg_n^{-1}K-\Kc}=0 ,
\label{eq:app_weak_residual}
\end{equation}
where $\Rreg_n^{-1}K-\Kc$ is the residual, the amount by which $K$ fails to solve the equation. 

The original BVP asks for a residual that vanishes at every point in time. Eq.~\eqref{eq:app_weak_residual} asks only that its time integral for each \emph{admissible} $\Phi$ vanish. In the continuous limit, these two statements are equivalent, but on a time grid the two are no longer equivalent. Discretizing Eq.~\eqref{eq:app_weak_form} therefore needs to represent $\ipt{\cdot}{\Rreg_n^{-1}\cdot}$ (shall be a matrix $\Sigma$) and the admissible set (via matrix $P$), as shown in the following. 

\subsubsection{Matrix representation of the regularizer}

The symmetric matrix $\Sigma$ is built by replacing $\ipt{A}{B}$ with $M_t$ and $\ipt{\partial_tA}{\partial_tB}$ with $D^{\mathsf T}M_tD$,
\begin{align}
\Sigma_2&=M_t+\gamma_2^2D^{\mathsf T}M_tD,\label{eq:app_sigma2}\\
\Sigma_4&=M_t+2\gamma_4^2D^{\mathsf T}M_tD+\gamma_4^4D_2^{\mathsf T}M_tD_2,
\end{align}
with $D_2=D^2$, and it therefore reproduces the right-hand sides of Eqs.~\eqref{eq:app_form2} and \eqref{eq:app_form4} exactly. 

The second matrix, $P$, keeps the solution inside the boundary conditions. The boundary conditions are $r$ independent linear relations for the $N$ grid values of $K$, so the admissible vectors form a subspace of dimension $N-r$. Let the columns of the $N\times(N-r)$ matrix $P$ be a basis of it, and let $\boldsymbol k$ be the grid values of $K$, so that every admissible vector is
\begin{equation}
\boldsymbol k=P\boldsymbol c
\label{eq:app_embedding}
\end{equation}
for a given vector $\boldsymbol c$. The two open-interval cases of Eqs.~\eqref{eq:bc_dirichlet} and \eqref{eq:bc_positivity_slope} use the grid of Appendix~\ref{app:reg:sbp}, $N=N_t$, with $r=2$ and $r=4$; the periodic case of Eq.~\eqref{eq:bc_periodic} uses the extended grid of Appendix~\ref{app:reg:periodic}, $N=N_t+1$, with $r=3$. 

The Dirichlet condition $k_1=k_N=0$ is the simplest case. The free values are the $N-2$ interior ones, $\boldsymbol c=(k_2,\ldots,k_{N-1})^{\mathsf T}$, and $P$ is 
\begin{equation}
P=\begin{pmatrix}0\\ \mathbb 1\\ 0\end{pmatrix},
\qquad
P\boldsymbol c=\begin{pmatrix}0\\ k_2\\ \vdots\\ k_{N-1}\\ 0\end{pmatrix},
\label{eq:app_P_dirichlet}
\end{equation}
with $\mathbb 1$ the $(N-2)\times(N-2)$ identity, and $0$ a row of zeros. 

The vanishing-slope condition of Eq.~\eqref{eq:bc_positivity_slope} adds $(D\boldsymbol k)_1=(D\boldsymbol k)_N=0$. Using the first row of $D$ from Appendix~\ref{app:reg:sbp} and $k_1=0$, 
\begin{equation}
(D\boldsymbol k)_1=\frac{1}{\Delta t}\left(\frac{59}{34}k_2-\frac{4}{17}k_3-\frac{3}{34}k_4\right)=0 ,
\end{equation}
which determines $k_2$, 
\begin{equation}
k_2=\frac{8}{59}k_3+\frac{3}{59}k_4 .
\end{equation}
From the bottom row, we can similarly express $k_{N-1}$ in terms of $k_{N-2}$ and $k_{N-3}$. The free values are now $\boldsymbol c=(k_3,\ldots,k_{N-2})^{\mathsf T}$, and $P$ becomes
\begin{equation}
P=\begin{pmatrix}
0&0&\cdots&0&0\\
\tfrac{8}{59}&\tfrac{3}{59}&\cdots&0&0\\
& &\mathbb 1& &\\
0&0&\cdots&\tfrac{3}{59}&\tfrac{8}{59}\\
0&0&\cdots&0&0
\end{pmatrix}.
\end{equation}
The periodic filter has three constraints of the same kind, Eq.~\eqref{eq:app_periodic_constraints}, as shown in Appendix~\ref{app:reg:periodic}. In every case the boundary conditions hold exactly for any $\boldsymbol c$, simply because $P$ was built from them. 

Restricting Eq.~\eqref{eq:app_reg_functional} to the admissible subspace by substituting Eq.~\eqref{eq:app_embedding} gives
\begin{equation}
\mathcal{J}_n(P\boldsymbol c)=\tfrac12\boldsymbol c^{\mathsf T}\left(P^{\mathsf T}\Sigma P\right)\boldsymbol c-\boldsymbol c^{\mathsf T}P^{\mathsf T}M_t\boldsymbol q ,
\end{equation}
where $\boldsymbol q$ is the vector of grid values of $\Kc$. This is a quadratic function of the free values in $\boldsymbol c$. Setting its gradient to zero,
\begin{equation}
\left(P^{\mathsf T}\Sigma P\right)\boldsymbol c=P^{\mathsf T}M_t\boldsymbol q ,
\label{eq:app_reduced_system}
\end{equation}
which is Eq.~\eqref{eq:app_weak_form} with $\Phi$ running over the columns of $P$. The matrix $P^{\mathsf T}\Sigma P$ is symmetric and positive definite, so Eq.~\eqref{eq:app_reduced_system} has a unique solution. Solving it for $\boldsymbol c$ and returning to $\boldsymbol k=P\boldsymbol c$ gives the discrete regularizer,
\begin{equation}
\Rreg=P\left(P^{\mathsf T}\Sigma P\right)^{-1}P^{\mathsf T}M_t .
\label{eq:app_discrete_regularizer}
\end{equation}

Read from right to left, each matrix plays the role of one step in the continuum: $M_t$ weights the source, $P^{\mathsf T}$ pairs it with the admissible basis, $(P^{\mathsf T}\Sigma P)^{-1}$ solves the restricted boundary value problem, and $P$ rebuilds a full grid vector. The symmetric matrix $P(P^{\mathsf T}\Sigma P)^{-1}P^{\mathsf T}$ is thus the discrete Green's function, with $M_t$ supplying the integration over $s$, so Eq.~\eqref{eq:app_discrete_regularizer} is the grid version of the convolution Eq.~\eqref{eq:app_convolution} and its symmetry is the counterpart of $\Green_n(t,s)=\Green_n(s,t)$.

We can now directly find the desired properties from the matrix form. The regularizer is self-adjoint, because $M_t\Rreg$ is symmetric. It is positive semidefinite, because $\ipt{A}{\Rreg A}=\boldsymbol y^{\mathsf T}(P^{\mathsf T}\Sigma P)^{-1}\boldsymbol y\ge0$ with $\boldsymbol y=P^{\mathsf T}M_t\boldsymbol a$; the inequality is not strict because $\Rreg$ annihilates whatever lies outside the range of $P$. And that range is the column space of $P$, so the essential boundary conditions hold exactly.

\subsection{Periodic time grid for Floquet protocols}
\label{app:reg:periodic}

The Haldane protocol is periodic, so there we define the time grid on a circle: $N_t$ points $t_n=(n-1)T/N_t$, with $t=T$ the same point as $t=0$. Then we have uniform $M_t=\Delta t \mathbb 1$. The discrete derivative $D$ is defined by
\begin{align}
(DA)_1&=\frac{A_2-A_1}{\Delta t},\\
(DA)_n&=\frac{A_{n+1}-A_{n-1}}{2\Delta t},\\
(DA)_{N_t}&=\frac{A_1-A_{N_t-1}}{2\Delta t},
\end{align}
for $n=2,\dots,N_t-1$. 

The two ends of Appendix~\ref{app:reg:sbp} have merged into the single point $t_1$, and so has the right-hand side of Eq.~\eqref{eq:sbp}. Once $K$ vanishes there, $K_1=0$, the only thing left of $\ipt{A}{\Liou K}+\ipt{\Liou A}{K}$ is
\begin{equation}
\tfrac12\tr\left[A_1\left(K_2+K_{N_t}\right)\right],
\end{equation}
which vanishes when $K_{N_t}=-K_2$, that is, when $K$ leaves and returns to $t_1$ with the same slope. This is the matched slope of the condition in Eq.~\eqref{eq:bc_periodic} on the grid. On a circle it is therefore not an extra requirement we impose to keep the protocol periodic: it is what keeps the contraction exact.

The regularizer itself follows Appendix~\ref{app:reg:discrete} unchanged, through Eq.~\eqref{eq:app_discrete_regularizer}. Because the fourth-order form of Eq.~\eqref{eq:app_form4} involves a second derivative, we solve its BVP on the extended grid that carries the endpoint twice, $t_1,\dots,t_{N_t},t_{N_t+1}{=}T$, so that both ends of the period are available. On this extended time grid of $N=N_t+1$ points, the weight becomes 
\begin{equation}
\widetilde M=\Delta t\,\mathrm{diag}\left(\tfrac12,1,\dots,1,\tfrac12\right),
\label{eq:app_periodic_Mt}
\end{equation}
and the first derivative becomes 
\begin{equation}
(\widetilde D\boldsymbol a)_j=
\begin{cases}
\dfrac{a_2-a_1}{\Delta t}, & j=1,\\[2ex]
\dfrac{a_{j+1}-a_{j-1}}{2\Delta t}, & 1<j<N,\\[2ex]
\dfrac{a_N-a_{N-1}}{\Delta t}, & j=N.
\end{cases}
\label{eq:app_periodic_D}
\end{equation}
The second derivative is 
\begin{equation}
(\widetilde D_2\boldsymbol a)_j=\frac{a_{j+1}-2a_j+a_{j-1}}{\Delta t^2},
\quad j=2,\dots,N-1 ,
\label{eq:app_periodic_D2}
\end{equation}
with the matching weight $\widetilde M_2=\Delta t\,\mathbb 1$ over those points. The fourth-order form of Eq.~\eqref{eq:app_form4} then gives
\begin{equation}
\widetilde\Sigma_4=\widetilde M+2\gamma_4^2\widetilde D^{\mathsf T}\widetilde M\widetilde D+\gamma_4^4\widetilde D_2^{\mathsf T}\widetilde M_2\widetilde D_2.
\label{eq:app_periodic_sigma}
\end{equation}
The matrix $P$ imposes the three relations
\begin{equation}
k_1=0,
\quad
k_{N_t+1}=0,
\quad
k_{N_t}=-k_2,
\label{eq:app_periodic_constraints}
\end{equation}
which gives $r=3$ in the counting of Appendix~\ref{app:reg:discrete}. The free values are $\boldsymbol c=(k_2,\ldots,k_{N_t-1})^{\mathsf T}$, so that
\begin{equation}
P=\begin{pmatrix}
0&0&\cdots&0\\
1&0&\cdots&0\\
&&\mathbb 1&\\
-1&0&\cdots&0\\
0&0&\cdots&0
\end{pmatrix},
\quad
P\boldsymbol c=\begin{pmatrix}0\\ k_2\\ \vdots\\ k_{N_t-1}\\ -k_2\\ 0\end{pmatrix},
\label{eq:app_P_periodic}
\end{equation}
where the $-1$ in row $N_t$ is the matched slope and $\mathbb 1$ is the $(N_t-3)\times(N_t-3)$ identity. 

\section{Haldane-model implementation and diagnostics}
\label{app:haldane}

\subsection{Lattice and Bloch Hamiltonians}

The three vectors from an $A$ site to its nearest neighbors are
\begin{equation}
\boldsymbol\delta_1=(1,0),\quad
\boldsymbol\delta_2=\left(-\tfrac12,\tfrac{\sqrt3}{2}\right),\quad
\boldsymbol\delta_3=\left(-\tfrac12,-\tfrac{\sqrt3}{2}\right),
\end{equation}
in units of the bond length, and the three next-nearest-neighbor vectors are
\begin{equation}
\mathbf b_1=(0,\sqrt3),\quad
\mathbf b_2=\left(-\tfrac32,-\tfrac{\sqrt3}{2}\right),\quad
\mathbf b_3=\left(\tfrac32,-\tfrac{\sqrt3}{2}\right).
\end{equation}
The term $c^\dagger_ic_{i'}$ with $\mathbf r_{i'}-\mathbf r_i=+\mathbf b_\alpha$ carries the amplitude $J'$ on sublattice $A$ and $J'^{*}$ on sublattice $B$. The three $\mathbf b_\alpha$ run anticlockwise around the $A$ triangle and clockwise around the $B$ triangle, so the two triangles enclose the fluxes $-3\arg J'$ and $+3\arg J'$. The Dirac points are $\mathbf K=\left(0,\tfrac{4\pi}{3\sqrt3}\right)$ and $\mathbf K'=-\mathbf K$. Plugging these vectors into Eq.~\eqref{eq:haldane_seed} gives Eq.~\eqref{eq:haldane_bloch_seed} up to a term proportional to $2 \mathrm{Re}(J')\sum_\alpha\cos(\mathbf k\cdot\mathbf b_\alpha)$ times the identity. Both targets of this work have purely imaginary $J'$, so the additional term vanishes and Eq.~\eqref{eq:haldane_bloch_seed} is exact. Since $f(\mathbf K)=f(\mathbf K')=0$, the Bloch Hamiltonian at the two Dirac points is only captured by the mass, equal to $f_z(\mathbf K)=M-3\sqrt3 \mathrm{Im}(J')$ and $f_z(\mathbf K')=M+3\sqrt3 \mathrm{Im}(J')$. These two masses change sign at
\begin{equation}
\frac{M}{|J'|}=\pm3\sqrt3\sin\left(\arg J'\right),
\end{equation}
which are the phase boundaries of the Haldane model~\cite{Haldane_1988_PRL} shown in Fig.~\ref{fig:haldane_flow}. It is convenient to collect the three schedules of Eq.~\eqref{eq:haldane_bloch_acc} into one complex amplitude,
\begin{equation}
\hacc(\mu,t,\mathbf k)=-\left[F(\mu,t,\mathbf k)\sigma_++\mathrm{h.c.}\right],
\end{equation}
where $F(\mu,t,\mathbf k)=\sum_{\alpha=1}^{3}J_\alpha(\mu,t) e^{i\mathbf k\cdot\boldsymbol\delta_\alpha}$. We write $\hNN(t,\mathbf k)=\hacc(\muf,t,\mathbf k)$ for the drive at the end of the flow.

\subsection{Control algebra and the symmetry}
\label{app:haldane:algebra}

Let $B^c_\alpha$ and $B^s_\alpha$ be the bond operators multiplying the real and the imaginary part of $J_\alpha$, so that together they span the range of $\Pi_\alpha$ in Eq.~\eqref{eq:haldane_regularizer}. The commutators that the flow generates from them are
\begin{align}
i\left[B^c_\alpha,B^c_{\alpha'}\right] & \propto\sin\left(\mathbf k\cdot(\boldsymbol\delta_\alpha-\boldsymbol\delta_{\alpha'})\right)\sigma_z, \label{eq:app_comm_odd}\\
i\left[B^c_\alpha,B^s_\alpha\right] & \propto\sigma_z . \label{eq:app_comm_even}
\end{align}
These operators $B^c_\alpha$ and $B^s_\alpha$ close $\mathfrak{su}(2)$ at each momentum, which is where the $3N_k^2$ components of the flow come from.

At $\mathbf K$, we have 
\begin{equation}
F(\mu,t,\mathbf K)\propto J_1+e^{2\pi i/3}J_2+e^{4\pi i/3}J_3.
\end{equation}
Three equal schedules in $J_{1,2,3}$ make this sum vanish at every time, so the drive gives $\Unn(\muf,\mathbf K)=\Unn(\muf,\mathbf K')=\mathbb 1$, while the target gives $e^{-iTf_z(\mathbf K)\sigma_z}$ and $e^{-iTf_z(\mathbf K')\sigma_z}$, respectively. They agree only if both Dirac masses are multiples of $2\pi/T$, otherwise the three schedules must take different values.

\subsection{Numerical setup and diagnostics}
\label{app:haldane:numerics}

We write two eigenvalues of $\Unn(\muf,\mathbf k)$ as $e^{-iT\varepsilon_\pm(\mathbf k)}$, which defines the quasienergies up to multiples of $2\pi/T$. Chern numbers $C$ are computed for the lower band with the lattice method of Ref.~\cite{Fukui_etal_2005_JPSJ} on a $50\times50$ grid. The Magnus diagnostics of Appendix~\ref{app:haldane:pert} use a $6\times6$ momentum grid after linearly interpolating each converged drive to $2400$ time points per period. Both converged drives have separated quasienergy bands and reproduce the Chern numbers $C=-1$ and $C=0$ of their respective targets. 

Both computations of flow use $JT=0.65$, on a $6\times6$ momentum grid and a periodic time grid of $100$ points, linearly interpolated to $5000$ points when computing the unitary. Both targets have $J'/J=0.4\, i$; the topological one has $M/J=0.4$, so $M/|J'|=1$, and the topologically trivial one has $M/J=2.4$, so $M/|J'|=6$. The two targets thus differ only in the sublattice offset. The regularizer is the periodic fourth-order filter of Appendix~\ref{app:reg:periodic} with $\gamma_4 J=0.1$ for every component and the prefactor of Eq.~\eqref{eq:haldane_regularizer} on the first bond direction. The flow runs from $\mu=0$ to $\muf=10^4$ with the implicit \emph{TRBDF2} solver at relative tolerance $10^{-4}$ and absolute tolerance $10^{-5}$.

\subsection{Smoothness of the converged drives}
\label{app:haldane:harmonics}

Here we ask how many Fourier components are required to implement the accessible controls we obtained in Fig.~\ref{fig:haldane_flow}. The schedules are periodic, and we expand them as
\begin{equation}
J_\alpha(\muf,t)=\sum_{m}J_{\alpha,m}e^{im\omega t},\quad \omega=\frac{2\pi}{T},
\label{eq:app_harmonics}
\end{equation}
where the index $m$ counts harmonics contained in the drive. Keeping the terms with $|m|\le \nc$ defines a band-limited drive with cutoff $\nc$.

\begin{figure}[t]
\centering
\includegraphics{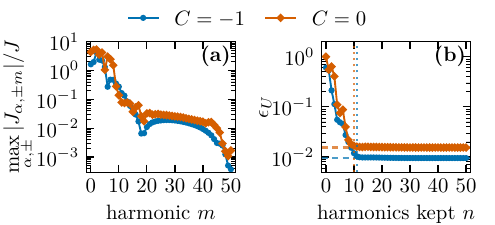}
\caption{
\textbf{Required harmonic components of the two converged Haldane drives.}
\textbf{(a)} Largest amplitude $|J_{\alpha,\pm m}|/J$ of the harmonic $m$ in Eq.~\eqref{eq:app_harmonics}. Most of the contributing modes lie below $m\approx20$.
\textbf{(b)} Unitary error $\epsU(\muf)$ of Eq.~\eqref{eq:haldane_epsU} for the converged drive truncated to harmonics $|m|\le \nc$. Dashed horizontal lines are the errors from the converged flow solutions, and the dotted vertical lines, at $\nc=11$ and $\nc=10$ for the topological and trivial targets, indicate the cutoffs defined in the text. See Appendix~\ref{app:haldane:harmonics} for details.
}
\label{fig:harmonic_content}
\end{figure}

The amplitudes in Fig.~\ref{fig:harmonic_content}(a), maximized over the three directions $\alpha$ and both signs of $m$, fall off quickly and drop by more than two orders of magnitude by $m\approx20$. To see whether it's enough to keep only the harmonics within $|m|\le \nc$, we truncate the converged drive and recompute the error $\epsU(\muf)$ of Eq.~\eqref{eq:haldane_epsU}. Fig.~\ref{fig:harmonic_content}(b) shows a sudden drop followed by a plateau. We define the cutoff as the smallest $\nc$ whose error lies within $10\%$ of the best truncation, $\min_{\nc}\epsU(\nc)$, which gives $\nc=11$ for the topological target and $\nc=10$ for the trivial one. Both cutoffs already match the accuracy of the untruncated flow solution, so the harmonics they discard are physically innocuous.

\subsection{Failure of the perturbative Magnus description}
\label{app:haldane:pert}

In a perturbative approach, one writes the Floquet unitary as a single exponential and expands the exponent,
\begin{equation}
\Unn(\muf,\mathbf k)=\exp\left[\sum_{n\ge1}\Omega_n(\mathbf k)\right],
\end{equation}
where the Magnus term $\Omega_n$ is an $n$-fold time integral of $n-1$ nested commutators of $\hNN$~\cite{Blanes_etal_2009_PhysRep}. The first two are
\begin{align}
\Omega_1&=-i\int_0^T\mathrm dt_1 \hNN(t_1),\\
\Omega_2&=-\frac12\int_0^T\mathrm dt_1\int_0^{t_1}\mathrm dt_2\left[\hNN(t_1),\hNN(t_2)\right],
\label{eq:app_magnus}
\end{align}
where we drop the momentum label and compute the series up to $\Omega_{32}$ by the recursion of Ref.~\cite{Blanes_etal_2009_PhysRep}. Keeping the first $n$ terms defines
\begin{equation}
H^{(n)}=\frac{i}{T}\sum_{m\le n}\Omega_m ,
\end{equation}
the static Hamiltonian from a perturbative expansion of order $n$. The first one, $H^{(1)}$, is the time average of the drive. The second adds the leading correction from the micromotion, and the full sum would be the Floquet Hamiltonian. These are what the inverse-frequency expansions of Refs.~\cite{Bukov_2015_AdvPhys,Eckardt_Anisimovas_2015_NJP,Mikami_etal_2016_PRB} compute, and a perturbative version of our result would mean that $H^{(n)}$ equals $\hseed$ for some small $n$.

Two features of Eq.~\eqref{eq:app_magnus} hold for whatever the target. First, $\hNN$ is a combination of $\sigma_x$ and $\sigma_y$. This means that $H^{(1)}$ has no $\sigma_z$ part. Hence, it cannot reproduce the Dirac masses $f_z(\mathbf K)$ and $f_z(\mathbf K')$ of the target, which are pure $\sigma_z$ terms. Moreover, its Bloch vector stays in the $\sigma_x$--$\sigma_y$ plane and cannot wrap the Bloch sphere, so its lower band has zero Chern number whenever $H^{(1)}$ is gapped. First order can therefore never reproduce the topology of the target. Second, the commutator in $\Omega_2$ is proportional to $\sigma_z$, so $\Omega_2$ is a mass term. The mass of the target is built by $\Omega_2$, and it disappears unless $\hNN$ changes its direction inside the $\sigma_x$--$\sigma_y$ plane, which is what the winding phases of Fig.~\ref{fig:haldane_flow}(a1,a2) do.

\begin{figure}[t]
\centering
\includegraphics{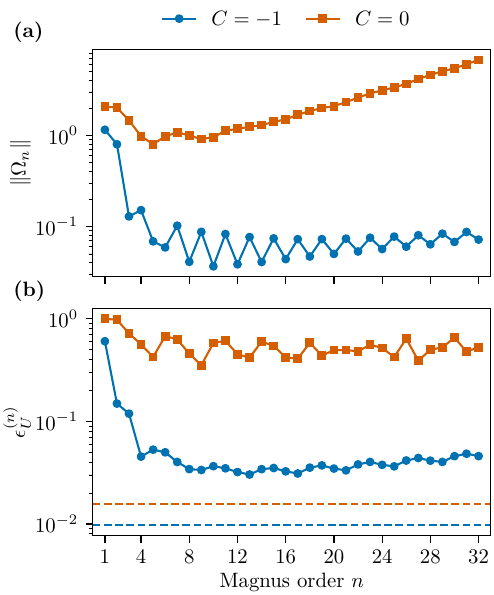}
\caption{
\textbf{Magnus expansion of the two converged drives.} \textbf{(a)} Norm $\lVert\Omega_n\rVert$ against the order $n$. After an initial decrease, the norms grow for the trivial target and stop decreasing for the topological one. \textbf{(b)} One-period error $\epsU^{(n)}$ of the truncated description, Eq.~\eqref{eq:app_magnus_err}; the dashed lines are the errors $\epsU(\muf)$ of the converged flow solutions (upper: trivial, lower: topological). The truncated errors remain above the corresponding flow errors at every order shown. See Appendix~\ref{app:haldane:pert} for details.}
\label{fig:magnus}
\end{figure}

To figure out whether the expansion actually describes our drives, we ask two questions. The first is the size $\lVert\Omega_n\rVert$ for increasing $n$, which tells us whether each term becomes smaller as $n$ increases. The second is the one-period error of the truncated description,
\begin{equation}
\epsU^{(n)}=\epsU\left[e^{-iTH^{(n)}}\right],
\label{eq:app_magnus_err}
\end{equation}
which is the error of Eq.~\eqref{eq:haldane_epsU} with $\Unn(\muf,\mathbf k)$ replaced by the propagator generated by $H^{(n)}$. Fig.~\ref{fig:magnus} shows the outcome, and the two targets fail the expansion in different ways. For the trivial one the terms reach their smallest norm at fifth order and then grow overall, reaching $\lVert\Omega_{32}\rVert=6.75$; the truncated error never falls below $0.35$, about $23$ times the error the flow reaches. For the topological target the terms first decrease, then fluctuate and turn upward at higher orders; the smallest truncated error is $0.030$ at order $13$, about three times the flow error, and it rises again by order $32$. Through the orders computed, neither perturbative description reaches the drive that the flow constructs.

\section{Ising-chain formulation and numerics} 
\label{app:ising}

\subsection{Jordan--Wigner decomposition}

The Jordan--Wigner transformation maps $\HTFIM(t)$ onto free fermions. Periodic boundary conditions split the Hilbert space into the two eigenspaces of the fermion parity $\prod_jZ_j=\pm1$, known as the Neveu--Schwarz ($+$) and Ramond ($-$) sectors, whose momentum grids are
\begin{equation}
{\rm BZ}^{-}=\left\{-\pi+\tfrac{2\pi m}{L}\right\},
\quad
{\rm BZ}^{+}=\left\{-\pi+\tfrac{\pi}{L}+\tfrac{2\pi m}{L}\right\},
\end{equation}
with $m=0,\dots,L-1$ and the lattice spacing set to unity. Each momentum contributes a two-level block,
\begin{align}
H_k(t)&=2\left[h(t)-J\cos k\right]\tau_z+2J\sin k \tau_x,\label{eq:app_hk}\\
\Delta_k(t)&=4\sqrt{\left[h(t)-J\cos k\right]^2+J^2\sin^2k},
\end{align}
where $\tau_x$, $\tau_y$, and $\tau_z$ are Pauli matrices in the two-dimensional space of the pair $\pm k$. The two eigenvalues of $H_k$ are $\pm\Delta_k/2$, so the level gap is $\Delta_k$, while the single-quasiparticle energy is $\Delta_k/2$. For an even chain in the Neveu--Schwarz sector the independent momenta are $k=(2m+1)\pi/L$ with $m=0,\dots,L/2-1$.

\subsection{Exact adiabatic gauge potential}
\label{app:ising:agp}

The exact gauge potential of the field is defined by its matrix elements in the instantaneous eigenbasis~\cite{Kolodrubetz_etal_2017_PhysRep},
\begin{equation}
\agp_h=i\sum_{m\neq n}\frac{|m\rangle\langle m|\partial_h H|n\rangle\langle n|}{E_n-E_m}.
\label{eq:app_agp_matrix}
\end{equation}
Within a momentum block $\partial_hH_k=2\tau_z$, and Eq.~\eqref{eq:app_agp_matrix} gives
\begin{equation}
\agp_{h,k}(t)=-\frac{8J\sin k}{\Delta_k(t)^2} \tau_y,
\label{eq:app_agp_k}
\end{equation}
so the gauge potential is purely off-diagonal in each block and has no accessible component. Transforming back to the spin chain, the weights of Eq.~\eqref{eq:app_agp_k} become the Fourier coefficients $\alpha_\ell^{\pm}(t)$ of Eq.~\eqref{eq:tfim_alpha} and the blocks assemble into the strings $A_\ell$ of Eq.~\eqref{eq:tfim_agp_string}, which reproduces Eq.~\eqref{eq:tfim_agp_sectors}. Every string has the same Frobenius norm,
\begin{equation}
\Fnorm{A_\ell}=\sqrt{2L 2^L}=\sqrt2 \mathcal N_L,
\end{equation}
for $L\ge3$ and independent of its range $\ell$, so a gauge potential with comparable $\alpha_\ell$ across $\ell$ is genuinely spread over the whole chain.

\subsection{Transverse-field schedule and protocol duration}
\label{app:ising:schedule}

We choose the schedule $h(t)$ so that the seed already meets the endpoint conditions the flow then preserves. The gauge-potential term of $\HCD$ has magnitude $|\dot h(t)|\Fnorm{\agp_{h(t)}}$, and we fix $h(t)$ by requiring that this follow a smooth window,
\begin{align}
|\dot h(t)| \Fnorm{\agp_{h(t)}}&=\frac{2\mathcal S(h_{\rm f})}{T}\sin^2 \left(\frac{\pi t}{T}\right),\label{eq:app_tfim_window}\\
\mathcal S(h)&=\int_{h_{\rm i}}^{h}\Fnorm{\agp_{h'}} \mathrm dh', \label{eq:app_tfim_arc}
\end{align}
The prefactor is fixed by the endpoint $\mathcal S(h(T))=\mathcal S(h_{\rm f})$ together with $\int_0^T\sin^2(\pi t/T)\mathrm dt=T/2$. Eq.~\eqref{eq:app_tfim_window} is a statement about $\mathcal S$, as its left-hand side is $\tfrac{\mathrm d}{\mathrm dt}\mathcal S(h(t))$ by the chain rule, so the condition integrates in closed form. Writing
\begin{equation}
Q(t)=\frac{t}{2}-\frac{T}{4\pi}\sin\!\left(\frac{2\pi t}{T}\right),
\quad
Q(T)=\frac{T}{2},
\label{eq:app_tfim_window_primitive}
\end{equation}
we have $Q'(t)=\sin^2(\pi t/T)$, integration from $0$ to $t$ gives $\mathcal S(h(t))=\mathcal S(h_{\rm f})Q(t)/Q(T)$, so that
\begin{equation}
h(t)=\mathcal S^{-1} \left(\mathcal S(h_{\rm f}) \frac{Q(t)}{Q(T)}\right).
\label{eq:app_tfim_schedule}
\end{equation}
In practice we evaluate $\mathcal S$ on a grid of $2001$ field values and invert it by linear interpolation.

Two properties follow. Since $\sin^2(\pi t/T)$ and its derivative both vanish at the two ends, $\dot h$ and $\ddot h$ vanish there as well, so every coefficient $\dot h(t)\alpha_\ell(t)$ of Eq.~\eqref{eq:tfim_agp_sectors} turns on and off smoothly, and the accessible seed schedules start and end with vanishing slope, consistent with the boundary conditions of Eq.~\eqref{eq:FF_BC}. And the field moves slowly wherever $\Fnorm{\agp_h}$ is large, in particular near the critical point $h=J$, which keeps the counterdiabatic amplitude bounded along the whole ramp. 

The duration is not chosen directly either. For each chain length we fix a target endpoint infidelity of the bare protocol $\HTFIM(t)$ alone and bisect $T$ until it is reached, which places protocols of different length on a common footing. The two calculations of the main text use $1-\Facc(0)=0.7$ at $L=4$ and $0.2$ at $L=50$, giving $JT=0.318488$ and $43.789062$, respectively; the component flow of Fig.~\ref{fig:tfim_cartoon} is the $L=50$ calculation.

\subsection{Closed operator algebra of the accessible controls}
\label{app:ising:basis}

The commutators of $\sum_jX_jX_{j+1}$ and $\sum_jZ_j$ close on the algebra $\mathfrak g$ of Eq.~\eqref{eq:tfim_algebra}, spanned by the three families
\begin{align}
\mathcal X_\ell&=\sum_{j=1}^{L}\left(\prod_{j<m<j+\ell}Z_m\right)X_jX_{j+\ell},\nonumber\\
\mathcal Y_\ell&=\sum_{j=1}^{L}\left(\prod_{j<m<j+\ell}Z_m\right)Y_jY_{j+\ell},
\end{align}
and the $A_\ell$ of Eq.~\eqref{eq:tfim_agp_string}, with $\ell=1,\dots,L-1$, together with the transverse field $\sum_jZ_j$ and its parity partner $\left(\prod_jZ_j\right)\sum_jZ_j$. This gives $3(L-1)+2=3L-1$ operators. The accessible sector is $\mathfrak c=\text{span}\{\mathcal X_1,\sum_jZ_j\}$, since $\mathcal X_1$ is the Ising interaction, the gauge-potential sector is $\mathfrak a=\text{span}\{A_\ell\}$, and the remaining $2L-2$ operators form $\mathfrak e=\text{span}\{E_r\}$, ordered by range as
\begin{align}
E_{2\ell-1}&=\mathcal X_{\ell+1},\quad \ell=1,\dots,L-2,\nonumber\\
E_{2\ell}&=\mathcal Y_\ell,\quad \ell=1,\dots,L-1,\nonumber\\
E_{2L-3}&=\left(\prod_jZ_j\right)\sum_jZ_j.
\end{align}
It is convenient to extend the $\mathcal X$ family to $\ell=L$ by identifying 
\begin{equation}
\mathcal X_L\equiv\left(\prod_jZ_j\right)\sum_jZ_j .
\label{eq:app_XL}
\end{equation}
Repeated commutators of the accessible pair $\mathcal X_1$ and $\sum_jZ_j$ then generate the whole set through
\begin{align}
i\left[\sum_jZ_j,\mathcal X_\ell\right] &= -2A_\ell, &&\ell=1,\dots,L-1,\nonumber\\
i\left[\mathcal X_1, A_\ell\right] &= -4\mathcal X_{\ell+1}+4\mathcal Y_{\ell-1}, &&\ell=1,\dots,L-1,
\label{eq:app_ising_ladder}
\end{align}
with the convention $\mathcal Y_0=-\sum_jZ_j$. The first relation produces the gauge-potential string of a given range, and the second raises the range by one, so that together they reach every $A_\ell$, every $\mathcal X_\ell$ up to $\mathcal X_L$, and every $\mathcal Y_\ell$ up to $\mathcal Y_{L-2}$. The last operator follows from
\begin{equation}
i\left[\mathcal X_L,A_1\right]=4\mathcal X_{L-1}-4\mathcal Y_{L-1} ,
\label{eq:app_ising_close}
\end{equation}
which completes $\mathfrak g$. 

The $E_\ell$ with increasing $\ell$ is the ordering of the bands in Fig.~\ref{fig:tfim_cartoon}(b). We work in the orthonormal basis $B_a=O_a/\Fnorm{O_a}$ built from these operators, so that coefficients measured in units of $\mathcal N_L$ give the physical magnitudes of the corresponding controls.

\subsection{Component flow equations}

In this basis a time-dependent control Hamiltonian is a vector of coefficients, $H(\mu,t)=\sum_ax_a(\mu,t)B_a$, and the algebra is encoded in the real structure constants
\begin{equation}
i\left[B_a,B_b\right]=\sum_cf_{abc}B_c.
\end{equation}
Writing $x^\perp_a=x_a$ for $a$ outside $\mathfrak c$ and $x^\perp_a=0$ otherwise, so that $x^\perp_a$ are the coefficients of $\Hper$, the canonical generator of Eq.~\eqref{eq:canonical_generator} has coefficients
\begin{equation}
\kappa^{(0)}_c=\sum_{a,b}f_{abc} x_a x^\perp_b+\partial_tx^\perp_c,
\end{equation}
where the sum collapses onto $a\in\mathfrak c$ because $[\Hper,\Hper]=0$. Applying the regularizer gives $\kappa=\Rreg\kappa^{(0)}$, and the flow equation Eq.~\eqref{eq:flow_equation} becomes
\begin{equation}
\partial_\mu x_c=\sum_{a,b}f_{abc} x_a\kappa_b+\partial_t\kappa_c.
\label{eq:app_component_flow}
\end{equation}
The flow starts from the counterdiabatic seed. Ordering the basis as $B_1=\mathcal X_1/\Fnorm{\mathcal X_1}$, $B_2=\sum_jZ_j/\Fnorm{\sum_jZ_j}$ and $B_{2+\ell}=A_\ell/\Fnorm{A_\ell}$, its coefficients at $\mu=0$ are
\begin{align}
x_1&=-J \mathcal N_L, \quad x_2=-h(t) \mathcal N_L,\nonumber \\
x_{2+\ell}&=\dot h(t) \frac{J}{2}\left[\alpha_\ell^{+}(t)+\alpha_\ell^{-}(t)\right]\sqrt2 \mathcal N_L,
\end{align}
with every extra-term coefficient vanishing, since the seed has no component there.

These are $3L-1$ coupled nonlinear partial differential equations in $\mu$ and $t$, $149$ of them at $L=50$, and the only $L$-dependent objects they require are $f_{abc}$ and the time grid. No object of dimension $2^L$ enters, which is what makes the large-chain calculation possible. After the time discretization of Appendix~\ref{app:reg:sbp} they become a stiff system of $(3L-1)N_t$ ordinary differential equations in $\mu$, $29949$ real coefficients at $L=50$ and $N_t=201$, which we integrate with the implicit \emph{TRBDF2} method at relative and absolute tolerances $10^{-4}$ and $10^{-5}$ for $L=4$, and $10^{-2}$ and $10^{-3}$ for $L=50$. We stop the flow once the ratio of the time-integrated weights of the inaccessible and accessible parts of $H(\mu,t)$ drops below $10^{-5}$.

\subsection{Adiabaticity and fidelity diagnostics}
\label{app:ising:diagnostics}

The adiabaticity $\epsad$ of Eq.~\eqref{eq:tfim_adiabaticity} is evaluated from Eq.~\eqref{eq:app_hk}, with $|g_k(t)\rangle$ and $|e_k(t)\rangle$ their instantaneous eigenstates. For even $L$ with periodic boundaries the ground state of $\HTFIM$ lies in the even-parity sector $+$, so both diagnostics are evaluated on its momenta $k=(2m+1)\pi/L$ with $m=0,\dots,L/2-1$. The endpoint fidelity $\Facc(\mu)$ of Eq.~\eqref{eq:tfim_fidelity} is evaluated by evolving the initial ground state under the accessible schedules alone. At small $L$ this is a dense propagation in the spin basis. At large $L$ we instead evolve the Majorana covariance matrix of the initial state under the quadratic form of $\Hacc(\mu,t)$ within each parity sector, which reproduces the dense result on small chains and avoids any $2^L$ object.

\subsection{Comparison with Floquet-engineered counterdiabatic driving}
\label{app:ising:floquet_cd}

An existing approach to accessible counterdiabatic driving uses Floquet engineering~\cite{Claeys_etal_2019_PRL}. Here we apply this construction to the Ising chain and compare it with the \ctrlRG solution presented in the main manuscript.

To formulate the construction in our terminology, we start from the counterdiabatic target
\begin{equation}
U_{\ast}=\mathcal{T}\exp\left[-i\int_0^T\mathrm dt\left(\HTFIM(t)+\dot h(t)\agp_h(t)\right)\right],
\label{eq:app_fe_target}
\end{equation}
where $\agp_h$ is the exact gauge potential of Eq.~\eqref{eq:app_agp_matrix}. This unitary follows the instantaneous eigenstates of $\HTFIM$ exactly, but the term $\dot h\agp_h$ contains the nonlocal strings of Eq.~\eqref{eq:tfim_agp_string}. The Floquet engineering approach tries to replace them by fast modulation of two accessible operators. To this end, one first represents the gauge potential by an odd nested-commutator series, and then engineers that series as the stroboscopic Hamiltonian.

\subsubsection{Variational ansatz}

The first step approximates the gauge potential as
\begin{equation}
\agp_h^{(\ell)} = i\sum_{n=1}^{\ell}\alpha_n\text{ad}_{\HTFIM}^{2n-1} \left(\partial_h\HTFIM\right),
\label{eq:app_fe_variational_agp}
\end{equation}
where $\text{ad}_{H}(O)\equiv[H,O]$, $\ell$ is the number of nested-commutator terms, and the coefficients $\alpha_n(h)$ are evaluated from the action 
\begin{equation}
S_{\ell}=\text{Tr} \left[G_{\ell}^{2}\right],
\quad
G_{\ell}=\partial_h\HTFIM-i\left[\HTFIM,\agp_h^{(\ell)}\right].
\label{eq:app_fe_action}
\end{equation}
To see how it works, let $\HTFIM|m\rangle=E_m|m\rangle$ and define the instantaneous level spacing $\omega_{mn}\equiv E_m-E_n$. Eq.~\eqref{eq:app_fe_variational_agp} gives
\begin{align}
\langle m|\agp_h^{(\ell)}|n\rangle &= iP_{\ell}(\omega_{mn})\langle m|\partial_h\HTFIM|n\rangle,\\
P_{\ell}(\omega) &\equiv \sum_{n=1}^{\ell}\alpha_n\omega^{2n-1}. \label{eq:app_fe_polynomial}
\end{align}
Each commutator with $\HTFIM$ gives one factor of $E_m-E_n$, so $P_{\ell}(\omega)$ is an odd polynomial. For example, $P_2(\omega)=\alpha_1\omega+\alpha_2\omega^3$. The numbers $\{\alpha_1,\ldots,\alpha_{\ell}\}$ are its coefficients and are precisely the variational parameters determined by minimizing $S_{\ell}$. In contrast, Eq.~\eqref{eq:app_agp_matrix} gives the exact prefactor $P_{\rm exact}(\omega)=-1/\omega$.

Up to the diagonal part of $\partial_h\HTFIM$, the action reads 
\begin{align}
S_{\ell}= &\sum_{m\ne n}v_{mn}\left[P_{\ell}(\omega_{mn})+\frac{1}{\omega_{mn}}\right]^2+\text{const.},\nonumber\\
v_{mn}\equiv &\omega_{mn}^2\left|\langle m|\partial_h\HTFIM|n\rangle\right|^2.
\label{eq:app_fe_action_spectral}
\end{align}
By minimizing this action, it chooses $P_{\ell}(\omega_{mn})$ to approach $-1/\omega_{mn}$, with weight $v_{mn}$. The approximate counterdiabatic Hamiltonian 
\begin{equation}
    H_{\mathrm{CD}}^{(\ell)}=\HTFIM+\dot h\agp_h^{(\ell)}
\end{equation}
approaches the exact seed as this fit improves with $\ell$. 

This polynomial representation is the first key ingredient of the construction. The second is that the high-frequency drive below produces another odd function of the same level spacing, now expressed as a Bessel series, whose coefficients can be chosen to reproduce $P_{\ell}(\omega)$ using only accessible controls.

\subsubsection{Accessible control via Floquet engineering}

The second step realizes $H_{\rm CD}^{(\ell)}$ stroboscopically by the accessible controls,
\begin{align}
H_{\rm FE}(t)= &\left[1+\frac{\omega}{\omega_0}\cos(\omega t)\right]\HTFIM(t) \nonumber\\
&+\dot h(t)f_{\ell}(t)\partial_h\HTFIM(t),\label{eq:app_fe_drive}\\
f_{\ell}(t)\equiv &\sum_{n=1}^{\ell}\beta_n(t)\sin\left((2n-1)\omega t\right),
\end{align}
where $\omega$ is the drive frequency, $\omega_0$ is a reference energy scale. This Hamiltonian contains only the Ising interaction and the transverse field, with
\begin{align}
J_{\rm FE}(t)&=J\left[1+\frac{\omega}{\omega_0}\cos(\omega t)\right],\\
h_{\rm FE}(t)&=h(t)\left[1+\frac{\omega}{\omega_0}\cos(\omega t)\right]+\dot h(t)f_{\ell}(t).
\end{align}
At stroboscopic times and in the high-frequency limit, Eq.~\eqref{eq:app_fe_drive} generates an effective Hamiltonian $H_{\mathrm{F}}=\HTFIM+\dot h\agp_{\rm F}$~\cite{Claeys_etal_2019_PRL}, where
\begin{align}
\langle m|\agp_{\rm F}|n\rangle = &iQ_{\ell}(\omega_{mn}) \langle m|\partial_h\HTFIM|n\rangle,\nonumber\\
Q_{\ell}(\epsilon)\equiv &\sum_{q=1}^{\ell}\beta_q \mathcal J_{2q-1} \left(\frac{\epsilon}{\omega_0}\right).
\label{eq:app_fe_bessel}
\end{align}
Here $\mathcal J_q$ is the Bessel function of the first kind. The coefficients $\beta_q$ are chosen so that $Q_{\ell}(\omega_{mn})$ reproduces the polynomial $P_{\ell}(\omega_{mn})$ of Eq.~\eqref{eq:app_fe_polynomial}. Thus, the accessible drive reproduces the stroboscopic evolution of the approximate counterdiabatic Hamiltonian. For level spacings small compared with $\omega_0$, there is 
\begin{equation}
H_{\rm F}=H_{\mathrm{CD}}^{(\ell)}+\mathcal O(\omega_0^{-2})
\label{eq:app_fe_reproduces_cd}
\end{equation}
in the limit $\omega\rightarrow\infty$~\cite{Claeys_etal_2019_PRL}.

\subsubsection{Drive frequency set by the flow solution}

The Hamiltonian of Eq.~\eqref{eq:app_fe_drive} introduces two new frequencies, $\omega$ and $\omega_0$. The drive frequency $\omega$ sets the rate at which the accessible controls are modulated, and is constrained by stroboscopic condition $\omega T=2\pi N$ with integer $N$, while the reference scale $\omega_0$ appears only through the dimensionless amplitude $\omega/\omega_0$ in Eq.~\eqref{eq:app_fe_drive}. This reference scale should be large enough, $\omega_{mn}/\omega_0\ll1$, as required by Eq.~\eqref{eq:app_fe_reproduces_cd}. At the same time, it cannot be too large, since the drive frequency must dominate every other scale, and matching $Q_{\ell}$ to $P_{\ell}$ gives $\beta_1\propto\omega_0$ and $\beta_2\propto\omega_0^3$. 

Because the construction improves as $\omega$ increases, a meaningful comparison needs a frequency fixed by the control problem rather than by hand. We therefore ask how many harmonics are required to implement the accessible schedules we obtained in Fig.~\ref{fig:tfim_history}. Unlike the periodic drives of Appendix~\ref{app:haldane:harmonics}, these schedules are fixed at the two temporal boundaries, so we subtract the straight line $c_\ast(t)$ through the two endpoint values and expand the remainder in the sine basis of the interval,
\begin{equation}
c(\muf,t)=c_\ast(t)+\sum_{m\ge1}c_m\sin\left(\frac{m\pi t}{T}\right),
\quad c=J,h.
\label{eq:app_fe_sine}
\end{equation}
Harmonic $m$ then has the frequency $m\pi/T$. Keeping the terms with $m\le\nc$ defines a band-limited protocol with cutoff $\nc$, and, as in Appendix~\ref{app:haldane:harmonics}, we take $\nc$ to be the smallest cutoff whose endpoint infidelity $1-\Facc(\muf)$ of Eq.~\eqref{eq:tfim_fidelity} lies within $10\%$ of the best truncation.

\begin{figure}[t]
\centering
\includegraphics{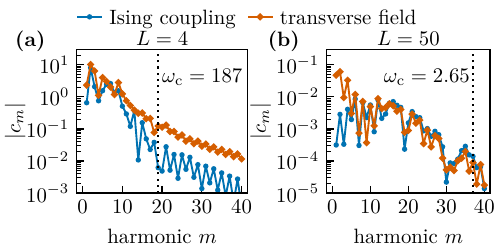}
\caption{\textbf{Required harmonic components of the two converged Ising protocols.} Amplitude $|c_m|$ of the harmonic $m$ in Eq.~\eqref{eq:app_fe_sine}, for the interaction schedule $J(\muf,t)$ and the transverse-field schedule $h(\muf,t)$. \textbf{(a)} The $L=4$ protocol with field ramp $h/J=0.2\to5.0$ and duration $JT=0.32$. \textbf{(b)} The $L=50$ protocol with $h/J=0.7\to1.4$ and $JT=43.79$. The dotted vertical lines indicate the cutoff frequencies defined in Eq.~\eqref{eq:app_fe_omega_cut}.
}
\label{fig:fe_spectrum}
\end{figure}

Fig.~\ref{fig:fe_spectrum} shows the amplitudes for the two Ising protocols of Fig.~\ref{fig:tfim_history}, i.e., the converged schedules of Figs.~\ref{fig:tfim_history}(a1,b1) for $L=4$ and Figs.~\ref{fig:tfim_history}(a2,b2) for $L=50$. The amplitudes fall quickly with increasing $m$, and both schedules are dominated by $m=2$, the protocol frequency $2\pi/T$. The cutoffs are $\nc=19$ at $L=4$ and $\nc=37$ at $L=50$, giving rise to
\begin{equation}
\omega_{\rm c}/J=\nc\pi/(JT)=187\ (L=4),
\quad
\omega_{\rm c}/J=2.65\ (L=50).
\label{eq:app_fe_omega_cut}
\end{equation}
We use $\omega_{\rm c}$ as the drive frequency at which the two constructions are compared.

\begin{figure}[t]
\centering
\includegraphics{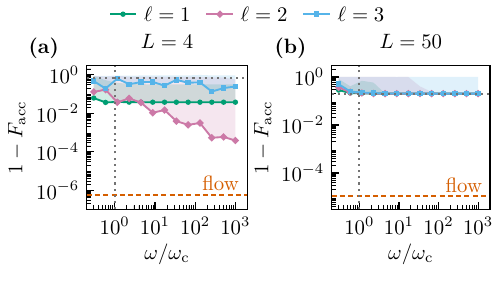}
\caption{\textbf{Floquet-engineered counterdiabatic driving of the two Ising protocols.} Endpoint infidelity $1-\Facc$ of Eq.~\eqref{eq:tfim_fidelity}, evaluated with the accessible schedules $J_{\rm FE}(t)$ and $h_{\rm FE}(t)$ of the drive in Eq.~\eqref{eq:app_fe_drive}, against the drive frequency $\omega$, for \textbf{(a)} $L=4$ and \textbf{(b)} $L=50$ at the orders $\ell=1,2,3$ of the ansatz in Eq.~\eqref{eq:app_fe_variational_agp}. Each curve is the best value over the window of Eq.~\eqref{eq:app_fe_omega0_window} for the reference scale, and the shadow behind it is that window's envelope. The dotted vertical lines indicate $\omega=\omega_{\rm c}$, the dashed horizontal lines give the endpoint infidelity of the converged \ctrlRG protocol, and the dotted horizontal lines that of the bare seed protocol. See Appendix~\ref{app:ising:floquet_cd} for details.
}
\label{fig:fe_frequency}
\end{figure}

\subsubsection{Numerical comparison}

We now apply the construction of Eqs.~\eqref{eq:app_fe_variational_agp} and \eqref{eq:app_fe_drive} to the two Ising protocols of Sec.~\ref{sec:ff_driving}. The coefficients $\alpha_n$ minimize the action $S_{\ell}$ of Eq.~\eqref{eq:app_fe_action}, and the $\beta_n$ are extracted from matching $Q_{\ell}$ to $P_{\ell}$ at small argument. 
The initial state is in the Neveu--Schwarz sector, so the level spacings $\omega_{mn}$ in Eq.~\eqref{eq:app_fe_action_spectral} are 
\begin{equation}
\omega_k=\Delta_k, \quad k=\frac{(2m+1)\pi}{L}, \quad m=0,\dots,\frac{L}{2}-1,
\label{eq:app_fe_spacings}
\end{equation}
with $\Delta_k$ the level gap of Eq.~\eqref{eq:app_hk}. Since $P_{\ell}$ has $\ell$ free coefficients, it can match $-1/\omega_k$ at $\ell$ of these spacings exactly. 
At $L=4$ there are two spacings, so $\ell=2$ already gives all $P_2(\omega_k)=-1/\omega_k$, and $\agp_h^{(2)}$ equals the exact gauge potential. 
At $L=50$ there are $25$ spacings instead, and a low-order ansatz cannot reach them all. Raising $\ell$ does not help either. At the critical point $h=J$, the spacings of Eq.~\eqref{eq:app_fe_spacings} span $\omega_{\max}/\omega_{\min}\approx32$ at $L=50$, so the largest term of $P_{\ell}$ varies by $32^{2\ell-1}$, which is orders of magnitude far beyond the numerical precision. 

The reference scale $\omega_0$ is still free. We scan it over the window
\begin{equation}
\frac{1}{10}\max_k\omega_k\le\omega_0\le4\max_k\omega_k ,
\label{eq:app_fe_omega0_window}
\end{equation}
which gives $\omega_0/J\in[2.30,92.0]$ at $L=4$ and $\omega_0/J\in[0.960,38.4]$ at $L=50$. The lower end keeps the Bessel argument $\omega_{mn}/\omega_0$ within the small-argument regime in which $Q_{\ell}$ was matched to $P_{\ell}$. The upper end is a numerical cutoff, and we observe that extending the window upwards never improves infidelity of the resulting dynamics.

With these choices of frequency, Fig.~\ref{fig:fe_frequency} shows the endpoint infidelity against $\omega/\omega_{\rm c}$. At the cutoff frequency of the flow solution, $\omega=\omega_{\rm c}$, the Floquet-engineered drive never reaches an infidelity as small as the flow does. At $L=4$ it improves with increasing drive frequency, but the infidelity still remains above $10^{-4}$ at a frequency $10^3$ times higher, where the optimal drive amplitude $\omega/\omega_0$ has grown by more than two orders of magnitude. At $L=50$ every $\ell$ saturates at the infidelity of the truncated counterdiabatic seed, limited by the accuracy of $\agp_h^{(\ell)}$.

The Floquet-engineered drive does reproduce $H_{\rm CD}^{(\ell)}$ in the regime where it was derived, namely the large-$\omega$ limit. But at the drive frequency from the flow solution, it remains four to five orders of magnitude less accurate than the flow. The saturation at $L=50$ motivates a comparison in which both constructions start from the same truncated gauge potential.

\begin{figure}[t]
\centering
\includegraphics{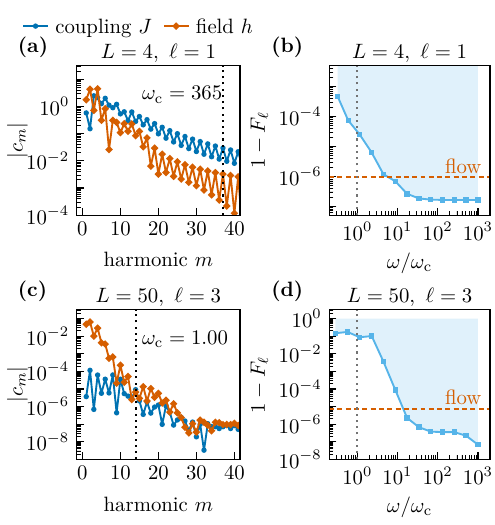}
\caption{\textbf{Floquet-engineered driving and \ctrlRG from the same truncated gauge potential.} Both constructions start from $H_{\rm CD}^{(\ell)}$, for $L=4$, $\ell=1$ (top row) and $L=50$, $\ell=3$ (bottom row). \textbf{(a,c)} Amplitude $|c_m|$ of the harmonic $m$ in Eq.~\eqref{eq:app_fe_sine} for the converged flow schedules $J(\muf,t)$ and $h(\muf,t)$; the dotted vertical lines mark the cutoff $\nc$ and the corresponding $\omega_{\rm c}$. \textbf{(b,d)} Infidelity $1-F_\ell$ from the Floquet-engineered drive in Eq.~\eqref{eq:app_fe_drive} against the drive frequency  $\omega$. The shaded region corresponds to the reference scale from the window of Eq.~\eqref{eq:app_fe_omega0_window}, with the minimum infidelity over that range in square markers. The dashed horizontal lines give $1-F_\ell$ of the converged flow, and the dotted vertical lines indicate $\omega=\omega_{\rm c}$. See Appendix~\ref{app:ising:floquet_cd} for details.
}
\label{fig:fe_truncated_seed}
\end{figure}

To this end, we also initiate the flow with the truncated counterdiabatic Hamiltonian, $\Hseed(t)=H_{\rm CD}^{(\ell)}(t)$. Both constructions now wish to achieve the state
\begin{equation}
|\psi_\ell\rangle = \mathcal{T} e^{-i\int_0^T\mathrm dt H_{\rm CD}^{(\ell)}(t)}|\psi_{\rm i}\rangle ,
\label{eq:app_fe_seed_state}
\end{equation}
and we measure each accessible drive by the fidelity $F_\ell=|\langle\psi_\ell|\psi(T)\rangle|^2$ of its final state $|\psi(T)\rangle$. The flow uses the regularizer, time grid, and solver tolerances used in the main text. We define $\omega_{\rm c}$ as in Eq.~\eqref{eq:app_fe_omega_cut}, using $1-F_\ell$ to select the harmonic cutoff. 

The truncated seed has ground-state infidelity $1-|\langle\psi_{\rm f}|\psi_\ell\rangle|^2=0.317$ for $L=4$ at $\ell=1$, and ground-state infidelity $0.2033$ for $L=50$ at $\ell=3$. This is slightly worse than the bare protocol, $0.2011$, so the truncated AGP at $\ell=3$ does not improve on the bare ramp, which explains the saturation in Fig.~\ref{fig:fe_frequency}(b).

Fig.~\ref{fig:fe_truncated_seed} compares the Floquet engineering construction with the \ctrlRG approach for $L=4$, $\ell=1$ and $L=50$, $\ell=3$. Near the cutoff frequencies of the converged flow schedules [Fig.~\ref{fig:fe_truncated_seed}(a,c)], the Floquet-engineered drive has higher infidelity for both protocols [Fig.~\ref{fig:fe_truncated_seed}(b,d)]. These Floquet-engineered drives reach the infidelity of the flow schedules only when the drive frequency exceeds the flow cutoff by about an order of magnitude.

\end{document}